\documentclass[twocolumn,aps,prc,floatfix,superscriptaddress]{revtex4-2}

\usepackage{graphicx}
\usepackage{dcolumn}
\usepackage{bm}
\usepackage{hyperref}
\usepackage{amsmath,bm}
\usepackage{amssymb}
\usepackage{longtable}
\usepackage{placeins}
\usepackage[mathlines]{lineno}
\usepackage{xcolor}

\newcommand{\mspl}{m_{\mathrm{n}\text{-}\mathrm{p}}^{\ast}}

\begin{document}

\title{Extended Skyrme effective interactions for transport model and neutron stars}

\author{Si-Pei Wang}
\affiliation{%
 School of Physics and Astronomy, Shanghai Key Laboratory for
Particle Physics and Cosmology, and Key Laboratory for Particle Astrophysics and Cosmology (MOE),
Shanghai Jiao Tong University, Shanghai 200240, China
}%

\author{Rui Wang}
\affiliation{%
INFN Sezione di Catania, 95123 Catania, Italy
}%

\author{Jun-Ting Ye}
\author{Lie-Wen Chen}
 \email{Correspinding author:lwchen@sjtu.edu.cn}
\affiliation{%
 School of Physics and Astronomy, Shanghai Key Laboratory for
Particle Physics and Cosmology, and Key Laboratory for Particle Astrophysics and Cosmology (MOE),
Shanghai Jiao Tong University, Shanghai 200240, China
}%

\date{\today}

\begin{abstract}
Interpreting the data of nuclear experiments and astrophysical observations requires advanced theoretical model. At this point, it is of particular importance to develop a unified theoretical framework to describe these experiments and observations based on the same effective nuclear interactions.
Based on the so-called Skyrme pseudopotential up to next-to-next-to-next-to-leading order, we construct a series of extended Skyrme interactions by modifying the density-dependent term and fitting the empirical nucleon optical potential up to above $1$ GeV, the empirical properties of isospin symmetric nuclear matter, the microscopic calculations of pure neutron matter and the properties of neutron stars from astrophysical observations.
The modification of
the density-dependent term in the extended Skyrme interactions follows the idea of Fermi momentum expansion and this leads to a highly flexible density behavior of the symmetry energy. In particular, the values of the density slope parameter $L$ of the symmetry energy for the new extended Skyrme interactions range from $L = -5$ MeV to $L = 125$ MeV by construction, to cover the large uncertainty of the density dependence of the symmetry energy.
Furthermore,
in order to consider the effects of isoscalar and isovector nucleon effective masses, we adjust the momentum dependence of the single-nucleon optical potential and the symmetry potential of these new extended Skyrme interactions and construct a parameter set family, by which we systematically study the impacts of the symmetry energy and the nucleon effective masses on the properties of nuclear matter and neutron stars.
The new extended Skyrme interactions
constructed in the present work will be useful to determine the equation of state of isospin asymmetric nuclear matter, especially the symmetry energy, as well as the nucleon effective masses and their isospin splitting, in transport model simulations for heavy-ion collisions, nuclear structure calculations and neutron star studies.
\end{abstract}

\maketitle

\section{INTRODUCTION}
Understanding the in-medium effective nuclear interactions is one of fundamental questions in nuclear physics.
The equation of state (EOS) of isospin asymmetric nuclear matter is an intuitive manifestation of the effective interactions, and is strongly connected to many important issues in various systems and processes in nuclear physics and
astrophysics~\cite{Li:1997px,Danielewicz:2002pu,Lattimer:2004pg,Baran:2004ih,Steiner:2004fi,Lattimer:2006xb,Li:2008gp,Trautmann:2012nk,Horowitz:2014bja,Li:2014oda,Wang:2014mra,Hebeler:2015hla,Baldo:2016jhp,Lattimer:2000kb,Oertel:2016bki,Li:2017nna,Li:2018lpy,Annala:2017llu,Dietrich:2020eud,Yasin:2018ckc,Sumiyoshi:2022uoj}, e.g., the properties of the nuclei close to the drip lines, the r-process nucleosynthesis in different astrophysical sites, the heavy-ion collisions~(HICs) induced by neutron-rich nuclei, the structure of neutron stars, the evolution in binaries and binary mergers, and the core-collapse supernovae dynamics.
For symmetric nuclear matter~(SNM) with same fraction of neutrons and protons, its EOS around the saturation density $\rho_0$ has been well constrained from the isoscalar giant monopole resonance of finite nuclei \cite{Youngblood:1999zza,Li:2007bp,Patel:2013uyt,Garg:2018uam}, and its EOS at suprasaturation densities up to approximately $5\rho_0$, has also been relatively well constrained by the experimental data on collective flows and kaon production in HICs~\cite{Danielewicz:2002pu,LeFevre:2015paj,Aichelin:1985rbt,Fuchs:2000kp,Hartnack:2005tr}.

While the EOS of SNM has been relatively well constrained, the isospin-dependent part of the EOS of isospin asymmetric nuclear matter, essentially described by the symmetry energy $E_\mathrm{sym}(\rho)$, is still largely uncertain, especially at suprasaturation densities.
Theoretically, based on ab initio chiral nuclear forces or realistic nuclear forces, microscopic many-body calculations, such as many-body perturbation theory~\cite{Tews:2012fj,Wellenhofer:2015qba,Drischler:2020hwi}, the quantum Monte Carlo methods~\cite{Gandolfi:2011xu,Wlazlowski:2014jna,Roggero:2014lga,Tews:2015ufa}, the variational many-body calculations~\cite{Akmal:1998cf}, the Bethe-Bruckner-Goldstone calculations~\cite{Baldo:2014rda} and the self-consistent Green's function approach~\cite{Carbone:2014mja}, have put relatively precise constraints on the EOS of pure neutron matter (PNM) ($E_{\mathrm{PNM}}(\rho)$) up to density $\rho \sim 0.2\, \mathrm{fm}^{-3}$ (see, e.g., Ref.~\cite{Zhang:2022bni}).
The combined constraint on the $E_{\mathrm{PNM}}(\rho)$ from these various microscopic calculations gives strong constraints on the symmetry energy below and around saturation density.
Experimentally,
significant progress on the determination of the symmetry energy at subsaturation densities has been made, mainly by analyzing the experimental data of finite nuclei (where the average density is around $2\rho_0/3$), e.g., the binding energy, charge radius, neutron skin thickness and isovector modes of resonances \cite{Zhang:2013wna,Brown:2013mga,Zhang:2014yfa,Zhang:2015ava,Roca-Maza:2018ujj}.
However, it should be mentioned that strong tension between the data recently reported by PREX-II~\cite{PREX:2021umo} and CREX~\cite{CREX:2022kgg} is observed for the extraction of neutron skin thickness and the symmetry energy within the non-relativistic and relativistic nuclear energy density functionals~\cite{Reed:2021nqk,Reinhard:2022inh,Yuksel:2022umn,Zhang:2022bni}.
In particular,
a large neutron skin in $^{208}\mathrm{Pb}$ obtained from PREX-II~\cite{PREX:2021umo} suggests a very stiff symmetry energy with a very large value for the density slope parameter $L$ of the symmetry energy, while the relatively small neutron skin in $^{48}\mathrm{Ca}$ obtained from CREX~\cite{CREX:2022kgg} suggests a soft symmetry energy with a much smaller value of $L$.
This tension makes the determination of the symmetry energy around saturation density remain elusive since the PREX-II and CREX are free from the strong interaction
uncertainties and thus they are believed to allow us to determine with minimal model
dependence the neutron skin thickness and the density dependence of the symmetry energy.

Although the nuclear structure data and microscopic theoretical calculations can relatively well constrain the symmetry energy at subsaturation densities, they can hardly constrain the high-density behavior of the symmetry energy.
On the other hand, the HICs experiments induced by neutron-rich nuclei at intermediate and high energies provide a possible way to extract information of the symmetry energy at high densities.
HICs experiments
perhaps are the only way in terrestrial laboratories to produce high-density nuclear matter.
Many radioactive beam facilities around the world, e.g., CSR/HIAF in China, SPIRAL2/GANIL in France, FAIR/GSI in Germany, RIBF/RIKEN in Japan, SPES/LNL in Italy, RAON in Korea and FRIB/NSCL in USA, provide a unique experimental tool to produce the neutron-rich radioactive nuclei and study the density dependence of the symmetry energy~\cite{Li:1997px,Danielewicz:2002pu,Baran:2004ih,Li:2008gp,Trautmann:2012nk,Li:2014oda,Li:2017nna}.
In order to
describe the dynamics of heavy-ion collisions and to extract the EOS of the hot and dense nuclear matter produced during the collisions, the microscopic transport models, e.g., the Boltzmann-Uehling-Uhlenbeck (BUU) equation \cite{Bertsch:1988ik} and the quantum molecular dynamics (QMD) model \cite{Aichelin:1991xy}, have been developed and extensively used.
In particular, in recent years, the Transport Model Evaluation Project~(TMEP) has been pursued to test the robustness of transport models and then try to narrow down the uncertainties of their predictions~\cite{TMEP:2016tup,TMEP:2017mex, TMEP:2019yci,TMEP:2021ljz,TMEP:2022xjg,TMEP:2023ifw}.
It should be noted that besides the applications in HICs, the transport models also provide an important approach to study the collective dynamics of finite nuclei such as giant or pygmy resonances \cite{Yildirim:2005ua,Gaitanos:2010fd,Urban:2011vb,Baran:2013gka,Baran:2014kla,Zheng:2016jrf,Kong:2017nil,Wang:2019ghr,Wang:2020ixf,Wang:2020xgk,Song:2021hyw}.
In the transport model (e.g., the BUU equation) simulations for the dynamics of non-equilibrium system, a direct and basic input is the single-nucleon potential (nuclear mean-field potential).
In the mean-field approximation, the single-nucleon potential is connected to the EOS through the corresponding energy-density functional~(EDF).
In previous works~\cite{Carlsson:2008gm,Raimondi:2011pz}, a Skyrme-like quasilocal EDF up to next-to-next-to-next-to-leading order~(N3LO) has been constructed, by including additional higher-order derivative terms (higher-power momentum dependence) in the conventional Skyrme interactions which is notorious with the incorrect high energy behavior of the nucleon optical potential when nucleon kinetic energy is above about 200 MeV/nucleon.
Based on the N3LO Skyrme pseudopotential, the extended Skyrme interactions have been built within the mean-field approximation in Ref.~\cite{Wang:2018yce} to reproduce the empirical results on the nucleon optical potential up to $1$ GeV obtained by Hama \textit{et al.} from analyzing the proton-nucleus eslatic scattering data~\cite{Hama:1990vr,Cooper:1993nx}, and very recently the extended Skyrme interactions have been applied in the lattice BUU transport model~\cite{Wang:2023gta} to successfully describe the FOPI data~\cite{FOPI:2010xrt} on the light-nuclei production in intermediate-energy HICs.

Besides nuclear experiments with finite nuclei and HICs in terrestrial labs, astrophysical observations of neutron stars and their mergers provide another way to extract information of the symmetry energy at high densities.
Neutron stars represents one kind of the densest objects in the universe, and they are regarded as an ideal site to explore the dense matter at high isospin asymmetry.
Indeed, the multimessenger data on the gravitational wave signal GW170817 of the binary neutron star merger detected by the LIGO-Virgo detectors~\cite{LIGOScientific:2018cki}, the discovery of heavy neutron stars with mass larger than two times solar mass by relativistic Shapiro delay measurements~\cite{NANOGrav:2019jur,Fonseca:2021wxt} as well as the X-ray emitted from hot millisecond pulsar detected by the Neutron Star Interior Composition Explorer (NICER) and X-ray Multi-Mirror~(XMM-Newton)~\cite{Miller:2019cac,Riley:2019yda,Miller:2021qha,Riley:2021pdl}, have put crucial constraints on the maximum mass ($M_{\mathrm{TOV}}$), the mass-radius~(M-R) and the tidal deformability ($\Lambda$) of neutron stars, and thus further on the high density behavior of the symmetry energy.
At this point,
we would like to mention that a series of works have been conducted to constrain the symmetry energy simultaneously by using the data of ground-state properties and giant monopole resonances (GMR) of finite nuclei,
the flow data in heavy-ion collisions as well as the multimessenger data on neutron stars and gravitational wave from the binary neutron star merger~\cite{Zhou:2019omw,Zhou:2019sci,Yue:2021yfx} based on a single unified framework of the extended Skyrme-Hartree-Fock~(eSHF) model~\cite{Zhang:2015vaa}.
Recently,
the central compact object (CCO) within the supernova remanant HESS J1731-347 is estimated to have an unusually low mass $M=0.77^{+0.20}_{-0.17}\,M_{\odot}$ and small radius $R=10.4^{+0.86}_{-0.78}\,\mathrm{km}$ from Gaia observations~\cite{2022NatAs...6.1444D}.
Assuming that this object is a neutron star, its mass-radius relation implies a relatively small value of $L$ and predicts a soft symmetry energy up to $2\rho_0$ (approximately corresponding to the center density of $0.77\,M_{\odot}$ neutron star).
The mass-radius relation of the CCO in HESS J1731-347~\cite{2022NatAs...6.1444D} together with the existence of the large mass neutron star PSR J0740+6620~\cite{NANOGrav:2019jur,Fonseca:2021wxt} suggest the presence of a soft symmetry energy at low to intermediate densities but a very stiff symmetry energy at high densities.
Relying solely on the magnitude $E_{\mathrm{sym}}(\rho_0)$ and the slope $L$ of the symmetry energy at saturation density seems to be insufficient to provide such a density dependence of the symmetry energy.
Therefore, higher-order coefficients of the symmetry energy, e.g., the curvature parameter $K_{\mathrm{sym}}$ and skewness parameter $J_{\mathrm{sym}}$ which characterize the high-density behaviors of the symmetry energy, should also be considered.

The above constrains/data obtained from various sources cover a wide range of densities as well as isospin asymmetries.
To probe the nuclear EOS using these constrains/data, it would be important to construct a single unified effective nuclear interaction that can be used to simultaneously describe the finite nuclei, neutron stars and HICs.
The N3LO Skyrme pseudopotential can be used to describe finite nuclei~\cite{Carlsson:2008gm,Raimondi:2011pz} and heavy-ion collisions at energy up to about $1$ GeV/nucleon~\cite{Wang:2018yce,Wang:2023gta}, but it is hard to describe the properties of neutron stars, especially the small mass and radius of the CCO in HESS J1731-347.
In the present work,
we demonstrate that a modification of the density-dependent~(DD) term of the N3LO Skyrme pseudopotential in Refs.~\cite{Carlsson:2008gm,Raimondi:2011pz,Wang:2018yce} will enable the pseudopotential to flexibly describe the finite nuclei, neutron stars and HICs simultaneously.
The DD term is usually introduced to mimic the effects of many-body forces in the non-relativistic models, e.g., the original Skyrme interaction~\cite{Skyrme:1959zz}, the Skyrme-Hartree-Fock calculations~\cite{Vautherin:1971aw} as well as the Gogny-Hartree-Fock calculations~\cite{Decharge:1979fa}.
A lot of efforts have been made to improve the conventional Skyrme interaction by extending the DD term, such as the eSHF model~\cite{Chamel:2009yx,Zhang:2015vaa} and the so-called KIDS model~\cite{Papakonstantinou:2016zpe}, to better describe the finite nuclei and neutron stars.
In the present work,
instead of considering the DD term as in Ref.~\cite{Wang:2018yce} (i.e., $\rho^{\alpha}$) with an adjustable parameter $\alpha$, we express it as three terms: $\rho^{1/3}$, $\rho^{3/3}$ and $\rho^{5/3}$.
Therefore, the resulting EOS can be exactly expressed as a power series in $\rho^{1/3}$ (equivalently the Fermi momentum $p_{F}$), from $\rho^{2/3}$ (kinetic energy contribution) to $\rho^{9/3}$, while the contributions from DD terms and momentum dependent~(MD) terms can be clearly distinguished.
Actually, expressing the EOS as a power series in $p_{F}$ is physically well motivated in the Brueckner theory for the nuclear matter with a realistic nuclear force as well as in the interacting hard-sphere Fermi system and the Galitskii equation~\cite{fetter2012quantum}. Sometimes, it is also considered to be a model-independent parameterization of the nuclear matter EOS~\cite{Patra:2022yqc}.
In such a way,
compared to the N3LO Skyrme pseudopotential in Ref.~\cite{Wang:2018yce},
an additional parameter related to the EOS of SNM and two more parameters related to the density dependence of the symmetry energy can be introduced.
Namely, three higher-order coefficients, i.e., $J_0$, $K_{\mathrm{sym}}$ and $J_{\mathrm{sym}}$, which are associated with the bulk properties of nuclear matter at high densities, can then be adjusted accordingly to more accurately describe the neutron stars.
Based on the Skyrme pseudopotential with the new extended DD terms, we construct eight Skyrme interaction parameter sets with $L$ values of $5$, $15$, $25$, $35$, $45$, $55$, $65$ and $75\,\mathrm{MeV}$, respectively.
These eight parameter sets are obtained by fitting the nucleon optical potential up to energy of $1\,\mathrm{GeV}$, a few selected empirical properties of nuclear matter, the microscopic many-body calculation results for PNM as well as the neutron star observations.
In addition, considering the significant uncertainties in the extraction of $L$, we provide four more parameter sets, for comparison, which give very soft or very stiff symmetry energy around the saturation density, with $L$ values of $-5$, $85$, $105$ and $125\,\mathrm{MeV}$, respectively.

Furthermore, we also explore several particularly interesting quantities, i.e.,
the isoscalar and isovector nucleon effective mass ($m^{\ast}_{s}$ and $m^{\ast}_{v}$)
as well as the neutron-proton effective mass splitting ($\mspl$), which characterize
the momentum dependence of the single-nucleon potential and the symmetry potential.
These nucleon effective masses are fundamentally connected to many interesting issues in both nuclear physics and astrophysics (see Ref.~\cite{Li:2018lpy} and the references therein).
Based on the constructed extended Skyrme interactions, we investigate the effects of nucleon effective masses on the properties of nuclear matter and neutron stars.
As a result,
we construct a family of parameter sets that can be used to simultaneously describe the properties of finite nuclei, neutron stars and HICs.
This framework can be applied to study the effects of the symmetry energy and nucleon effective masses with more transparent ways.
Utilizing these new extended Skyrme interactions,
we demonstrate the following points:
(i) To simultaneously satisfy constraints from microscopic calculations on $E_{\mathrm{PNM}}(\rho)$ and astrophysical observations of neutron stars with masses of $1.4\,M_{\odot}$ and $2.0\,M_{\odot}$, $L$ must lie within the range of $5\,\mathrm{MeV} \leq L \leq 75\,\mathrm{MeV}$;
(ii) A range of $-5\,\mathrm{MeV} \leq L \leq 35\,\mathrm{MeV}$ is necessary to describe the CCO in HESS J1731-347;
(iii) A peak structure of the squared sound speed for neutron star matter arises for interactions with the soft symmetry energy around saturation density, especially when  $-5\,\mathrm{MeV} \leq L \leq 25\,\mathrm{MeV}$.

This paper is organized as follow:
In Sec.~\ref{sec:framework}, we introduce the Skyrme pseudopotential up to N3LO with the extended DD terms, and display the corresponding Hamiltonian density and single-nucleon potential.
In Sec.~\ref{sec:fitting}, we present the fitting strategy as well as the experimental data and constraints used in our fitting, and we give the eight new parameter sets of the extended Skyrme interactions.
The bulk properties of cold nuclear matter, the single-nucleon potential behaviors and neutron stars structures of the eight interactions are presented in Sec.~\ref{sec:properties}.
The interactions with supersoft and superstiff symmetry energy around saturation density are introduced in Sec.~\ref{sec:soft-stiff}.
In Sec.~\ref{sec:EffMasses}, the extended Skyrme interactions with different symmetry energy behaviors are combined with different momentum dependencies (i.e., different nucleon effective masses), and we obtain a parameter set family consisting of 144 parameter sets.
Based on the parameter set family, we systematically explore the impacts of the symmetry energy and nucleon effective masses on the properties of nuclear matter and neutron stars.
Finally, we summarize this work and make a brief outlook in Sec.~\ref{sec:summary}.

For completeness, we include several Appendixes.
In Appendix~\ref{sec:Appendix_Exp},
we present the macroscopic quantities at arbitrary density as linear combinations of the model parameters and give the representation matrix and its inverse.
In Appendix~\ref{sec:Appendix_Exp_rho0},
we provide the representation matrix and its inverse at the saturation density.
Using these matrices, readers can conveniently obtain the corresponding set of parameters based on the values of macroscopic quantities at saturation density, or vice versa.
Expressions for the fourth-order symmetry energy, the kurtosis coefficients (corresponding to the fourth-order density derivative of the SNM EOS and the symmetry energy), the linear isospin splitting coefficient of the nucleon effective mass as well as the isoscalar and isovector nucleon effective masses are detailed in Appendix~\ref{sec:Appendix_Other}, where we also establish the relations between the fourth-order symmetry energy as well as the linear isospin splitting coefficient and the isoscalar and isovector nucleon effective masses.

\section{THEORETICAL FRAMEWORK}
\label{sec:framework}
\subsection{A new extended Skyrme interaction based on N3LO pseudopotential}
Conventionally, we refer to the effective interactions with quasilocal operators depending on spatial derivatives as pseudopotential, and the standard Skyrme interaction could be recognized as pseudopotential up to NLO.
In the previous works \cite{Carlsson:2008gm,Raimondi:2011pz}, nuclear EDFs in terms of derivatives of densities up to sixth order have been constructed and mapped to Skyrme interaction with additional fourth and sixth-order derivative terms.
The expressions of Hamiltonian density and single-nucleon potential of the N3LO Skyrme inteaction have been derived within the Hartree-Fock approximation \cite{Wang:2018yce}.

The full Skyrme pseudopotential generally contains spin-independent, spin-orbit, and tensor components (see, e.g., Refs.~\cite{Carlsson:2008gm,Raimondi:2011pz,Davesne:2014wya,Davesne:2014rva}).
Since we are focusing only on the spin-averaged quantities, we ignore the spin-orbit and tensor components, which have no contribution in our present case for transport model simulation and the study of nuclear matter and neutron stars~\cite{Wang:2018yce}.
The central term of the N3LO Skyrme pseudopotential is written as
\begin{widetext}
\begin{equation}
\label{eq:VN3LO}
\small
\begin{aligned}
V_{\mathrm{N} 3 \mathrm{LO}}^C  =
& t_0\left(1+x_0 \hat{P}_\sigma\right)+t_1^{[2]}\left(1+x_1^{[2]} \hat{P}_\sigma\right) \frac{1}{2}\left(\hat{\vec{k}}^{\prime 2}+\hat{\vec{k}}^2\right)+t_2^{[2]}\left(1+x_2^{[2]} \hat{P}_\sigma\right) \hat{\vec{k}}^{\prime} \cdot \hat{\vec{k}}+t_1^{[4]}\left(1+x_1^{[4]} \hat{P}_\sigma\right)\left[\frac{1}{4}\left(\hat{\vec{k}}^{\prime 2}+\hat{\vec{k}}^2\right)^2+\left(\hat{\vec{k}}^{\prime} \cdot \hat{\vec{k}}\right)^2\right] \\
& +t_2^{[4]}\left(1+x_2^{[4]} \hat{P}_\sigma\right)\left(\hat{\vec{k}}^{\prime} \cdot \hat{\vec{k}} \right)\left(\hat{\vec{k}}^{\prime 2}+\hat{\vec{k}}^2\right)+t_1^{[6]}\left(1+x_1^{[6]} \hat{P}_\sigma\right)\left(\hat{\vec{k}}^{\prime 2}+\hat{\vec{k}}^2\right)\left[\frac{1}{2}\left(\hat{\vec{k}}^{\prime 2}+\hat{\vec{k}}^2\right)^2+6\left(\hat{\vec{k}}^{\prime} \cdot \hat{\vec{k}}\right)^2\right] \\
& +t_2^{[6]}\left(1+x_2^{[6]} \hat{P}_\sigma\right)\left(\hat{\vec{k}}^{\prime} \cdot \hat{\vec{k}}\right)\left[3\left(\hat{\vec{k}}^{\prime 2}+\hat{\vec{k}}^2\right)^2+4\left(\hat{\vec{k}}^{\prime} \cdot \hat{\vec{k}}\right)^2\right] ,
\end{aligned}
\end{equation}
\end{widetext}
where $\hat{P}_\sigma$ is the spin-exchange operator;
$ \hat{\vec{k}}=-i\left( \hat{\vec{\nabla}}_1-\hat{\vec{\nabla}}_2 \right)/2 $
is the relative momentum operator and $ \hat{\vec{k}}^{\prime} $ is the conjugate operator of $ \hat{\vec{k}}$ acting on the left.

Usually, the contribution from many-body force is effectively replaced by a DD term $\frac{1}{6} t_3\left(1+x_3 \hat{P}_\sigma\right) \rho^\alpha(\vec{R})$ \cite{Chabanat:1997qh,Chabanat:1997un}, where $t_3$, $x_3$ and the power index $\alpha$ are determined through a fitting process.
In the present work, in order to impart greater flexibility to the symmetry energy at varying densities, we rewrite the density-dependent term into three terms as follows:
\begin{equation}
\label{eq:Vdd_new}
V^{\mathrm{DD}}_{\mathrm{N} 1 \mathrm{LO}}= \sum_{n=1,3,5} \frac{1}{6} t_{3}^{\left[ n \right] }\left(1+x_{3}^{\left[ n \right] } \hat{P}_\sigma\right) \rho^{n/3}(\vec{R})
\end{equation}
where $ \vec{R} = \left( \vec{r}_1 + \vec{r}_2 \right)/2  $.

For brevity, the factor $\hat{\delta}\left(\vec{r_1}-\vec{r_1} \right)$ is omitted from Eq.~(\ref{eq:VN3LO}) and Eq.~(\ref{eq:Vdd_new}).
In Eq.~(\ref{eq:Vdd_new}), we introduce two additional $t_{3}$-parameters and two additional $x_{3}$-parameters, while eliminating the $\alpha$.
The three extra parameters in Eq.~(\ref{eq:Vdd_new}) can be used to freely adjust $J_0$, $K_{\mathrm{sym}}$ and $J_{\mathrm{sym}}$ of nuclear matter, in contrast to those in Ref.~\cite{Wang:2018yce}.
Here, three indices of density are set to $1/3$, $3/3$ and $5/3$.
As a result, the nuclear matter EOS can be exactly expressed as a power series in $p_F$ (see Eq.~(\ref{eq:EOS})).

The $t_0$, $x_0$; $t_{i}^{[n]}$, $x_{i}^{[n]}$ ($n=2,4,6$ and $i=1,2$); $t_{3}^{[n]}$, $x_{3}^{[n]}$ ($n=1,3,5$) are Skyrme parameters, and the total number of these parameters is 20 for the new Skyrme effective interaction.
The Skyrme interaction used in this work is then written as
\begin{equation}
\label{eq:Vsk}
v_{sk}= V_{\mathrm{N} 3 \mathrm{LO}}^C + V_{\mathrm{N} 1 \mathrm{LO}}^{\mathrm{DD}}.
\end{equation}

\subsection{Hamiltonian density and single-nucleon potential in one-body transport model}
During the heavy-ion collision process, the nucleons are generally far from thermal equilibrium.
In transport models, these nucleons are described by the phase space distribution function (Wigner function) $f_{\tau} \left( \vec{r},\vec{p} \right)$, with $\tau=1$ [or n] for neutrons and $-1$ [or p] for protons.
Therefore, we need to express the single-nucleon potentials $U_{\mathrm{\tau}}\left( \vec{r},\vec{p} \right)$ in terms of $f_{\tau} \left( \vec{r},\vec{p} \right)$, and then the Hamiltonian density $\mathcal{H} \left( \vec{r} \right)$ of the collision system can also be expressed in terms of $f_{\tau} \left( \vec{r},\vec{p} \right)$.
With the Hartree-Fock method, the expression of $U_{\mathrm{\tau}}\left( \vec{r},\vec{p} \right)$ and $\mathcal{H} \left( \vec{r} \right)$ of the N3LO Skyrme pseudopotential have been derived in Ref.~\cite{Wang:2018yce}.
The Hamiltonian density takes the following form:
\begin{equation}
\label{eq:Hdensity}
\mathcal{H} \left( \vec{r} \right) =
\mathcal{H} ^{ \mathrm{kin}  } \left( \vec{r} \right)
+ \mathcal{H}^{ \mathrm{loc}  } \left( \vec{r} \right)
+ \mathcal{H}^{ \mathrm{MD}  } \left( \vec{r} \right)
+ \mathcal{H}^{ \mathrm{grad}  } \left( \vec{r} \right)
+ \mathcal{H}^{ \mathrm{DD}  } \left( \vec{r} \right),
\end{equation}
where $\mathcal{H} ^{ \mathrm{kin}  } \left( \vec{r} \right)$, $\mathcal{H}^{ \mathrm{loc}  } \left( \vec{r} \right)$, $\mathcal{H}^{ \mathrm{MD}  } \left( \vec{r} \right)$, $\mathcal{H}^{ \mathrm{grad}  } \left( \vec{r} \right)$ and $\mathcal{H}^{ \mathrm{DD}  } \left( \vec{r} \right)$ are the kinetic, local, momentum-dependent (MD), gradient and density-dependent (DD) terms, respectively.
The kinetic term and the local term are the same as those in standard Skyrme interaction (see, e.g., Refs.~\cite{Chabanat:1997qh,Chabanat:1997un}), where they are expressed as
\begin{equation}
\mathcal{H} ^{ \mathrm{kin}  } \left( \vec{r} \right)
=\sum_{\tau=n,p} \int d^3 p \frac{p^2}{2 m_{\tau}} f_\tau \left( \vec{r}, \vec{p} \right)
\end{equation}
and
\begin{equation}
\mathcal{H}^{ \mathrm{loc}  } \left( \vec{r} \right)
= \frac{1}{4} t_0 \left[ \left( 2+x_0 \right) \rho^2 - \left( 2 x_0+1 \right) \sum_{\tau=n, p} \rho_\tau^2\right],
\end{equation}
 respectively.
The $\rho_{\tau} \left( \vec{r} \right) = \int f_\tau \left( \vec{r}, \vec{p} \right) d^3 p $ is the nucleon density and the $\rho \left( \vec{r} \right) = \rho_n \left( \vec{r} \right) + \rho_p \left( \vec{r} \right) $ is the total nucleon density.

The MD and gradient term include the contributions from additional derivative terms in Eq.~(\ref{eq:VN3LO}).
Their expressions have been derived in Ref.~\cite{Wang:2018yce}, and we include them here for completeness.
The MD term can be expressed as
\begin{widetext}
\begin{equation}
\begin{aligned}
\mathcal{H}^{\mathrm{MD}}(\vec{r})=
& \frac{C^{[2]}}{16 \hbar^2} \mathcal{H}^{\mathrm{md}[2]}(\vec{r})+\frac{D^{[2]}}{16 \hbar^2} \sum_{\tau=n, p} \mathcal{H}_\tau^{\mathrm{md}[2]}(\vec{r})
 +\frac{C^{[4]}}{32 \hbar^4} \mathcal{H}^{\mathrm{md}[4]}(\vec{r})+\frac{D^{[4]}}{32 \hbar^4} \sum_{\tau=n, p} \mathcal{H}_\tau^{\mathrm{md}[4]}(\vec{r}) \\
& +\frac{C^{[6]}}{16 \hbar^6} \mathcal{H}^{\mathrm{md}[6]}(\vec{r})+\frac{D^{[6]}}{16 \hbar^6} \sum_{\tau=n, p} \mathcal{H}_\tau^{\mathrm{md}[4]}(\vec{r}),
\end{aligned}
\end{equation}
where $\mathcal{H}^{\mathrm{md}\left[ n \right]}(\vec{r})$ and $\mathcal{H}_{\tau}^{\mathrm{md}\left[ n \right]}(\vec{r})$ are defined as
\begin{align}
\mathcal{H}^{\mathrm{md}[n]}(\vec{r})&= \int d^3 p d^3 p^{\prime}\left(\vec{p}-\vec{p}^{\, \prime}\right)^n f(\vec{r}, \vec{p}) f\left(\vec{r}, \vec{p}^{\, \prime}\right), \\
\mathcal{H}_{\tau}^{\mathrm{md}[n]}(\vec{r})&= \int d^3 p d^3 p^{\prime}\left(\vec{p}-\vec{p}^{\, \prime}\right)^n f_{\tau}(\vec{r}, \vec{p}) f_{\tau}\left(\vec{r}, \vec{p}^{\, \prime}\right),
\end{align}
with $f(\vec{r}, \vec{p})=f_{n}(\vec{r}, \vec{p})+f_{p}(\vec{r}, \vec{p})$.
The gradient term takes the form as follows:
\begin{equation}
\label{eq:Hgrad}
\begin{aligned}
\mathcal{H}^{\mathrm{grad}}(\vec{r})=
& \frac{1}{16} E^{[2]}\left\{2 \rho(\vec{r}) \nabla^2 \rho(\vec{r})-2[\nabla \rho(\vec{r})]^2\right\}+\frac{1}{16} F^{[2]} \sum_{\tau=n, p}\left\{2 \rho_\tau(\vec{r}) \nabla^2 \rho_\tau(\vec{r})-2\left[\nabla \rho_\tau(\vec{r})\right]^2\right\} \\
& +\frac{1}{32} E^{[4]}\left\{2 \rho(\vec{r}) \nabla^4 \rho(\vec{r})-8 \nabla \rho(\vec{r}) \nabla^3 \rho(\vec{r})+6\left[\nabla^2 \rho(\vec{r})\right]^2\right\} \\
& +\frac{1}{32} F^{[4]} \sum_{\tau=n, p}\left\{2 \rho_\tau(\vec{r}) \nabla^4 \rho_\tau(\vec{r})-8 \nabla \rho_\tau(\vec{r}) \nabla^3 \rho_\tau(\vec{r})+6\left[\nabla^2 \rho_\tau(\vec{r})\right]^2\right\} \\
& +\frac{1}{16} E^{[6]}\left\{2 \rho(\vec{r}) \nabla^6 \rho(\vec{r})-12 \nabla \rho(\vec{r}) \nabla^5 \rho(\vec{r})+30 \nabla^2 \rho(\vec{r}) \nabla^4 \rho(\vec{r})-20\left[\nabla^3 \rho(\vec{r})\right]^2\right\} \\
& +\frac{1}{16} F^{[6]} \sum_{\tau=n, p}\left\{2 \rho_\tau(\vec{r}) \nabla^6 \rho_\tau(\vec{r})-12 \nabla \rho_\tau(\vec{r}) \nabla^5 \rho_\tau(\vec{r})+30 \nabla^2 \rho_\tau(\vec{r}) \nabla^4 \rho_\tau(\vec{r})-20\left[\nabla^3 \rho_\tau(\vec{r})\right]^2\right\}.
\end{aligned}
\end{equation}
\end{widetext}
In the above expressions, for convenience, we have recombined the Skryme parameters as follows:
\begin{align}
\label{eq:defineC}
C^{\left[ n \right]} =& t_{1}^{\left[ n \right]} \left( 2 + x_{1}^{\left[ n \right]} \right) + t_{2}^{\left[ n \right]} \left( 2 + x_{2}^{\left[ n \right]} \right), \\
\label{eq:defineD}
D^{\left[ n \right]} =& -t_{1}^{\left[ n \right]} \left( 2 x_{1}^{\left[ n \right]} +1 \right) + t_{2}^{\left[ n \right]} \left( 2 x_{2}^{\left[ n \right]} +1 \right), \\
\label{eq:defineE}
E^{\left[ n \right]} =& \frac{i^{n}}{2^{n}} \left[
t_{1}^{\left[ n \right]} \left(2 + x_{1}^{\left[ n \right]} \right) - t_{2}^{\left[ n \right]} \left( 2 + x_{2}^{\left[ n \right]} \right)
\right], \\
\label{eq:defineF}
F^{\left[ n \right]} =& - \frac{i^{n}}{2^{n}} \left[
t_{1}^{\left[ n \right]} \left( 2 x_{1}^{\left[ n \right]} +1 \right) + t_{2}^{\left[ n \right]} \left( 2 x_{2}^{\left[ n \right]} +1 \right)
\right],
\end{align}
with $i$ being the imaginary unit.
Based on Eq.~(\ref{eq:Vdd_new}), the DD term can be expressed as
\begin{equation}
\begin{aligned}
\mathcal{H}^{\mathrm{DD}}(\vec{r}) = &
\sum_{n=1,3,5} \frac{1}{24} t_3^{[n]} \left[ \left(2+x_3^{[n]}\right) \rho^2 \right. \\
& \left. -\left(2 x_3^{[n]}+1\right) ( \rho_n^2 + \rho_p^2 )\right] \rho^{n / 3}.
\end{aligned}
\end{equation}

\begin{widetext}
Within the framework of Landau Fermi liquid theory, the single-nucleon potential can be obtained by taking the variation of $\mathcal{H}\left( \vec{r} \right)$ with respect to $f_{\tau} \left( \vec{r},\vec{p} \right)$.
Due to the presence of the gradient operators in $\mathcal{H}\left( \vec{r} \right)$, the single-nucleon potential can be calculated as described in \cite{Kolomietz:2017qkb}:
\begin{equation}
\label{eq:U=dHdf}
\begin{aligned}
U_{\tau} \left( \vec{r}, \vec{p} \right)
=& \, \frac{\delta \mathrm{H}^{\mathrm{pot}}}{ \delta n_\tau\left( \vec{r}, \vec{p} \right)}
= \, \frac{ \partial \left[ \mathcal{H}^{\mathrm{loc}} \left( \vec{r} \right) + \mathcal{H}^{\mathrm{DD}} \left( \vec{r} \right) + \mathcal{H}^{\mathrm{grad}} \left( \vec{r} \right) \right] }
{\partial \rho_\tau(\vec{r})}
 + \sum_{n} (-1)^n \nabla^n \frac{\partial \mathcal{H}^{\mathrm{grad}} \left( \vec{r} \right) }{ \partial \left[ \nabla^n \rho_\tau(\vec{r})\right]}
 + \frac{\delta \mathrm{H}^{\mathrm{MD}}}{ \delta n_\tau\left( \vec{r}, \vec{p} \right)},
\end{aligned}
\end{equation}
where $\mathrm{H}^{\mathrm{pot}}= \int d \vec{r} \left[ \mathcal{H}^{\mathrm{loc}}(\vec{r})+\mathcal{H}^{\mathrm{DD}}(\vec{r})+\mathcal{H}^{\mathrm{MD}}(\vec{r})+\mathcal{H}^{\mathrm{grad}}(\vec{r}) \right]$ is the potential part of the Hamiltonian with $\mathrm{H}^{\mathrm{MD}}= \int d \vec{r} \mathcal{H}^{\mathrm{MD}}(\vec{r}) $ being the MD part and $n_{\tau}(\vec{r}, \vec{p})=\frac{ (2\pi\hbar)^3}{2} f_{\tau}(\vec{r}, \vec{p})$ being the occupation probability function.
Substitute Eq.~(\ref{eq:Hdensity}) into Eq.~(\ref{eq:U=dHdf}), and this yields:
\begin{equation}
\label{eq:U_general}
\begin{aligned}
U_\tau(\vec{r}, \vec{p})= &
\frac{1}{2} t_0\left[\left(2+x_0\right) \rho\left(\vec{r}\right)-\left(2 x_0+1\right) \rho_\tau \left(\vec{r}\right) \right] \\
&+\sum_{n=1,3,5} \left\{ \frac{t_{3} ^{\left[n \right]}}{24} \frac{n}{3} \left[\left(2+x_{3} ^{\left[n \right]}\right) \rho(\vec{r})^2-\left(2 x_{3} ^{\left[n \right]}+1\right) \sum_{\tau=n, p} \rho_\tau(\vec{r})^2\right] \rho(\vec{r})^{ \frac{n}{3}-1 }  \right\}  \\
& + \sum_{n=1,3,5} \left\{ \frac{1}{12} t_{3} ^{\left[n \right]}\left[\left(2+x_{3} ^{\left[n \right]}\right) \rho(\vec{r})-\left(2 x_{3} ^{\left[n \right]}+1\right) \rho_\tau(\vec{r})\right] \rho(\vec{r})^{\frac{n}{3}}  \right\}
+\frac{1}{8 \hbar^2} C^{[2]} U^{\mathrm{md}[2]}(\vec{r}, \vec{p})+\frac{1}{8 \hbar^2} D^{[2]} U_\tau^{\mathrm{md}[2]}(\vec{r}, \vec{p})   \\
& +\frac{1}{16 \hbar^4} C^{[4]} U^{\mathrm{md}[4]}(\vec{r}, \vec{p})+\frac{1}{16 \hbar^4} D^{[4]} U_\tau^{\mathrm{md}[4]}(\vec{r}, \vec{p})+\frac{1}{8 \hbar^6} C^{[6]} U^{\mathrm{md}[6]}(\vec{r}, \vec{p})+\frac{1}{8 \hbar^6} D^{[6]} U_\tau^{\mathrm{md}[6]}(\vec{r}, \vec{p}) \\
& +\frac{1}{2} E^{[2]} \nabla^2 \rho(\vec{r})+\frac{1}{2} F^{[2]} \nabla^2 \rho_\tau(\vec{r})+  E^{[4]} \nabla^4 \rho(\vec{r})+ F^{[4]} \nabla^4 \rho_\tau(\vec{r})+ 8 E^{[6]} \nabla^6 \rho(\vec{r})+ 8 F^{[6]} \nabla^6 \rho_\tau(\vec{r}),
\end{aligned}
\end{equation}
\end{widetext}
where the MD terms $U^{\mathrm{md}[n]}(\vec{r}, \vec{p})$ and $U_{\tau}^{\mathrm{md}[n]}(\vec{r}, \vec{p})$ are defined as
\begin{align}
 U^{\operatorname{md}[n]}(\vec{r}, \vec{p})&=\int d^3 p^{\prime}\left(\vec{p}-\vec{p}^{\, \prime}\right)^n f\left(\vec{r}, \vec{p}^{\, \prime}\right), \\
 U_\tau^{\operatorname{md}[n]}(\vec{r}, \vec{p})&=\int d^3 p^{\prime}\left(\vec{p}-\vec{p}^{\, \prime}\right)^n f_\tau\left(\vec{r}, \vec{p}^{\, \prime}\right).
\end{align}
Based on the above expressions, one can see that the Hamiltonian density $\mathcal{H} \left( \vec{r} \right)$ depends explicitly on $f_{\tau} \left( \vec{r},\vec{p} \right)$, $ \rho_{\tau} \left( \vec{r} \right) $ and the derivatives of $ \rho_{\tau} \left( \vec{r} \right) $, while the single-nucleon potentials additionally depend on the nucleon momentum.

\subsection{The equation of state of nuclear matter}
The EOS of isospin asymmetric nuclear matter with total nucleon density $\rho=\rho_{n} + \rho_{p}$ and isospin asymmetry $\delta = (\rho_{n}-\rho_{p})/\rho$ are defined as its binding energy per nucleon.
In uniform infinite system, all the gradient terms in the Hamiltonian density (Eq.~(\ref{eq:Hdensity})) vanish.
At zero temperature, $f_{\tau} ( \vec{r},\vec{p} )$ becomes a step function, i.e., $f_{\tau} ( \vec{r},\vec{p} )=\frac{2}{(2 \pi \hbar)^3} \theta(p_{F_{\tau}}-|\vec{p}|)$, with $p_{F_{\tau}}=\hbar(3 \pi^2 \rho_\tau)^{1/3}$ being the Fermi momentum of nucleons with isospin $\tau$.
In this case, the EOS of isospin asymmetric nuclear can be analytically expressed as
\begin{widetext}
\begin{equation}
\label{eq:EOS}
\begin{aligned}
E(\rho, \delta)  = & \, \frac{3}{5} \frac{\hbar^2 a^2}{2 m}  F_{5 / 3} ~ \rho^{2 / 3}
 +\frac{1}{8} t_0^{[0]}\left[2\left(x_0^{[0]}+2\right)-\left(2 x_0^{[0]}+1\right) F_2\right] \rho^{3/3}
 +\frac{1}{48} t_3^{[1]}\left[2\left(x_3^{[1]}+2\right)-\left(2 x_3^{[1]}+1\right) F_2\right] \rho^{4/3}  \\
& +\frac{9 a^2}{64}\left[\frac{8}{15} C^{[2]} F_{5 / 3}+\frac{4}{15} D^{[2]} F_{8 / 3}\right] \rho^{5 / 3}
 +\frac{1}{48} t_3^{[3]}\left[2\left(x_3^{[3]}+2\right)-\left(2 x_3^{[3]}+1\right) F_2\right] \rho^{6/3}  \\
& + \frac{9 a^4}{128} \left[
C^{[4]}\left(\frac{68}{105} F_{7 / 3}+\frac{4}{15} \delta G_{7 / 3}+\frac{4}{15} H_{5 / 3}\right) \right.
 \left. + \frac{16}{35} D^{[4]} F_{10 / 3}
\right] \rho^{7/3}
 +\frac{1}{48} t_3^{[5]}\left[2\left(x_3^{[5]}+2\right)-\left(2 x_3^{[5]}+1\right) F_2\right] \rho^{8/3} \\
&+\frac{9 a^6}{64}\left[
C^{[6]}\left(\frac{148}{135} F_3+\frac{4}{5} \delta G_3+\frac{4}{5} H_{5 / 3} F_{2 / 3}\right) \right.
 \left. +\frac{128}{135} D^{[6]} F_4
\right] \rho^{9/3},
\end{aligned}
\end{equation}
\end{widetext}
where $a=(3\pi^2/2)^{1/3}$, and $m$ is nucleon rest mass in vacuum.
In Eq.~(\ref{eq:EOS}), $F_x$, $G_x$ and $H_x$ are defined as
\begin{align*}
F_x & =\left[(1+\delta)^x+(1-\delta)^x\right] /2 ,  \\
G_x & =\left[(1+\delta)^x-(1-\delta)^x\right] / 2 ,\\
H_x & =\left[(1+\delta)(1-\delta)\right]^x.
\end{align*}

The EOS can be expanded as a power series in $\delta$, i.e.,
\begin{equation}
E\left(\rho, \delta\right) = E_{0}\left(\rho\right) + E_{\mathrm{sym}}\left(\rho\right)\delta^2 + E_{\mathrm{sym},4}\left(\rho\right)\delta^4 + \mathcal{O}\left(\delta^6\right),
\end{equation}
where $E_0(\rho)$ is the EOS of the SNM.
The symmetry energy $E_{\mathrm{sym}}(\rho)$ and the fourth-order symmetry energy $E_{\mathrm{sym},4}(\rho)$ are defined as
\begin{equation}
E_{\mathrm{sym}}(\rho)  =\left.\frac{1}{2 !} \frac{\partial^2 E(\rho, \delta)}{\partial \delta^2}\right|_{\delta=0},
\end{equation}
and
\begin{equation}
\label{eq:Esym4}
E_{\mathrm{sym}, 4}(\rho)  =\left.\frac{1}{4 !} \frac{\partial^4 E(\rho, \delta)}{\partial \delta^4}\right|_{\delta=0}.
\end{equation}
The expression of $E_{\mathrm{sym}, 4}(\rho)$ is shown in Appendix~\ref{sec:Appendix_Other}.
The value of $E_{\mathrm{sym}, 4}(\rho_0)$ is usually very small, as indicated by microscopic many-body approaches and predictions from phenomenological models ($E_{\mathrm{sym}, 4}(\rho_0) \lesssim 2\,\mathrm{MeV}$)~\cite{Chen:2009wv,Cai:2022uld}.
Specially, for non-relativistic mean-field models, estimates for $E_{\mathrm{sym}, 4}(\rho_0)$ are around $1.02 \pm 0.49\, \mathrm{MeV}$, $1.02 \pm 0.50\, \mathrm{MeV}$, $0.70 \pm 0.60\, \mathrm{MeV}$ and $0.74 \pm 0.63\, \mathrm{MeV}$ in the SHF, eSHF, Gogny-Hartree-Fock and momentum-dependent interaction models, respectively~\cite{Pu:2017kjx}.
However, $E_{\mathrm{sym}, 4}(\rho)$ could significantly impact the properties of nuclear matter with large isospin asymmetry at suprasaturation densities, i.e., the cooling mechanism \cite{Zhang:2000my,Steiner:2006bx} and the core-crust transition density \cite{Xu:2008vz,Xu:2009vi} of neutron stars.

The pressure of the isospin asymmetric nuclear matter can be expressed as
\begin{equation}
\label{eq:press}
P \left(\rho,\delta \right) = \rho^2 \frac{\partial E(\rho, \delta)}{\partial \rho}.
\end{equation}
The saturation density $\rho_{0}$ is defined where the pressure of the SNM is zero (except for $\rho=0$), i.e.,
\begin{equation}
\label{eq:rho_0}
\left. P \left(\rho_0, \delta=0 \right) = \rho_{0}^2 \frac{d E(\rho, 0)}{d \rho} \right| _{\rho=\rho_0}=0 .
\end{equation}

Around the saturation density $\rho_{0}$, both $E_0(\rho)$ and $E_{\mathrm{sym}}(\rho)$ can be expanded as power series in a dimensionless variable $\chi \equiv \frac{\rho-\rho_0}{3 \rho_0}$, i.e.,
\begin{equation}
E_0(\rho)=E_0\left(\rho_0\right)+L_0 \chi+\frac{K_0}{2 !} \chi^2+\frac{J_0}{3 !} \chi^3 +\frac{I_0}{4 !} \chi^4+ \mathcal{O}\left(\chi^5\right),
\end{equation}
and
\begin{equation}
\begin{aligned}
E_{\mathrm{sym}}(\rho)= & \, E_{\mathrm{sym}}\left(\rho_0\right)+L \chi+\frac{K_{\mathrm{sym}}}{2 !} \chi^2 \\ &+\frac{J_{\mathrm{sym}}}{3 !} \chi^3
+\frac{I_{\mathrm{sym}}}{4 !} \chi^4
+\mathcal{O}\left(\chi^5\right).
\end{aligned}
\end{equation}
The first four coefficients of $\chi^n$ in the two expansions are
\begin{align*}
L_0 &  =\left.3 \rho_0 \frac{d E_0(\rho)}{d \rho}\right|_{\rho=\rho_0},
L  =\left.3 \rho_0 \frac{d E_{\mathrm{sym}}(\rho)}{d \rho}\right|_{\rho=\rho_0},\\
K_0 &  =\left.9 \rho_0^2 \frac{d^2 E_0(\rho)}{d \rho^2}\right|_{\rho=\rho_0},
K_{\mathrm{sym}}  =\left.9 \rho_0^2 \frac{d^2 E_{\mathrm{sym}}(\rho)}{d \rho^2}\right|_{\rho=\rho_0},\\
J_0 &  =\left.27 \rho_0^3 \frac{d^3 E_0(\rho)}{d \rho^3}\right|_{\rho=\rho_0},
J_{\mathrm{sym}}=\left.27 \rho_0^3 \frac{d^3 E_{\mathrm{sym}}(\rho)}{d \rho^3}\right|_{\rho=\rho_0},\\
I_0 & =\left.81 \rho_0^4 \frac{d^4 E_0(\rho)}{d \rho^4}\right|_{\rho=\rho_0},
I_{\mathrm{sym}}=\left.81 \rho_0^4 \frac{d^4 E_{\mathrm{sym}}(\rho)}{d \rho^4}\right|_{\rho=\rho_0},
\end{align*}
respectively.
Obviously, we have $L_0=0$ by the definition of $\rho_0$ in Eq.~(\ref{eq:rho_0}).
$K_0$ is the incompressibility coefficient of SNM which characterizes the curvature of $E_0(\rho)$ at $\rho_0$.
$J_0$ and $I_0$ represent higher-order contributions and are commonly referred to as the skewness and kurtosis coefficients of SNM.
$L$, $K_{\mathrm{sym}}$, $J_{\mathrm{sym}}$ and $I_{\mathrm{sym}}$ are the slope coefficient, curvature coefficient, skewness coefficient and kurtosis coefficient of the symmetry energy at $\rho_0$.

The quantities $E_0(\rho_0)$, $K_0$, $J_0$, as well as $L$, $K_{\mathrm{sym}}$ and $J_{\mathrm{sym}}$ are utilized in the following fitting procedure, with their expressions provided in Appendix~\ref{sec:Appendix_Exp}.
In addition, we also present the expressions of $I_0$ and $I_{\mathrm{sym}}$ in Appendix~\ref{sec:Appendix_Other} for completeness.

\subsection{Single-nucleon potential, symmetry potential and nucleon effective masses in cold nuclear matter}
In the case of zero-temperature and uniform nuclear matter, the single-nucleon potential (Eq.~(\ref{eq:U_general})) reduces to an analytical function of $\rho$, $\delta$ and the magnitude of nucleon momentum $p=\left|\vec{p}\right|$ :
\begin{widetext}
\begin{equation}
\label{eq:U_analytic}
\begin{aligned}
U_\tau(\rho, \delta,p)= & \frac{1}{4} t_0\left[2\left(x_0+2\right)-\left(2 x_0+1\right)(1+\tau \delta)\right] \rho \\
&+ \sum_{n=1,3,5} \frac{1}{24} t_3^{[n]}\left[(\frac{n}{3}+2)\left(x_3^{[n]}+2\right)-\left(2 x_3^{[n]}+1\right)\left(\frac{1}{2} \frac{n}{3} F_2+1+\tau \delta\right)\right] \rho^{\frac{n}{3}+1} \\
& +\frac{1}{4} C^{[2]}\left[\frac{1}{3} \frac{k_F^3}{\pi^2}\left(\frac{p}{\hbar}\right)^2+\frac{1}{5} \frac{k_F^5}{\pi^2} F_{5 / 3}\right]+\frac{1}{8} D^{[2]}\left[\frac{1}{3} \frac{k_F^3}{\pi^2}\left(\frac{p}{\hbar}\right)^2(1+\tau \delta)+\frac{1}{5} \frac{k_F^5}{\pi^2}(1+\tau \delta)^{5 / 3}\right] \\
& +\frac{1}{8} C^{[4]}\left[\frac{1}{3} \frac{k_F^3}{\pi^2}\left(\frac{p}{\hbar}\right)^4+\frac{2}{3} \frac{k_F^5}{\pi^2}\left(\frac{p}{\hbar}\right)^2 F_{5 / 3}+\frac{1}{7} \frac{k_F^7}{\pi^2} F_{7 / 3}\right] \\
& +\frac{1}{16} D^{[4]}\left[\frac{1}{3} \frac{k_F^3}{\pi^2}\left(\frac{p}{\hbar}\right)^4(1+\tau \delta)+\frac{2}{3} \frac{k_F^5}{\pi^2}\left(\frac{p}{\hbar}\right)^2(1+\tau \delta)^{5 / 3}+\frac{1}{7} \frac{k_F^7}{\pi^2}(1+\tau \delta)^{7 / 3}\right] \\
& +\frac{1}{4} C^{[6]}\left[\frac{1}{3} \frac{k_F^3}{\pi^2}\left(\frac{p}{\hbar}\right)^6+\frac{7}{5} \frac{k_F^5}{\pi^2}\left(\frac{p}{\hbar}\right)^4 F_{5 / 3}+\frac{k_F^7}{\pi^2}\left(\frac{p}{\hbar}\right)^2 F_{7 / 3}+\frac{1}{9} \frac{k_F^9}{\pi^2} F_3\right] \\
& +\frac{1}{8} D^{[6]}\left[\frac{1}{3} \frac{k_F^3}{\pi^2}\left(\frac{p}{\hbar}\right)^6(1+\tau \delta)+\frac{7}{5} \frac{k_F^5}{\pi^2}\left(\frac{p}{\hbar}\right)^4(1+\tau \delta)^{5 / 3}+\frac{k_F^7}{\pi^2}\left(\frac{p}{\hbar}\right)^2(1+\tau \delta)^{7 / 3}+\frac{1}{9} \frac{k_F^9}{\pi^2}(1+\tau \delta)^3\right],
\end{aligned}
\end{equation}
\end{widetext}
where $\tau$ equals $1$ [$-1$] for neutrons [proton] and $k_F=\left(3 \pi^2 \rho / 2\right)^{1/3}$ is the Fermi wave number of nucleons in the SNM.

Expanding $U_\tau(\rho, \delta,p)$ as a power series in $\tau\delta$, we obtain
\begin{equation}
\label{eq:Utau_series}
\begin{aligned}
U_\tau(\rho, \delta, p) = & \, U_0(\rho,p)+\sum_{i=1,2, \cdots} U_{\mathrm{sym}, i}(\rho,p)(\tau \delta)^i \\
= & \, U_0(\rho,p)+U_{\mathrm{sym}, 1}(\rho,p)(\tau \delta) \\
&+U_{\mathrm{sym}, 2}(\rho, p)(\tau \delta)^2+\cdots,
\end{aligned}
\end{equation}
where
\begin{equation}
\begin{aligned}
\label{eq:U0}
U_0(\rho,p) \equiv & \, U_{\tau}(\rho,0,p)\\
=& \, \frac{3}{4}t_{0}\rho + \sum_{n=1,3,5} \frac{t_3^{[n]}}{16} \left( \frac{n}{3}+2 \right) \rho^{\frac{n}{3}+1} \\
& +\frac{1}{8} \left(2C^{[2]} +D^{[2]} \right) \left[\frac{1}{3} \frac{k_F^3}{\pi^2}\left(\frac{p}{\hbar}\right)^2+\frac{1}{5} \frac{k_F^5}{\pi^2}\right] \\
& +\frac{1}{16} \left(2C^{[4]}+D^{[4]}\right) \left[\frac{1}{3} \frac{k_F^3}{\pi^2}\left(\frac{p}{\hbar}\right)^4 \right. \\
&+ \left. \frac{2}{3} \frac{k_F^5}{\pi^2}\left(\frac{p}{\hbar}\right)^2 +\frac{1}{7} \frac{k_F^7}{\pi^2} \right] \\
&+\frac{1}{8} \left(2C^{[6]}+D^{[6]}\right) \left[\frac{1}{3} \frac{k_F^3}{\pi^2}\left(\frac{p}{\hbar}\right)^6 \right. \\
&+ \left. \frac{7}{5} \frac{k_F^5}{\pi^2}\left(\frac{p}{\hbar}\right)^4 +\frac{k_F^7}{\pi^2}\left(\frac{p}{\hbar}\right)^2 +\frac{1}{9} \frac{k_F^9}{\pi^2} \right]
\end{aligned}
\end{equation}
is the single-nucleon potential in SNM and $U_{\mathrm{sym}, i}$ can be expressed as
\begin{equation}
\begin{aligned}
U_{\mathrm{sym}, i}(\rho,p) & \left.\equiv \frac{1}{i !} \frac{\partial^i U_n(\rho, \delta,p)}
{\partial \delta^i}\right|_{\delta=0} \\
& =\left.\frac{(-1)^i}{i !} \frac{\partial^i U_p(\rho, \delta,p)}{\partial \delta^i}\right|_{\delta=0}.
\end{aligned}
\end{equation}
The $U_{\mathrm{sym}, 1}(\rho,p)$ is the well-known first-order symmetry potential \cite{Li:2008gp}.
And $U_{\mathrm{sym}, 2}(\rho,p)$ represents the second-order symmetry potential, the contribution of which to $L(\rho)$ could be significant based on the single-nucleon potential decomposition of $L(\rho)$ \cite{Chen:2011ag}.
However, there is currently no experimental or empirical information available regarding $U_{\mathrm{sym}, 2}$.
Neglecting higher-order terms $(\delta^2,\delta^3,\cdots)$ in Eq.~(\ref{eq:Utau_series}) leads to the well-known Lane potential \cite{Lane:1962zz} :
\begin{equation}
U_\tau(\rho, \delta, p) \approx U_0(\rho,p)+U_{\mathrm{sym }}(\rho,p)(\tau \delta).
\end{equation}
In the following, we abbreviate the first-order symmetry potential $U_{\mathrm{sym}, 1}$ as $U_{\mathrm{sym}}$.
For the new Skyrme pseudopotential interaction in Eq.~(\ref{eq:Vsk}), the symmetry potential can be expressed as
\begin{equation}
\label{eq:Usym}
\begin{aligned}
U_{\mathrm{sym}}(\rho,p)=
&-\frac{1}{4} t_0\left(2 x_0+1\right) \rho \\
&-\sum_{n=1,3,5}  \frac{1}{24} t_{3}^{[n]}\left(2 x_{3}^{[n]}+1\right) \rho^{\frac{n}{3}+1} \\
& +\frac{D^{[2]}}{8}\left[\frac{1}{3} \frac{k_F^3}{\pi^2}\left(\frac{p}{\hbar}\right)^2+\frac{1}{3} \frac{k_F^5}{\pi^2}\right] \\
& +\frac{D^{[4]}}{16}\left[\frac{1}{3} \frac{k_F^3}{\pi^2}\left(\frac{p}{\hbar}\right)^4+\frac{10}{9} \frac{k_F^5}{\pi^2}\left(\frac{p}{\hbar}\right)^2+\frac{1}{3} \frac{k_F^7}{\pi^2}\right] \\
& +\frac{D^{[6]}}{8}\left[\frac{1}{3} \frac{k_F^3}{\pi^2}\left(\frac{p}{\hbar}\right)^6+\frac{7}{3} \frac{k_F^5}{\pi^2}\left(\frac{p}{\hbar}\right)^4\right. \\
& \left.+\frac{7}{3} \frac{k_F^7}{\pi^2}\left(\frac{p}{\hbar}\right)^2+\frac{1}{3} \frac{k_F^9}{\pi^2}\right] .
\end{aligned}
\end{equation}

One important quantity related to the single-nucleon potential is the nucleon effective mass. From Eq.~(\ref{eq:U=dHdf}), it is seen that the single-nucleon potential represents the net effect of the nuclear medium, defined as the single-nucleon energy subtracting the kinetic energy part, i.e.,
$U_{\tau} \equiv E_{\tau} - p^{2}/2m$ in non-relativistic models.
Thus the single-nucleon potential is generally dependent on single-particle energy and momentum, which is also evident from the observed energy/momentum dependence of nucleon optical potential.
The nucleon effective mass obtained from the momentum dependence of the single-nucleon potential, i.e., the p-mass $m_{\tau}^{\ast,p}$, is different from the E-mass $m_{\tau}^{\ast,E}$ derived from the energy dependence, by their definitions (see Eq.~(1) in Ref.~\cite{Li:2018lpy}).
They respectively reflect the spatial and the time non-locality of the underlying nuclear interactions \cite{Jaminon:1989wj}.
Once the non-relativistic on-shell single-nucleon spectrum $E_{\tau} = p^{2}/2m + U_{\tau}$ is given, the E-mass and p-mass are connected to the (total) nucleon effective mass $m_{\tau}^{\ast}$ by the well-known relation $\frac{m_{\tau}^{\ast}}{m} = \frac{m_{\tau}^{\ast,E}}{m} \cdot \frac{m_{\tau}^{\ast,p}}{m}$.
Thus the total nucleon effective mass $m_{\tau}^{\ast}$, which is considered in the present work, can be expressed as~\cite{Li:2018lpy}
\begin{equation}
\frac{m_{\tau}^{\ast}(\rho,\delta)}{m}=\left[1+\left.\frac{m}{p} \frac{d U_\tau(\rho, \delta,p)}{d p}\right|_{p=p_{F_{\tau}}}\right]^{-1}.
\end{equation}
The isoscalar nucleon effective mass $m_{s}^{\ast}$ is the nucleon effective mass in SNM, and the isovector nucleon effective mass $m_{v}^{\ast}$ is the effective mass of proton (neutron) in pure neutron (proton) matter.
Additionally, a subscript ``0" denotes that the nucleon effective mass is defined at the saturation density $\rho_0$, e.g., $m_{s,0}^{\ast}$ and $m_{v,0}^{\ast}$, and their expressions are shown in Appendix~\ref{sec:Appendix_Other}.
The nucleon effective mass splitting, denoted as $\mspl(\rho,\delta) \equiv \left[m_{n}^{\ast}(\rho,\delta)-m_{p}^{\ast}(\rho,\delta) \right]/m$, is extensively used in nuclear physics.
$\mspl(\rho,\delta)$ can be expanded as a power series in $\delta$, i.e.,
\begin{equation}
\label{eq:spl_coes}
\mspl(\rho,\delta) =\sum_{n=1}^{\infty}
\Delta m_{2 n-1}^{\ast}(\rho) \delta^{2 n-1},
\end{equation}
where $\Delta m_{2 n-1}^{\ast}(\rho)$ are the isospin splitting coefficients (of the nucleon effective mass), and the first coefficient $\Delta m_{1}^{\ast}(\rho)$ is usually referred to as the linear isospin splitting coefficient.
By analyzing experimental data, significant progress has been made in determining the linear isospin splitting coefficient in recent years \cite{Li:2015pma,Li:2018lpy}, including $\mspl(\rho_0,\delta)=(0.41 \pm 0.15) \delta$ from the optical model analysis of nucleon-nucleus scatterings \cite{Li:2014qta}, $\mspl(\rho_0,\delta)=(0.27 \pm 0.15) \delta$ from the SHF+RPA calculations of the isovector giant dipole resonance (IVGDR) and the electric dipole polarizability ($\alpha_{\mathrm{D}}$) in $^{208}\mathrm{Pb}$ \cite{Zhang:2015qdp} as well as $\mspl(\rho_0,\delta)=(0.216 \pm 0.114) \delta$ from the IBUU transport model simulations of the IVGDR and $\alpha_{\mathrm{D}}$ in $^{208}\mathrm{Pb}$ \cite{Kong:2017nil}.
It should be noted that
a linear isospin splitting of $\mspl(\rho_0,\delta)=0.187 \delta$ is predicted very recently with the relativistic Brueckner-Hartree-Fock (RBHF) theory in the full Dirac space~\cite{Wang:2023owh}.

The very interesting and simple relations between the isoscalar and isovector nucleon effective masses $m_{s}^{\ast}$, $m_{v}^{\ast}$, and the fourth-order symmetry energy $E_{\mathrm{sym},4}(\rho)$ as well as the isospin splitting coefficients $\Delta m_{2 n-1}^{\ast}(\rho)$ have been discovered (see Eq.~(33) in Ref.~\cite{Pu:2017kjx} and Eq.~(8) in Ref.~\cite{Zhang:2015qdp}, respectively), and these relations hold true in both standard SHF and eSHF models.
Given that the single-nucleon potential in N3LO Skyrme pseudopotential contains higher-order momentum terms, $m_{s}^{\ast}$ and $m_{v}^{\ast}$ are momentum dependent.
Consequently, the connections between $E_{\mathrm{sym},4}(\rho)$ as well as $\Delta m_{2 n-1}^{\ast}(\rho)$ and $m_{s}^{\ast}$, $m_{v}^{\ast}$ explicitly involve the derivatives of $m_{s}^{\ast}$ and $m_{v}^{\ast}$ with respect to momentum, and we demonstrate this in Appendix~\ref{sec:Appendix_Other}.

\section{FITTING STRATEGY AND NEW INTERACTIONS}
\label{sec:fitting}
Since the gradient operator makes no contribution toward the uniform nuclear matter, the gradient terms of the Hamitonian density in Eq.~(\ref{eq:Hgrad}) vanish.
As a result, the six coefficients $E^{[n]}$ and $F^{[n]}$ ($n=2,4,6$) are irrelevant to the properties of the nuclear matter, but they are important for transport model and nuclear structure.
$E^{[n]}$ and $F^{[n]}$ ($n=2,4,6$) could be obtained by the properties of finite nuclei, and thus the 12 Skyrme parameters $t_{1}^{[n]}$, $x_{1}^{[n]}$, $t_{2}^{[n]}$, $x_{2}^{[n]}$ ($n=2,4,6$) can be totally determined together with $C^{[n]}$ and $D^{[n]}$ ($n=2,4,6$) from Eqs.~(\ref{eq:defineC})-(\ref{eq:defineF}).
To determine $E^{[n]}$ and $F^{[n]}$ ($n=2,4,6$) from finite nuclei calculations is beyond the scope of this work and will be pursued in future research.
In this context, the total number of the present extended Skyrme parameters is reduced from $20$ to $14$, i.e., $t_0$, $t_{3}^{[1]}$, $t_{3}^{[3]}$, $t_{3}^{[5]}$, $x_0$, $x_{3}^{[1]}$, $x_{3}^{[3]}$, $x_{3}^{[5]}$, $C^{[2]}$, $C^{[4]}$, $C^{[6]}$, $D^{[2]}$, $D^{[4]}$ and $D^{[6]}$.

One of our main goals in developing this new extended Skyrme interaction in the recent work is its application in two areas: the one-body transport model for HICs, and its future applications in the EOS of warm nuclear matter for protoneutron stars, as well as in the numerical simulations of supernovae and neutron star mergers.
In transport equations (such as BUU equation or Vlasov equation), the single-nucleon potential is a basic input.
At finite temperature, even in thermal equilibrium, the single-nucleon potential is necessary to determine the nucleons distribution.
The new extended Skyrme interaction is therefore necessary to accurately describe the well-known empirical momentum dependence of single-nucleon potential in SNM at saturation density $U_{0}(\rho_0,p)$, with nucleon momentum up to $1.5 \, \mathrm{GeV}/c$ (approximately corresponding to nucleon kinetic energy of $1 \, \mathrm{GeV}$).
We use the data of the real part of the nucleon optical potential (Schr\"{o}dinger equivalent potential) obtained by Hama \textit{et al.} \cite{Hama:1990vr,Cooper:1993nx} in the model parameter optimization.
Furthermore, the momentum dependence of the symmetry potential at saturation density $U_{\mathrm{sym}}(\rho_0,p)$, derived from the new extended Skyrme interaction should be comparable to that from microscopic calculations, such as the Brueckner-Hartree-Fock (BHF) calculation \cite{vanDalen:2005sk} and relativistic impulse approximation \cite{Chen:2005hw,Li:2006nd} (still for nucleon momentum up to $1.5 \, \mathrm{GeV}/c$).

Firstly, we rewrite $U_{0}(\rho_0,p)$ and $U_{\mathrm{sym}}(\rho_0,p)$ in the forms we employ frequently in the subsequent discussion, i.e.,
\begin{align}
\label{eq:U0_a0246}
U_{0}(\rho_0,p) & = a_{0} + a_2 \left(\frac{p}{\hbar}\right)^2  + a_{4} \left(\frac{p}{\hbar}\right)^4
+ a_{6} \left(\frac{p}{\hbar}\right)^6, \\
\label{eq:Usym_b0246}
U_{\mathrm{sym}}(\rho_0,p) & = b_{0} + b_{2} \left(\frac{p}{\hbar}\right)^2  + b_{4} \left(\frac{p}{\hbar}\right)^4
+ b_{6} \left(\frac{p}{\hbar}\right)^6,
\end{align}
where $a_0$, $a_2$, $a_4$, $a_6$ and $b_0$, $b_2$, $b_4$, $b_6$ take the following following forms:
\begin{equation}
\begin{aligned}
 a_0 = & \, \frac{3}{4} t_0 \rho_0+\sum_{n=1,3,5} \frac{t_3^{[n]}}{16}\left(\frac{n}{3}+2\right) \rho_0^{\frac{n}{3}+1} \\
 & +\frac{k_{F_{0}}^5}{40\pi^2}\left(2 C^{[2]}+D^{[2]}\right)
 + \frac{k_{F_{0}}^7}{112\pi^2}\left(2 C^{[4]}+D^{[4]}\right) \\
 & + \frac{k_{F_{0}}^9}{72\pi^2}\left(2 C^{[6]}+D^{[6]}\right),
\end{aligned}
\end{equation}
\begin{equation}
\begin{aligned}
a_2 = &\, \frac{k_{F_{0}}^3}{24\pi^2}\left(2 C^{[2]}+D^{[2]}\right)
+ \frac{k_{F_{0}}^5}{24\pi^2}\left(2 C^{[4]}+D^{[4]}\right) \\
& + \frac{k_{F_{0}}^7}{8\pi^2}\left(2 C^{[6]}+D^{[6]}\right),
\end{aligned}
\end{equation}
\begin{equation}
a_4 = \frac{k_{F_{0}}^3}{48\pi^2}\left(2 C^{[4]}+D^{[4]}\right) +\frac{7 k_{F_{0}}^5}{40\pi^2} \left(2C^{[6]}+D^{[6]}\right),
\end{equation}
\begin{equation}
a_6 = \frac{k_{F_{0}}^3}{24\pi^2}\left(2 C^{[6]}+D^{[6]}\right),
\end{equation}
and
\begin{equation}
\begin{aligned}
b_0 = &  -\frac{1}{4} t_0\left(2 x_0+1\right) \rho_0-\sum_{n=1,3,5} \frac{1}{24} t_3^{[n]}\left(2 x_3^{[n]}+1\right) \rho_0^{\frac{n}{3}+1} \\
& +\frac{k_{F_{0}}^5}{24\pi^2} D^{[2]} +\frac{k_{F_{0}}^7}{48\pi^2} D^{[4]}+\frac{k_{F_{0}}^9}{24\pi^2} D^{[6]},
\end{aligned}
\end{equation}
\begin{equation}
b_2 = \frac{k_{F_{0}}^3}{24\pi^2} D^{[2]} +\frac{5k_{F_{0}}^5}{72\pi^2} D^{[4]}+\frac{7k_{F_{0}}^7}{24\pi^2} D^{[6]},
\end{equation}
\begin{equation}
b_4 = \frac{k_{F_{0}}^3}{48\pi^2} D^{[4]} +\frac{7k_{F_{0}}^5}{24\pi^2} D^{[6]},
\end{equation}
\begin{equation}
b_6 = \frac{k_{F_{0}}^3}{24\pi^2} D^{[6]}.
\end{equation}

Secondly, we take the values of $\rho_0$, $E_0(\rho_0)$ and $K_0$ to be $0.16 \, \mathrm{fm}^{-3}$, $-16\,\mathrm{MeV}$ and $230\,\mathrm{MeV}$, respectively.
The thermodynamic relationship gives
\begin{equation}
\label{eq:th_relation}
\rho E\left(\rho, \delta \right) + P\left(\rho, \delta \right) =  \sum_{\tau=n,p} \mu_{\tau} \rho_{\tau},
\end{equation}
where $\mu_{\tau}$ is the chemical potentials of nucleons with isospin $\tau$.
According to the Hugenholtz-Van Hove (HVH) theorem \cite{Hugenholtz:1958zz,SATPATHY199985}, $\mu_{\tau}$ can be expressed as
\begin{align}
\label{eq:HVH}
\mu_{\tau} = \frac{p^{2}_{F_{\tau}}}{2m} + U_{\tau}\left(\rho,\delta,p_{F_{\tau}}  \right),
\end{align}
where $p_{F_{\tau}}=\hbar(3 \pi^2 \rho_\tau)^{1/3}$ is the Fermi momentum of nucleons with isospin $\tau$.
Substituting Eq.~(\ref{eq:HVH}) into Eq.~(\ref{eq:th_relation}) at saturation density in SNM, we obtain
\begin{equation}
\label{eq:U0_constraint}
E_{0}\left(\rho_0\right) = \frac{p_{F_0}^{2}}{2m} +U_{0} \left(\rho_0,p_{F_0}\right),
\end{equation}
where $p_{F_0}=\hbar\left(3 \pi^2 \rho_0 / 2\right)^{1 / 3}$ is the Fermi momentum of nucleons in the SNM at saturation density.

Thirdly, we use the GEKKO optimization suite \cite{beal2018gekko} to minimize the weighted sum of squared difference between $U_{0}$ in Eq.~(\ref{eq:U0_a0246}) and the nucleon optical potential data $U_{\mathrm{opt}}$ \cite{Hama:1990vr,Cooper:1993nx},
\begin{equation}
\chi^{2} = \sum_{i=1}^{N_{d}} \left( \frac{U_{0,i}-U_{\mathrm{opt},i}}{\sigma_{i}} \right)^{2},
\end{equation}
with the constraint of Eq.~(\ref{eq:U0_constraint}),
where $N_{d}$ is the number of the experimental data points.
Since there are actually no practical errors $\sigma_{i}$ here, we assign equal weights to each data point within the range of the nucleon momentum up to $1.5 \, \mathrm{GeV}/c$.
We obtain $a_0=-64.03448\,\mathrm{MeV}$, $a_2=6.517778\,\mathrm{MeV}\,\mathrm{fm}^{2}$,$a_4=-0.1259551\,\mathrm{MeV}\,\mathrm{fm}^{4}$ and $a_6=8.133124\times10^{-4}\,\mathrm{MeV}\,\mathrm{fm}^{6}$.
The last independent quantity in the new extended Skyrme interactions related to SNM, $J_0$, is constrained by the flow data in HICs \cite{Danielewicz:2002pu}.
We set $J_0$ to its maximum allowed value by the flow data, which is $-383\,\mathrm{MeV}$.
The values of the seven macroscopic quantities of SNM, namely, $\rho_0$, $E_{0}(\rho_0)$, $K_{0}$, $J_{0}$, $a_2$, $a_4$ and $a_6$ (due to the constraint of Eq.~(\ref{eq:U0_constraint}), $a_0$ is redundant), uniquely determine the values of the parameters $t_0$, $t_{3}^{[1]}$, $t_{3}^{[3]}$, $t_{3}^{[5]}$ as well as the parameters combinations $2C^{[2]}+D^{[2]}$, $2C^{[4]}+D^{[4]}$ and $2C^{[6]}+D^{[6]}$.
The isoscalar nucleon effective mass at $\rho_0$, denoted by $m^{\ast}_{s,0}$, can be obtained as $0.773m$.

Furthermore, we set the values of $b_2$, $b_4$ and $b_6$ in Eq.~(\ref{eq:Usym_b0246}) to be $-3\,\mathrm{MeV}\,\mathrm{fm}^{2}$, $0.078\,\mathrm{MeV}\,\mathrm{fm}^{4}$ and $-7\times10^{-4}\,\mathrm{MeV}\,\mathrm{fm}^{6}$, respectively.
This choice ensures that the momentum dependence of the symmetry potential at saturation density,  $U_{\mathrm{sym}}(\rho_0,p)$, is consistent with the microscopic calculations \cite{Zuo:2006nz,vanDalen:2005sk,Chen:2005hw,Li:2006nd}.
The values of parameters $D^{[2]}$, $D^{[4]}$ and $D^{[6]}$ are solely determined by $b_2$, $b_4$ and $b_6$, which can be seen by comparing Eq.~(\ref{eq:Usym}) with Eq.~(\ref{eq:Usym_b0246}).
Combined with the parameters of the SNM that have been previously determined, the values of $t_0$, $t_{3}^{[1]}$, $t_{3}^{[3]}$, $t_{3}^{[5]}$, $C^{[2]}$, $C^{[4]}$, $C^{[6]}$, $D^{[2]}$, $D^{[4]}$ and $D^{[6]}$ can then be obtained.
The isovector nucleon effective mass at $\rho_0$, denoted by $m^{\ast}_{v,0}$, can be obtained as $0.691m$.
The values of $b_0$ can be obtained through the widely used theorem of the symmetry energy decomposition \cite{Brueckner:1964zz,Dabrowski:1972mbb,Dabrowski:1973zz,Xu:2010fh,Xu:2010kf,Chen:2011ag}:
\begin{equation}
\label{eq:HVH_Cor}
E_{\mathrm{sym}}(\rho_0)=\frac{1}{3} \frac{p_{F_0}^{2}}{2m^{\ast}_{s,0}}+\frac{1}{2} U_{\mathrm{sym}}\left(\rho_0,p_{F_0}\right),
\end{equation}
once the value of $E_{\mathrm{sym}}(\rho_0)$ is given.

Finally, we construct eight interaction parameter sets with $L$ values of $5$, $15$, $25$, $35$, $45$, $55$, $65$ and $75\,\mathrm{MeV}$, respectively, combined with three additional parameters of the symmetry energy: $E_{\mathrm{sym}}(\rho_0)$, $K_{\mathrm{sym}}$ and $J_{\mathrm{sym}}$, and they are required to strictly satisfy the following constraints:
(1) the EOS of PNM predicted by combined results from various microscopic calculations \cite{Zhang:2022bni};
(2) the largest mass of neutron stars reported so far from PSR J0740+6620 \cite{NANOGrav:2019jur,Fonseca:2021wxt};
(3) the limit of $\Lambda_{1.4} \leqslant 580$ for the dimensionless tidal deformability of canonical $1.4\,M_{\odot}$ neutron star from the gravitational wave signal GW170817 \cite{LIGOScientific:2018cki};
(4) the mass-radius determinations from NICER for PSR J0030+0451 \cite{Miller:2019cac,Riley:2019yda} with a mass around $1.4\,M_{\odot}$ as well as for PSR J0740+6620 \cite{Miller:2021qha,Riley:2021pdl} with a mass around $2.0\,M_{\odot}$.
The details on the calculations of neutron stars can be found in Section~\ref{subsec:NSs}.
Therefore, based on the preceding discussion, in the present work, the 14 parameters of the new extended Skyrme interactions---namely, $t_0$, $t_{3}^{[1]}$, $t_{3}^{[3]}$, $t_{3}^{[5]}$, $x_0$, $x_{3}^{[1]}$, $x_{3}^{[3]}$, $x_{3}^{[5]}$, $C^{[2]}$, $C^{[4]}$, $C^{[6]}$, $D^{[2]}$, $D^{[4]}$ and $D^{[6]}$---are uniquely determined by 14 macroscopic quantities: $\rho_0$, $E_{0}(\rho_0)$, $K_0$, $J_0$, $a_2$, $a_4$, $a_6$, $b_2$, $b_4$, $b_6$, $E_{\mathrm{sym}}(\rho_0)$, $L$, $K_{\mathrm{sym}}$ and $J_{\mathrm{sym}}$.

We name these parameter sets as SP6X, where ``SP6'' indicates the framework of the Skyrme pseudopotential with momentum up to the sixth order and X denotes their $L$ values, i.e., SP6L5, SP6L15, SP6L25, SP6L35, SP6L45, SP6L55, SP6L65 and SP6L75.
In addition to these eight parameter sets, referred to as the default-SP6X interactions, we also develop four interactions: SP6Lm5, SP6L85, SP6L105 and SP6L125, representing supersoft and superstiff symmetry energies around saturation density, and we will discuss them in detail in Section~\ref{sec:soft-stiff}.
It should be noted that all the interactions predict exactly the same properties of SNM, including the pressure $P_{\mathrm{SNM}}(\rho)$ and the single-nucleon potential $U_{0}(\rho,p)$.
In Table~\ref{tab:10SetsParas}, we list the 14 Skyrme parameters for these new interactions.
\begin{table*}
\caption{\label{tab:10SetsParas}
Parameters of the Skyrme interactions SP6X.
Here the recombination of Skyrme parameters defined in Eq.~(\ref{eq:defineC})
and Eq.~(\ref{eq:defineD}) are used.
The units of parameters:
$t_0$: $\mathrm{MeV}\, \mathrm{fm}^3$; $t_{3}^{[n]}$ ($n=1,3,5$), $C^{[n]}$ and $D^{[n]}$ ($n=2,4,6$): $\mathrm{MeV}\, \mathrm{fm}^{n+3}$; $x_0$ and $x_{3}^{[n]}$ ($n=1,3,5$) are dimensionless.
}
\begin{ruledtabular}
\begin{tabular}{ccccccccccccc}
X&Lm5&L5&L15&L25&L35&L45&L55&L65&L75&L85&L105&L125 \\ \hline
$t_0$ &
-1840.45  & -1840.45  & -1840.45  & -1840.45  & -1840.45  & -1840.45  & -1840.45  & -1840.45 & -1840.45  & -1840.45 &
-1840.45  & -1840.45
\\
$x_0$&
-0.530286 & -0.378151 & -0.0684472& 0.0420416& 0.150705  & 0.241260  & 0.128971  & 0.239451  & 0.284729 & 0.333631  &
0.480330  & 0.565458
\\
$t_{3}^{[1]}$ &
13010.2   & 13010.2   & 13010.2   & 13010.2   & 13010.2   & 13010.2   & 13010.2   & 13010.2   & 13010.2   & 13010.2 &
13010.2   & 13010.2
\\
$x_{3}^{[1]}$&
-3.00102  & -2.13561  & -1.08968  & -0.680843 & -0.123375 & 0.221724  & -0.224254 & 0.184563  & 0.450028  & 0.577453 &
1.09245   & 1.48535
\\
$t_3^{[3]}$ &
-4036.41  & -4036.41  & -4036.41  & -4036.41  & -4036.41  & -4036.41  & -4036.41  & -4036.41  & -4036.41  & -4036.41 &
-4036.41  & -4036.41
\\
$x_{3}^{[3]}$&
-38.8861  & -23.9051  & -14.4988  & -10.5117  & -4.12450  & -1.60843  & -4.93746  & -0.950337 & 1.75934   & 3.15290  &
6.75277   & 11.2433
\\
$t_3^{[5]}$ &
2386.36   & 2386.36   & 2386.36   & 2386.36   & 2386.36   & 2386.36   & 2386.36   & 2386.36   & 2386.36   & 2386.36  &
2386.36   & 2386.36
\\
$x_{3}^{[5]}$&
-84.9734  & -43.7168  & -26.3216  & -19.5904  & -7.26000  & -5.59387  & -11.7256  & -4.99398  & -1.66162  & 0.471173 &
1.87073   & 8.66907
\\
$C^{[2]}$&
523.869   & 523.869   & 523.869   & 523.869   & 523.869   & 523.869   & 523.869   & 523.869   & 523.869   & 523.869  &
523.869   & 523.869
\\
$D^{[2]}$&
-349.811  & -349.811  & -349.811  & -349.811  & -349.811  & -349.811  & -349.811  & -349.811  & -349.811  & -349.811 &
-349.811  & -349.811
\\
$C^{[4]}$&
-21.8732  & -21.8732  & -21.8732  & -21.8732  & -21.8732  & -21.8732  & -21.8732  & -21.8732  & -21.8732  & -21.8732 &
-21.8732  & -21.8732
\\
$D^{[4]}$&
17.3414   & 17.3414   & 17.3414   & 17.3414   & 17.3414   & 17.3414   & 17.3414   & 17.3414   & 17.3414   & 17.3414 &
17.3414   & 17.3414
\\
$C^{[6]}$&
0.07567   & 0.07567   & 0.07567   & 0.07567   & 0.07567   & 0.07567   & 0.07567   & 0.07567   & 0.07567   & 0.07567 &
0.07567   & 0.07567
\\
$D^{[6]}$&
-0.07000  & -0.07000  & -0.07000  & -0.07000  & -0.07000  & -0.07000  & -0.07000  & -0.07000  & -0.07000  & -0.07000 &
-0.07000  & -0.07000
\\
\end{tabular}
\end{ruledtabular}
\end{table*}

\section{THE PROPERTIES OF COLD NUCLEAR MATTER WITH THE NEW EXTENDED SKYRME INTERACTIONS}
\label{sec:properties}
\subsection{Bulk properties of cold nuclear matter}
Table~\ref{tab:10SetsQuants} summarizes the macroscopic characteristic quantities of nuclear matter obtained with these new extended Skyrme interactions SP6X.
We would like to mention that the subsaturation cross density $\rho_{\mathrm{sc}}=2/3\rho_0$ is approximately the average density of heavy nuclei.
$E_{{\mathrm{sym}}}(\rho_{\mathrm{sc}})$ and $L(\rho_{\mathrm{sc}})$ have been commonly used to describe the subsaturation properties of the symmetry energy, and they are strongly correlated with many properties of finite nuclei \cite{Zhang:2013wna,Brown:2013mga,Zhang:2014yfa,Zhang:2015ava}.
One can see from Table~\ref{tab:10SetsQuants} that for the default-SP6X interactions (with $5\,\mathrm{MeV} \leq L \leq 75\,\mathrm{MeV}$), the values of $E_{\mathrm{sym}}(2\rho_0)$ vary from about $28\,\mathrm{MeV}$ to $61\,\mathrm{MeV}$, which is consistent with the $E_{\mathrm{sym}}(2\rho_0)=47^{+23}_{-22}\,\mathrm{MeV}$ \cite{Xie:2020tdo}, obtained by averaging essentially all the existing constraints.
For SP6X interactions with the superstiff symmetry energies, i.e., with X being L85, L105 and L125, we may have much larger $E_{\mathrm{sym}}(2\rho_0)$ values, i.e., $E_{\mathrm{sym}}(2\rho_0)= 66.87\,\mathrm{MeV}$, $82.70\,\mathrm{MeV}$ and $93.57\,\mathrm{MeV}$, respectively.
\begin{table*}
\caption{\label{tab:10SetsQuants}
Macroscopic characteristic quantities of nuclear matter with these new extended Skyrme interactions SP6X.
Note: $\rho_{\mathrm{sc}}=2/3 \rho_0$.}
\begin{ruledtabular}
\begin{tabular}{ccccccccccccc}
X&Lm5&L5&L15&L25&L35&L45&L55&L65&L75&L85&L105&L125 \\ \hline
$\rho_0\,(\rm{fm}^{-3})$ &
0.160   & 0.160  & 0.160   & 0.160   & 0.160   & 0.160  & 0.160   & 0.160   & 0.160   & 0.160 & 0.160   & 0.160
\\
$E_0(\rho_0)\,(\rm{MeV})$ &
-16.0   & -16.0  & -16.0   & -16.0   & -16.0   & -16.0  & -16.0   & -16.0   & -16.0   & -16.0 & -16.0   & -16.0
\\
$K_0\,(\rm{MeV})$ &
230.0   & 230.0  & 230.0   & 230.0   & 230.0   & 230.0  & 230.0   & 230.0   & 230.0   & 230.0 & 230.0   & 230.0
\\
$J_0\,(\rm{MeV})$ &
-383.0  & -383.0 & -383.0  & -383.0  & -383.0  & -383.0 & -383.0  & -383.0  & -383.0  & -383.0 & -383.0  & -383.0
\\
$I_0\,(\rm{MeV})$ &
1819  & 1819  & 1819  & 1819  & 1819  & 1819  &
1819  & 1819  & 1819  & 1819  & 1819  & 1819
\\
$m^\ast_{s,0}/m$ &
 0.773  & 0.773  & 0.773   & 0.773   & 0.773   & 0.773  & 0.773   & 0.773   & 0.773   & 0.773  & 0.773   & 0.773
\\
$m^\ast_{v,0}/m$ &
0.691   & 0.691  & 0.691   & 0.691   & 0.691   & 0.691  & 0.691   & 0.691   & 0.691   & 0.691  & 0.691   & 0.691
\\
$E_{{\mathrm{sym}}}(\rho_{\mathrm{sc}})\,(\rm{MeV})$ &
23.51   & 25.42  & 25.51   & 25.64   & 24.77   & 24.14  & 26.05   & 26.18   & 25.46   & 26.49  & 26.09   & 26.36
\\
$L(\rho_{\mathrm{sc}})\,(\rm{MeV})$ &
15.01   & 30.50  & 33.96   & 37.45   & 40.54   & 41.51  & 48.16   & 51.66   & 53.53   & 58.41  & 62.01   & 69.16
\\
$E_{{\rm{sym}}}(\rho_0)\,(\rm{MeV})$ &
24.00   & 28.00  & 29.00   & 30.00   & 30.00   & 30.00  & 33.00   & 34.00   & 34.00   & 36.00  & 37.00   & 39.00
\\
$L\,(\rm{MeV})$ &
-5.000  & 5.000  & 15.00   & 25.00   & 35.00   & 45.00  & 55.00   & 65.00   & 75.00   & 85.00  & 105.0   & 125.0
\\
$E_{{\rm{sym}}}(1.5\rho_0)\,(\rm{MeV})$ &
26.26   & 26.97  & 29.25   & 32.13   & 33.66   & 36.45   & 41.42  & 44.30   & 46.72   & 50.60  & 56.94   & 63.04
\\
$L(1.5\rho_0)\,(\rm{MeV})$ &
76.94   & -11.80 & -8.534  & 8.503   & 17.28   & 52.44   & 74.47  & 91.50   & 119.2   & 137.4  & 204.0   & 245.8
\\
$E_{{\rm{sym}}}(2\rho_0)\,(\rm{MeV})$ &
47.27   & 28.47  & 29.10   & 33.47   & 34.59   & 42.37   & 50.63  & 54.99   & 60.95   & 66.87  & 82.70   & 93.57
\\
$L(2\rho_0)\,(\rm{MeV})$ &
437.6   & 69.87  & 19.22   & 29.61   & 3.155   & 75.87   & 127.6  & 137.8   & 184.9   & 208.7  & 347.8   & 403.8
\\
$K_{{\rm{sym}}}\,(\rm{MeV})$ &
-10.00  & -250.0 & -240.0  & -210.0  & -190.0  & -110.0  & -100.0 & -70.00  & -10.00  & 10.00  & 150.0   & 220.0
\\
$J_{{\rm{sym}}}\,(\rm{MeV})$ &
4250    & 2100   & 1450    & 1200    & 670.0   & 700.0   & 900.0  & 650.0   & 550.0   & 470.0  & 580.0   & 320.0
\\
$I_{{\rm{sym}}}\,(\rm{MeV})$ &
-1140  & -293.9  & -1220  & -1586  & -1896  & -2446  &
-1974  & -2340   & -2640  & -2752  & -3638  & -3970
\\
$E_{{\rm{sym}},4}(\rho_0)\,(\rm{MeV})$ &
0.7471 & 0.7471 & 0.7471 & 0.7471 & 0.7471 & 0.7471 &
0.7471 & 0.7471 & 0.7471 & 0.7471 & 0.7471 & 0.7471
\\
$\Delta m_1^{\ast}(\rho_0)$ &
0.1740 & 0.1740 & 0.1740 & 0.1740 & 0.1740 & 0.1740 &
0.1740 & 0.1740 & 0.1740 & 0.1740 & 0.1740 & 0.1740
\\
\end{tabular}
\end{ruledtabular}
\end{table*}

\begin{figure}[ht]
    \centering
    \includegraphics[width=\linewidth]{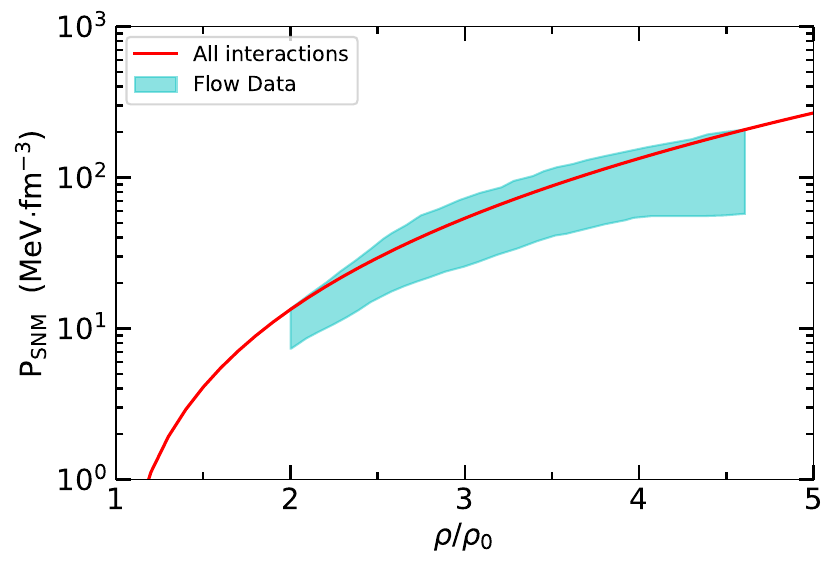}
    \caption{The pressure of SNM ($P_{\mathrm{SNM}}(\rho)$) as a function of nucleon density given by the new interactions SP6X. Also included are the constraints from flow data in HICs \cite{Danielewicz:2002pu}.}
    \label{fig:flow1}
\end{figure}

Shown in Fig.~\ref{fig:flow1} is the pressure of the SNM, $P_{\mathrm{SNM}}(\rho)$, as a function of density of these new interactions, as well as the constraint on the $P_{\mathrm{SNM}}(\rho)$ in the density region from $2\rho_0$ to $4.6\rho_0$ obtained from analyzing the flow data in HICs \cite{Danielewicz:2002pu}.
The $P_{\mathrm{SNM}}(\rho)$ predicted by these new interactions are identical and it is seen that they all conform to the flow data as required in the construction of the model parameters.

\begin{figure}[ht]
    \centering
    \includegraphics[width=\linewidth]{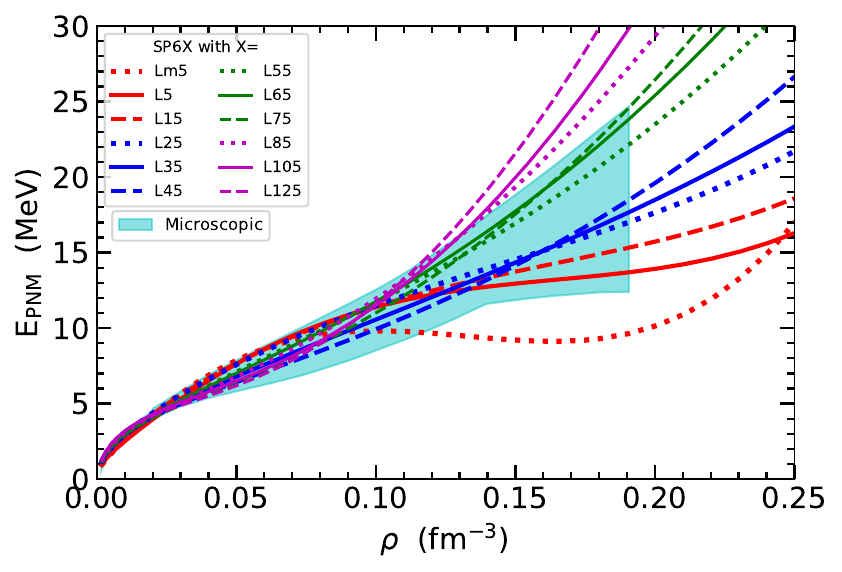}
    \caption{The EOS of the PNM ($E_{\mathrm{PNM}}$) predicted by the new interactions SP6X, with X = Lm5, \dots, L125. The band represents the results from microscopic calculations~\cite{Zhang:2022bni} (see text for the details).}
    \label{fig:Epnm_10Ls}
\end{figure}

Figure~\ref{fig:Epnm_10Ls} displays the EOS of PNM as a function of density determined by these new interactions SP6X and the results from combined microscopic calculations \cite{Zhang:2022bni}, including many-body perturbation theory using N3LO chiral interactions by Tews \textit{et al.} \cite{Tews:2012fj}, Wellenhofer \textit{et al.} \cite{Wellenhofer:2015qba}, and Drischler \textit{et al.} \cite{Drischler:2020hwi}, the quantum Monte Carlo methods by Gandolfi \textit{et al.} \cite{Gandolfi:2011xu}, Wlaz\l{}owski \textit{et al.} \cite{Wlazlowski:2014jna}, Roggero \textit{et al.} \cite{Roggero:2014lga}, and Tews \textit{et al.} \cite{Tews:2015ufa}, the variational calculations by Akmal-Pandharipande-Ravenhall (APR) \cite{Akmal:1998cf}, the Bethe-Bruckner-Goldstone calculations (BBG-QM 3h-gap and BBG-QM 3h-con) \cite{Baldo:2014rda}, and the self-consistent Green's function approach (SCGF-N3LO+N2LOdd) \cite{Carbone:2014mja}.
It is seen that the $E_{\mathrm{PNM}}$ predicted by the default-SP6X interactions (with $5\,\mathrm{MeV} \leq L \leq 75\,\mathrm{MeV}$) are in perfect agreement with the microscopic calculations.
On the other hand, for the superstiff SP6X interactions, with X=L85, L105 and L125, they predict too large $E_{\mathrm{PNM}}$ above $0.12\small{\sim}0.14\,\mathrm{fm}^{-3}$, while the supersoft interaction SP6X, with X = Lm5, predicts too small $E_{\mathrm{PNM}}$ above $0.12\,\mathrm{fm}^{-3}$.

\begin{figure}[ht]
    \centering
    \includegraphics[width=\linewidth]{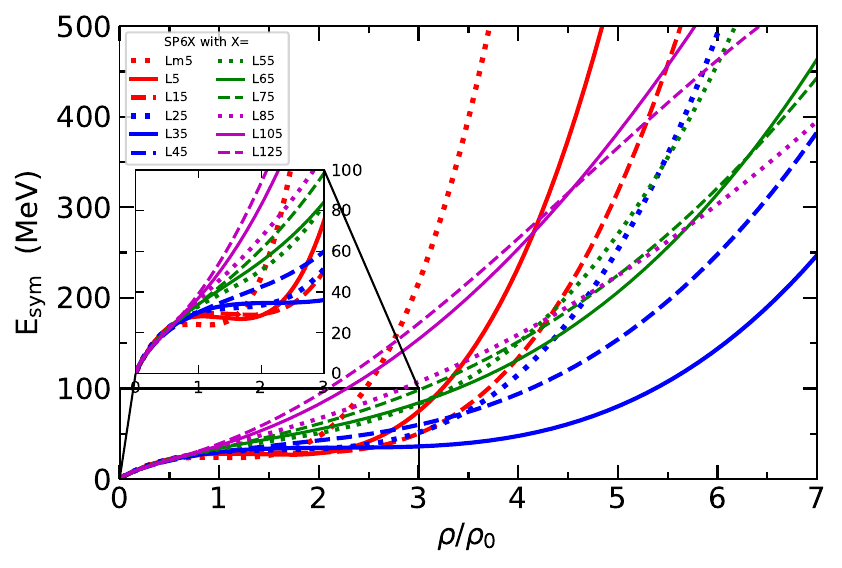}
    \caption{The density dependence of the symmetry energy predicted by the new interactions SP6X, with X = Lm5, \dots, L125.}
    \label{fig:Esym_10Ls}
\end{figure}

Shown in Fig.~\ref{fig:Esym_10Ls} is the density dependence of the symmetry energy predicted by these new interactions SP6X, with X = Lm5, \dots, L125.
These interactions with smaller values of $L$ also have smaller values of $E_{\mathrm{sym}}(\rho_0)$, and this is a consequence of the constraints on $E_{\mathrm{PNM}}$ from microscopic calculations \cite{Zhang:2022bni}.
Although the higher-order characteristic parameters of the symmetry energy, $K_{\mathrm{sym}}$ and $J_{\mathrm{sym}}$, may have effects on $E_{\mathrm{PNM}}$ at subsaturation densities, their values mainly influence the high-density behaviors of the symmetry energy, and thus determined by the properties of neutron stars.
It is seen from Fig.~\ref{fig:Esym_10Ls} that all of these new interactions SP6X have relatively stiff symmetry energy at suprasaturation densities, especially for those with smaller values of $L$, which are necessary to predict the maximum mass of neutron stars that matching the astrophysical observations on the maximum mass of neutron stars \cite{NANOGrav:2019jur,Fonseca:2021wxt}.

\subsection{Single-nucleon potential and symmetry potential}

\begin{figure}[ht]
    \centering
    \includegraphics[width=\linewidth]{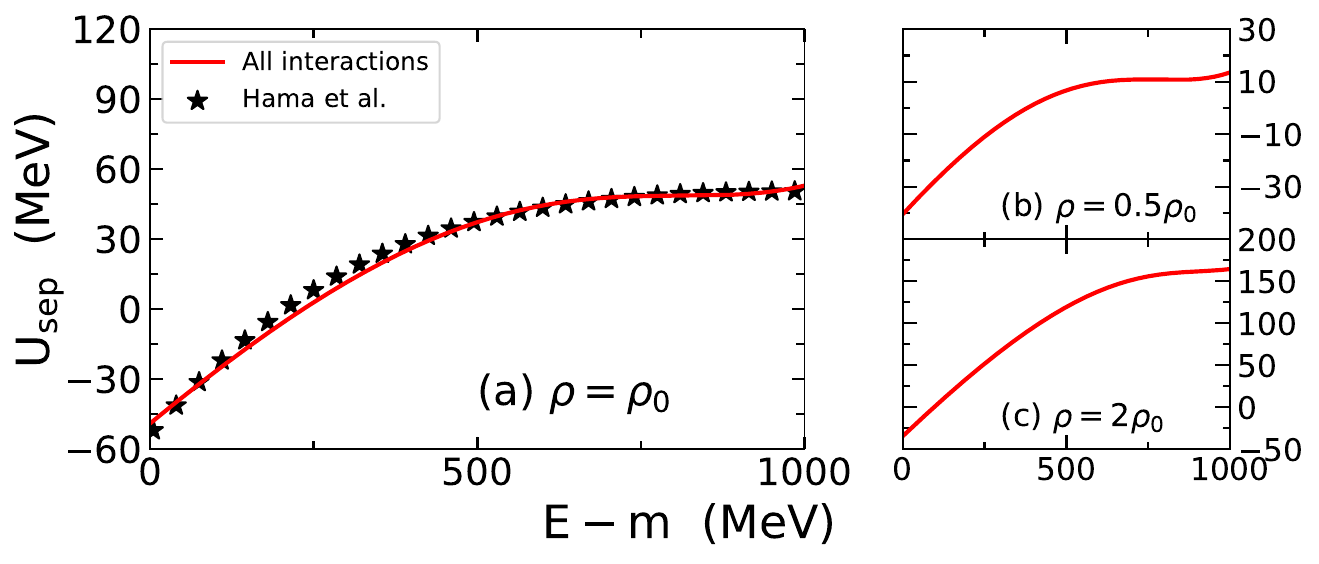}
    \caption{The energy dependence of the single-nucleon potential in cold SNM predicted by the new interactions SP6X.
    The nucleon optical potential (Schr\"{o}dinger equivalent potential, $\mathrm{U}_{\mathrm{sep}}$) in SNM at $\rho_0$ obtained by Hama \textit{et al.} \cite{Hama:1990vr,Cooper:1993nx} is also shown.}
    \label{fig:U0_1}
\end{figure}

Shown in Fig.~\ref{fig:U0_1} is the single-nucleon potential $U_{0}(\rho,p)$ in cold SNM, predicted by these new interactions, as a function of nucleon kinetic energy $E-m=\sqrt{p^2+m^2}+U_{0}(\rho,p)-m$, at $\rho=0.5\rho_0$, $\rho_0$ and $2\rho_0$, respectively.
Also shown in Fig.~\ref{fig:U0_1}(a) is the real part of nucleon optical potential (Schr\"{o}dinger equivalent potential) in SNM at saturation density $\rho_0$ obtained by Hama \textit{et al.} \cite{Hama:1990vr,Cooper:1993nx}, from Dirac phenomenology of nucleon-nucleus scattering data.
As a result of the model parameter optimization process we have performed in Section~\ref{sec:fitting}, $U_{0}(\rho_0,p)$ conforms very well to the empirical nucleon optical potential obtained by Hama \textit{et al.} \cite{Hama:1990vr,Cooper:1993nx} for nucleon kinetic energy up to $1\,\mathrm{GeV}$.

\begin{figure*}[ht]
    \centering
    \includegraphics[width=\linewidth]{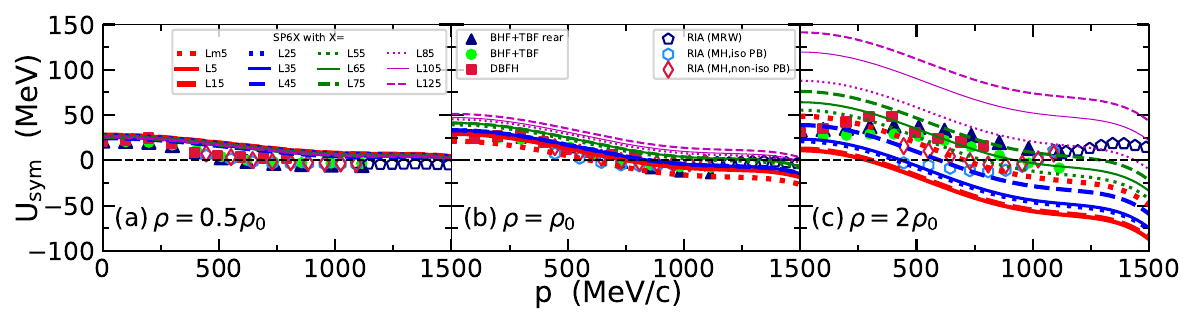}
    \caption{The momentum dependence of the symmetry potential in cold nuclear matter predicted by the interactions SP6X, with X = Lm5, \dots, L125.
    The microscopic calculations from BHF method \cite{Zuo:2006nz}, relativistic Dirac-BHF method \cite{vanDalen:2005sk} and relativistic impulse approximation \cite{Chen:2005hw,Li:2006nd} are also included for comparison.}
    \label{fig:Usym_1}
\end{figure*}

Shown in Fig.~\ref{fig:Usym_1} is the symmetry potential $U_{\mathrm{sym}}(\rho,p)$ of cold nuclear matter, predicted by these new interactions SP6X, with X = Lm5, \dots, L125, as function of nucleon momentum $p$, at $\rho=0.5\rho_0$, $\rho_0$ and $2\rho_0$, respectively.
Also shown in Fig.~\ref{fig:Usym_1} are the corresponding results from several microscopic calculations: the nonrelativistic BHF theory with and without rearrangement contribution from the three-body force \cite{Zuo:2006nz}; the relativistic Dirac-BHF theory \cite{vanDalen:2005sk}; the relativistic impulse approximation \cite{Chen:2005hw,Li:2006nd} using the empirical nucleon-nucleon scattering amplitude determined in Refs.~\cite{Murdock:1986fs,McNeil:1983yi}.
Since the parameters $D^{[2]}$, $D^{[4]}$ and $D^{[6]}$ are the same in these new interactions, the momentum dependence of $U_{\mathrm{sym}}(\rho,p)$ is identical for all interactions, as shown in Fig.~\ref{fig:Usym_1}.
The momentum dependence of the symmetry potential at $\rho_0$, $U_{\mathrm{sym}}(\rho_0,p)$, predicted by these new interactions, is in good agreement with the microscopic calculations, as shown in Fig.~\ref{fig:Usym_1}(b).
The upward and downward translation of $U_{\mathrm{sym}}(\rho,p)$ is determined by the value of $E_{\mathrm{sym}}(\rho)$ through Eq.~(\ref{eq:HVH_Cor}).
The difference in the values of $E_{\mathrm{sym}}(0.5\rho_0)$ among these new interactions are very small, resulting in nearly identical $U_{\mathrm{sym}}(0.5\rho_0,p)$.
However, there are substantial differences in the values of $E_{\mathrm{sym}}(2\rho_0)$, resulting in a significant variation in $U_{\mathrm{sym}}(2\rho_0,p)$, as shown in Fig.~\ref{fig:Usym_1}(c), which may influence the isospin dynamic in HICs at intermediate and high energies.

\subsection{Neutron star properties}
\label{subsec:NSs}
It is generally believed that neutron stars consist of three parts from the inside out: the core, the inner crust and the outer crust.
In the present work, we assume the core consists of free neutrons, protons, electrons and possible muons ($npe\mu$ matter) without phase transition and other degrees of freedom at high densities.
The $\beta$-equilibrium between neutrons, protons, electrons and muons requires
\begin{equation}
\label{eq:npemu}
\mu_n-\mu_p=\mu_e=\mu_{\mu},
\end{equation}
and the appearance of muons requires $\mu_e>m_{\mu}$.
Eq.~(\ref{eq:npemu}) together with the charge neutral condition,
\begin{equation}
\label{eq:charge}
\rho_p=\rho_e+\rho_{\mu},
\end{equation}
are sufficient to determine the fraction of every component as a function of nucleon density, as well as the EOS of the $\beta$-equilibrium $npe\mu$ matter in neutron star core.

\begin{figure*}[ht]
    \centering
    \includegraphics[width=\linewidth]{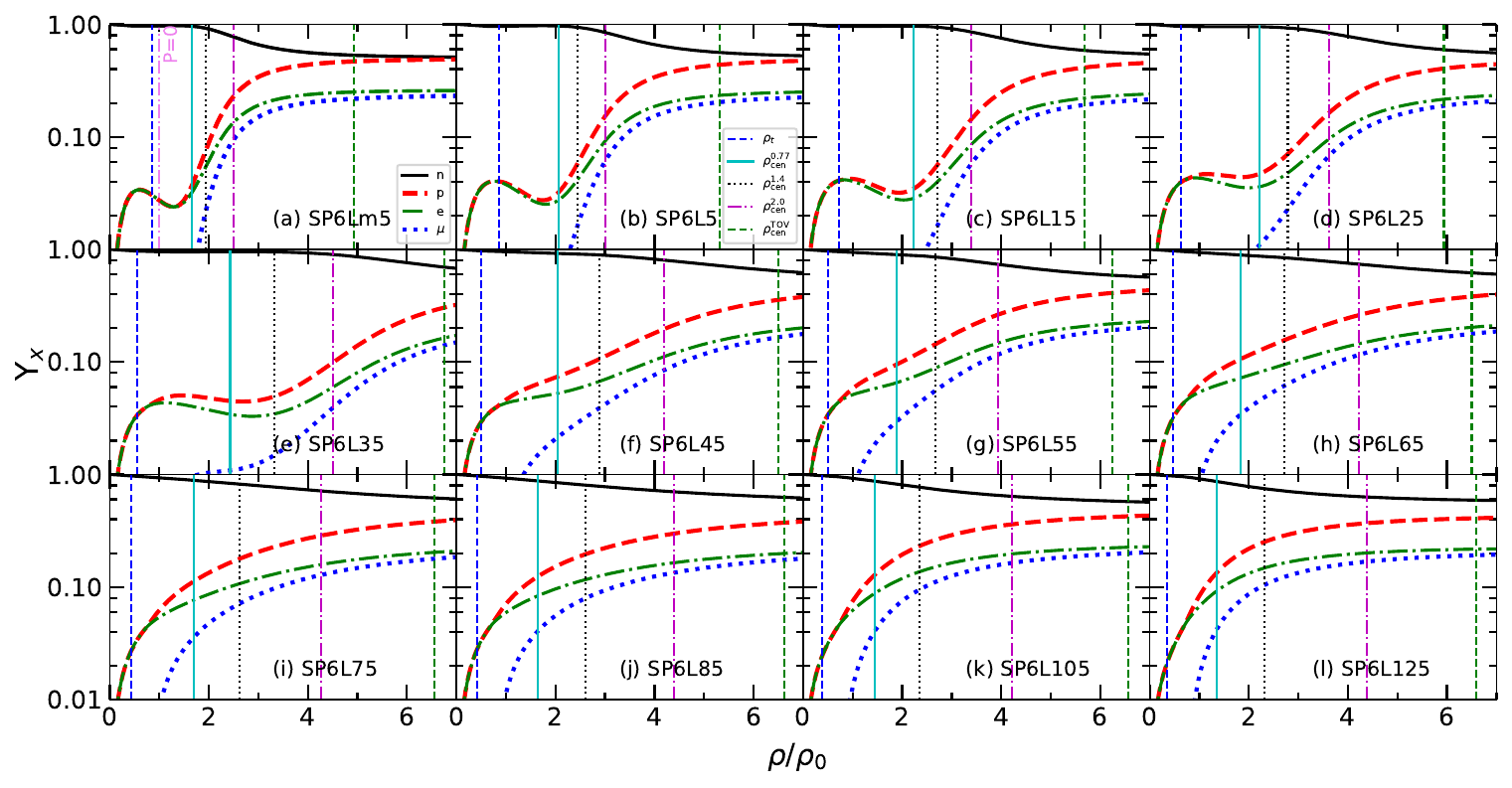}
    \caption{Particle fractions as function of nucleon density in $\beta$-equilibrium matter predicted by the interactions SP6X, with X = Lm5, \dots, L125.
    The vertical lines indicate the core-crust transition density, as well as the center densities of neutron stars with masses of $0.77M_{\odot}$, $1.4M_{\odot}$, $2.0M_{\odot}$ and the maximum mass, respectively.
    For interaction SP6Lm5, the density at the zero-pressure point is also shown.}
    \label{fig:Fractions}
\end{figure*}

Shown in Fig.~\ref{fig:Fractions} is the particle fraction $Y_x \equiv \rho_x / \rho$ ($\rho$ is the nucleon density and $x=n,p,e,\mu$) as functions of nucleon density in $\beta$-equilibrium $npe\mu$ matter predicted by these new interactions SP6X, with X = Lm5, \dots, L125.
Also shown in Fig.~\ref{fig:Fractions} is the core-crust transition density $\rho_t$, which separates the liquid core from the nonuniform inner crust, and is obtained self-consistently through the so-called dynamical method of Ref.~\cite{Xu:2009vi}.
In the present work, the critical density between the inner and the outer crust is taken to be $\rho_{\mathrm{out}}=2.46\times10^{-4}\,\mathrm{fm}^{-3}$ \cite{Carriere:2002bx,Xu:2008vz,Xu:2009vi}.
For the outer crust, where $\rho<\rho_{\mathrm{out}}$, we use the EOS of BPS (FMT) \cite{1971ApJ...170..299B}; for the inner crust, where $\rho_{\mathrm{out}}<\rho<\rho_t$, we construct the EOS by interpolation with the form \cite{Carriere:2002bx,Xu:2008vz,Xu:2009vi}
\begin{equation}
\label{eq:innerCrust}
P=a+b \epsilon^{4/3} .
\end{equation}
Fig.~\ref{fig:Fractions} also displays the center densities of neutron stars corresponding to different masses, namely, $\rho_{\mathrm{cen}}^{0.77}$, $\rho_{\mathrm{cen}}^{1.4}$, $\rho_{\mathrm{cen}}^{2.0}$ and $\rho_{\mathrm{cen}}^{\mathrm{TOV}}$.
One can see that the proton fraction in $\beta$-equilibrium nuclear matter is strongly correlated with the value of $E_{\mathrm{sym}}(\rho)$.
For example, considering SP6Lm5, which possesses the smallest $E_{\mathrm{sym}}(\rho_{0})$ among the SP6X interactions, it predicts the lowest proton fraction at $\rho_0$.
Nevertheless, at high nucleon densities, SP6Lm5 predicts the stiffest symmetry energy, leading to the most isospin symmetric nuclear matter at $\rho_{\mathrm{cen}}^{\mathrm{TOV}}$.
Additionally, it is worth noting that SP6Lm5, SP6L5, SP6L15, SP6L25 and SP6L35 predict a bump structure of proton fraction, which roughly spans between $\rho_t$ and $\rho_{\mathrm{cen}}^{0.77}$.

In Table~\ref{tab:10SetsNSs}, we present various properties of neutron stars obtained with interactions SP6X, with X = Lm5, \dots, L125.
These properties include the core-crust transition density $\rho_t$, the transition pressure $P_t$, the transition energy density $\epsilon_t$, the center density $\rho_{\mathrm{cen}}^{0.77}$ and radius $R_{0.77}$ of $0.77M_{\odot}$ neutron star,
the center density $\rho_{\mathrm{cen}}^{1.4}$ and radius $R_{1.4}$ of $1.4M_{\odot}$ neutron star, the center density $\rho_{\mathrm{cen}}^{2.0}$ and radius $R_{2.0}$ of $2.0M_{\odot}$ neutron star, the center density $\rho_{\mathrm{cen}}^{\mathrm{TOV}}$ and radius $R_{\mathrm{TOV}}$ of the maximum mass neutron star configuration, the dimensionless tidal deformability of $1.4M_{\odot}$ neutron star $\Lambda_{1.4}$ and the maximum mass $M_{\mathrm{TOV}}$.
\begin{table*}
\caption{\label{tab:10SetsNSs}
Core-crust transition density ($\rho_t$), the transition pressure ($P_t$), the transition energy density ($\epsilon_t$), the center densitiy ($\rho_{\mathrm{cen}}^{0.77}$) and radius ($R_{0.77}$) of $0.77M_{\odot}$ neutron star, the center densitiy ($\rho_{\mathrm{cen}}^{1.4}$) and radius ($R_{1.4}$) of $1.4M_{\odot}$ neutron star, the center densitiy ($\rho_{\mathrm{cen}}^{2.0}$) and radius ($R_{2.0}$) of $2.0M_{\odot}$ neutron star, the center density ($\rho_{\mathrm{cen}}^{\mathrm{TOV}}$) and radius ($R_{\mathrm{TOV}}$) of the maximum mass neutron star configuration, the dimensionless tidal deformability of $1.4M_{\odot}$ neutron star ($\Lambda_{1.4}$) and the maximum mass of neutron star ($M_{\mathrm{TOV}}$) for the interactions SP6X, with X = Lm5, \dots, L125.
}
\begin{ruledtabular}
\begin{tabular}{ccccccccccccc}
X&Lm5&L5&L15&L25&L35&L45&L55&L65&L75&L85&L105&L125 \\ \hline
$\rho_t\,(\rm{fm}^{-3})$&
 0.136   & 0.139   & 0.115   & 0.100   & 0.0874  & 0.0804  & 0.0797  & 0.0741  & 0.0691  & 0.0667  & 0.0618 & 0.0568
\\
$P_t\,(\mathrm{MeV}\,\mathrm{fm}^{-3})$ &
-0.244   & 0.423   & 0.551   & 0.548   & 0.464   & 0.372   & 0.476   & 0.423   & 0.344   & 0.353   & 0.267  & 0.217
\\
$\epsilon_t\,(\mathrm{MeV}\,\mathrm{fm}^{-3})$ &
 129.1   & 132.0   & 109.2   & 95.40   & 82.83   & 76.11   & 75.54   & 70.20   & 65.39   & 63.14   & 58.45  & 53.69
\\
$\rho_{\mathrm{cen}}^{0.77}\,(\rm{fm}^{-3})$ &
 0.265   & 0.331   & 0.356   & 0.354   & 0.389   & 0.327   & 0.303   & 0.293   & 0.273   & 0.263   & 0.232  & 0.215
\\
$R_{0.77}\,(\mathrm{km})$ &
 9.929   & 10.66   & 11.08   & 11.61   & 11.72   & 12.35   & 13.19   & 13.52   & 13.71   & 14.13   & 14.32  & 14.70
\\
$\rho_{\mathrm{cen}}^{1.4}\,(\rm{fm}^{-3})$ &
 0.310   & 0.391   & 0.433   & 0.446   & 0.531   & 0.461   & 0.426   & 0.436   & 0.419   & 0.418   & 0.376  & 0.370
\\
$R_{1.4}\,(\mathrm{km})$ &
 11.70   & 11.67   & 11.67   & 11.88   & 11.54   & 12.23   & 12.80   & 12.91   & 13.15   & 13.36   & 13.80  & 14.06
\\
$\rho_{\mathrm{cen}}^{2.0}\,(\rm{fm}^{-3})$ &
 0.400   & 0.482   & 0.544   & 0.580   & 0.721   & 0.671   & 0.629   & 0.676   & 0.683   & 0.703   & 0.674  & 0.700
\\
$R_{2.0}\,(\mathrm{km})$ &
 12.68   & 12.22   & 11.96   & 11.95   & 11.22   & 12.80   & 12.28   & 12.11   & 12.20   & 12.20   & 12.53  & 12.49
\\
$\rho_{\mathrm{cen}}^{\mathrm{TOV}}\,(\rm{fm}^{-3})$ &
 0.80    & 0.85    & 0.91    & 0.95    & 1.08    & 1.04    & 1.00    & 1.04    & 1.05    & 1.06    & 1.05   & 1.06
\\
$R_{\mathrm{TOV}}\,(\mathrm{km})$ &
 11.87   & 11.63   & 11.30   & 11.14   & 10.38   & 10.73   & 11.06   & 10.86   & 10.87   & 10.83   & 11.00  & 11.00
\\
$\Lambda_{1.4}$ &
 714.0   & 433.9   & 359.5   & 355.7   & 264.5   & 390.2   & 468.9   & 475.9   & 546.7   & 571.7   & 751.4  & 837.6
\\
$M_{\mathrm{TOV}}/M_{\odot}$ &
 2.26    & 2.32    & 2.31    & 2.28    & 2.24    & 2.22    & 2.21    & 2.19    & 2.17    & 2.16    & 2.15   & 2.13
\\
\end{tabular}
\end{ruledtabular}
\end{table*}
One can see from Table~\ref{tab:10SetsNSs}, there is a strong correlation between $\rho_t$ and $L$.
For the default-SP6X interactions, with $5\mathrm{MeV}\leq L \leq 75\mathrm{MeV}$, $\rho_t$ exhibits a nearly linear decrease with increasing $L$ \cite{Oyamatsu:2006vd,Xu:2009vi,Chen:2010qx}.
$R_{0.77}$ increases as $L$ increases, and the same trend is observed for $R_{1.4}$, except for SP6Lm5 and SP6L35.
It is seen from Table~\ref{tab:10SetsNSs} that all the default-SP6X interactions satisfy the constraint of $\Lambda_{1.4} \leqslant 580$ as required in the construction of them.
$M_{\mathrm{TOV}}$, $R_{\mathrm{TOV}}$ and $\rho_{\mathrm{cen}}^{\mathrm{TOV}}$ are mainly affected by the high-density behaviors of the symmetry energy, and thus correlated with $K_{\mathrm{sym}}$ and $J_{\mathrm{sym}}$.
Greater values of $K_{\mathrm{sym}}$ and $J_{\mathrm{sym}}$ indicate a stiffer symmetry energy at high densities, leading to larger $M_{\mathrm{TOV}}$, $R_{\mathrm{TOV}}$ and smaller $\rho_{\mathrm{cen}}^{\mathrm{TOV}}$.

\begin{figure}[ht]
    \centering
    \includegraphics[width=\linewidth]{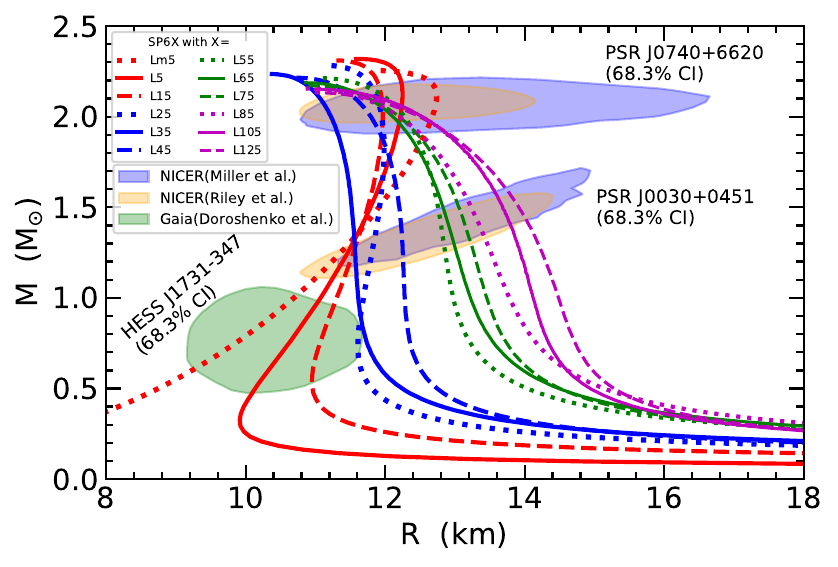}
    \caption{M-R relation for static neutron stars from the interactions SP6X, with X = Lm5, \dots, L125.
    The NICER (XMM-Newton) constraints for PSR J0030+0451 \cite{Miller:2019cac,Riley:2019yda}, PSR J0740+6620 \cite{Miller:2021qha,Riley:2021pdl} and Gaia constraint for the CCO in HESS J1731-347 \cite{2022NatAs...6.1444D} are also included for comparison.
    All contours are plotted for 68.3\% CI.}
    \label{fig:MR}
\end{figure}

Shown in Fig.~\ref{fig:MR} are the mass-radius relations of neutron stars obtained using interactions SP6X, with X = Lm5, \dots, L125.
For comparison, we also show in Fig.~\ref{fig:MR} the simultaneous mass-radius determinations for PSR J0030+0451 \cite{Miller:2019cac,Riley:2019yda} with a mass around $1.4\,M_{\odot}$ and PSR J0740+6620 \cite{Miller:2021qha,Riley:2021pdl} with a mass around $2.0\,M_{\odot}$ obtained from NICER (XMM-Newton).
We also present in Fig.~\ref{fig:MR} the mass-radius determinations for the CCO within HESS J1731-347 from Gaia  \cite{2022NatAs...6.1444D} and all contours are plotted for 68.3\% credible intervals (CI).
As shown in Fig.~\ref{fig:MR}, all of these SP6X interactions, with X = Lm5, \dots, L125, align with the astrophysical observations and measurements for both PSR J0030+0451 and PSR J0740+6620, falling within the 68.3\% CI.
In addition, SP6Lm5, SP6L5, SP6L15, SP6L25 and SP6L35 are compatible with the constraint for the CCO in HESS J1731-347 within 68.3\% CI.
It is interesting to note from Fig.~\ref{fig:MR} that the interactions SP6L5, SP6L15 and SP6L25 predict a ``Z''-shaped mass-radius relation of neutron stars, and these interactions predict large radii for heavy neutron stars with masses around $2.0\,M_{\odot}$, while exhibiting  smaller radii for light neutron stars with masses around $0.5\,M_{\odot}$.

\begin{figure}[ht]
    \centering
    \includegraphics[width=\linewidth]{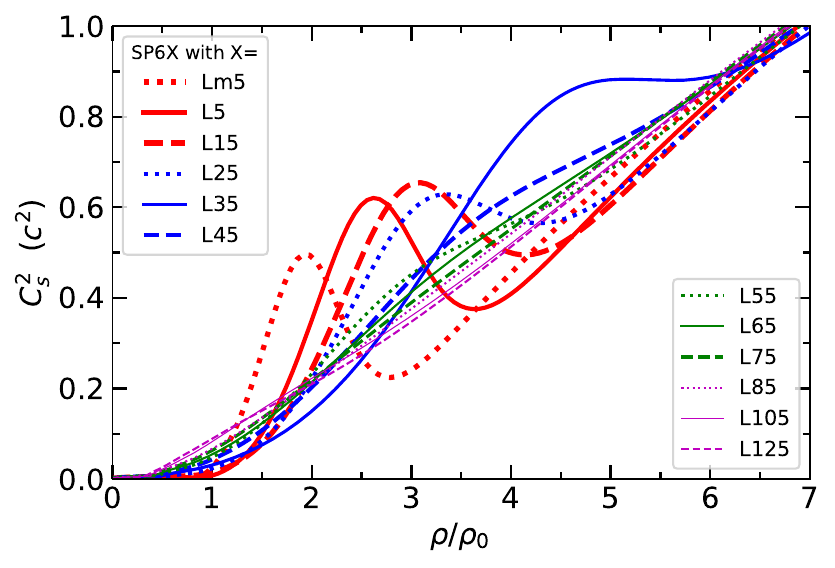}
    \caption{The squared sound speed ($C_{s}^{2}\equiv dP/d\epsilon$) of neutron star matter as a function of nucleon density predicted by the interactions SP6X, with X = Lm5, \dots, L125.}
    \label{fig:Cs2}
\end{figure}

The sound speed is an important quantity to characterize the EOS of dense matter.
A peak structure of the squared sound speed $C_{s}^{2}\equiv dP/d\epsilon$ for neutron star matter,
with the peak value being around $0.5c^2$ ($c$ is the speed of light in vacuum) at density around $3.5\rho_0$, has been observed in recent studies by utilizing Bayesian model-agnostic analyses of multimessenger observations combined with ab-initio theoretical calculations based on chiral effective field and perturbative QCD \cite{Gorda:2022jvk,Han:2022rug,Cao:2023rgh}.
This peak structure could be considered as a possible indication for the existence of the quarkyonic matter \cite{McLerran:2018hbz}.
Or alternatively, it might be associated with the density dependence of the symmetry energy, particularly its high-density behavior \cite{Zhang:2022sep}.
In Fig.~\ref{fig:Cs2}, we display the $C_{s}^{2}$ as function of nucleon density for the SP6X interactions, with X = Lm5, \dots, L125.
The causality condition $C_{s}^{2}\leq c^2 $ is satisfied by all the interactions used in neutron star calculations.
It is worth noting that the interactions SP6Lm5, SP6L5, SP6L15 and SP6L25 predict a clear peak in $C_{s}^{2}$ between $2\rho_0$ and $3.5\rho_0$, with peak values approximately ranging from $0.5c^2$ to $0.7c^2$.
The interactions SP6Lm5, SP6L5, SP6L15 and SP6L25 exhibit relatively softer symmetry energies around saturation density, resulting in lower $C_{s}^{2}$ around $\rho_0$.
However, for the most massive neutron star observed to date \cite{NANOGrav:2019jur,Fonseca:2021wxt}, the symmetry energy at high densities must be sufficiently stiff to support its existence, leading to a rapid rise in $C_{s}^{2}$ for these interactions.
Yet, this upsurge is not sustainable due to the causality condition.
Consequently, we observe a sharp rise followed by a decline in $C_{s}^{2}$ for SP6Lm5, SP6L5, SP6L15 and SP6L25.

\section{Interactions with supersoft and superstiff symmetry energy around saturation density}
\label{sec:soft-stiff}
The density dependence of the symmetry energy is still largely uncertain, even around the nuclear saturation density.
The charge-weak form factor differences in $^{48}\mathrm{Ca}$ and $^{208}\mathrm{Pb}$, extracted through parity-violating electron scattering measurements by PREX-2 \cite{PREX:2021umo} and CREX \cite{CREX:2022kgg} collaborations, are considered to be model-independent probes for neutron skin thickness and further used to constrain the density dependence of the symmetry energy.
The rather thick neutron skin in $^{208}\mathrm{Pb}$, as observed in the PREX experiment \cite{PREX:2021umo}, leads to a significantly large value of $L$, e.g., a prediction of $L=106 \pm 37 \, \mathrm{MeV}$ from a relativistic EDF \cite{Reed:2021nqk}.
However, the CREX experiment \cite{CREX:2022kgg} indicates a rather thin neutron skin in $^{48}\mathrm{Ca}$.
The data from PREX and CREX result in significant tension for constraining the value of $L$ \cite{Zhang:2022bni,Reinhard:2022inh,Yuksel:2022umn} within nuclear EDFs.
In Ref.~\cite{Zhang:2022bni}, a Bayesian analysis is performed on the charge-weak form factor differences in $^{48}\mathrm{Ca}$ and $^{208}\mathrm{Pb}$ by PREX-2 \cite{PREX:2021umo} and CREX \cite{CREX:2022kgg} collaborations, together with some well-determined properties of doubly magic nuclei, and it is shown that the standard Skyrme EDF can be consistent with both the PREX and CREX data for 90\% CI, which suggests the symmetry energy could be supersoft with negative vaule of $L$, i.e., $L=17.1_{-22.3(36.0)}^{+23.8(39.3)} \, \mathrm{MeV}$ at 68.3\% (90\%) CI.
To establish more rigorous constraints on $L$, further experimental data and theoretical investigations are necessary.

In this paper, based on the significant uncertainty of $L$, we construct four additional parameter sets with $L$ values of $-5$, $85$, $105$ and $125\,\mathrm{MeV}$ for comparison, and they are denoted as SP6Lm5, SP6L85, SP6L105 and SP6L125, respectively.
These interactions with supersoft and superstiff symmetry energies could be applied in transport models and tested against experimental data from HICs at intermediate and high energies.
We present the corresponding Skyrme parameters in Table~\ref{tab:10SetsParas} and the macroscopic characteristic quantities of nuclear matter in Table~\ref{tab:10SetsQuants} with these four interactions.
The properties of the SNM, including the pressure of the SNM and the single-nucleon potential, predicted by these four interactions are identical to those given by the default-SP6X interactions, which can be seen from Fig.~\ref{fig:flow1} and Fig.~\ref{fig:U0_1}.
The density dependence of the symmetry energy and the momentum dependence of the symmetry potential with these four interactions are shown in Fig.~\ref{fig:Esym_10Ls} and Fig.~\ref{fig:Usym_1}, respectively.
The EOS of the PNM predicted by these four interactions are presented in Fig.~\ref{fig:Epnm_10Ls}, and it can be seen that they hardly fit the microscopic calculations results at $\rho>0.12\,\mathrm{fm}^{-3}$.
It should be mentioned that SP6Lm5 predicts that the $E_{\mathrm{PNM}}$ will decrease with increasing density at $0.10\,\mathrm{fm}^{-3}$ to $0.17\,\mathrm{fm}^{-3}$.
This leads to negative pressure in PNM at these densities, as seen in Eq.~(\ref{eq:press}), suggesting the potential existence of a quasibound state for PNM \cite{Oyamatsu:2017qzv}.
The small values of $L$ and $E_{\mathrm{sym}}(\rho_{\mathrm{sc}})$ of SP6Lm5 may have substantial impact on the neutron drip line location as well as the r-process paths in the nuclear landscape \cite{Oyamatsu:2017qzv,Wang:2014mra}.
Moreover, the core-crust transition pressure of a neutron star could be negative in this scenario, suggesting the neutron star may have no crust structure, which could significantly affect the structure of neutron stars.

\begin{figure}[ht]
    \centering
    \includegraphics[width=\linewidth]{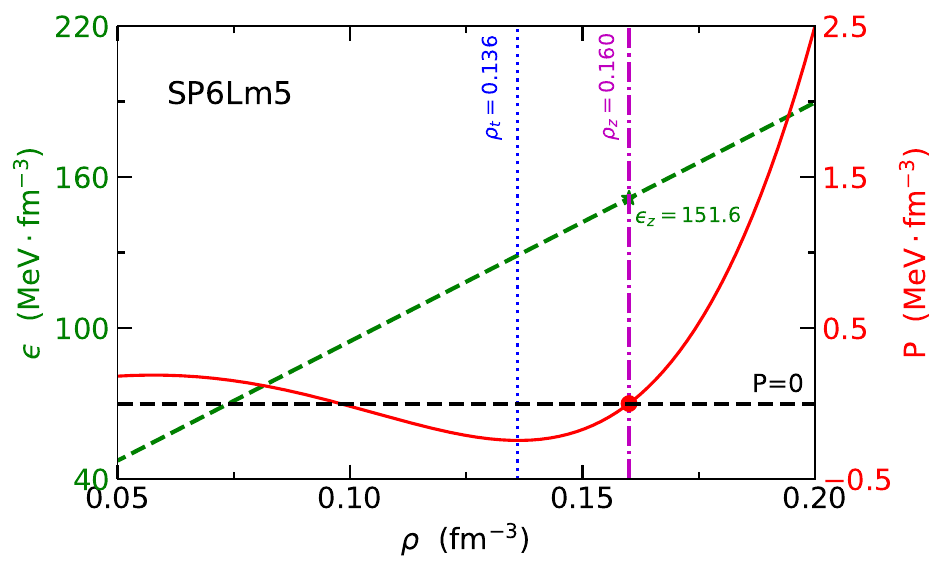}
    \caption{The energy density and the pressure of the $\beta$-equilibrium $npe\mu$ matter with the SP6Lm5 interaction.
    The vertical lines indicate the transition density $\rho_t = 0.136\,\mathrm{fm}^{-3}$ obtained through the dynamical method \cite{Xu:2009vi} and the zero-pressure density $\rho_z = 0.160\,\mathrm{fm}^{-3}$, respectively.
    $\rho_z$ corresponds to the density at neutron star surface, where the energy density $\epsilon_z$ is 151.6 $\mathrm{MeV}\,\mathrm{fm}^{-3}$.}
    \label{fig:Lm5EOS}
\end{figure}

Shown in Fig.~\ref{fig:Lm5EOS} is the energy density and the pressure as functions of nucleon density of the $\beta$-equilibrium $npe\mu$ matter predicted by the SP6Lm5 interaction.
Also shown in Fig.~\ref{fig:Lm5EOS} is the transition density $\rho_t = 0.136\,\mathrm{fm}^{-3}$ obtained through the dynamical method.
One can see SP6Lm5 predicts a negative transition pressure, implying that the pressure of the $\beta$-equilibrium nuclear matter in the uniform core decreases to zero before becoming dynamically unstable.
This interesting feature means that the neutron star matter would be against clusterization and the neutron stars could be composed of only uniform liquid core without crust.
The nucleon density $\rho_z$ and energy density $\epsilon_z$ at the zero-pressure point, which corresponds to the surface of the neutron star, are 0.160 $\mathrm{fm}^{-3}$ and 151.6 $\mathrm{MeV}\,\mathrm{fm}^{-3}$, respectively.

The particle fractions as functions of nucleon density in the $\beta$-equilibrium matter given by SP6Lm5, SP6L85, SP6L105 and SP6L125 are presented in Fig.~\ref{fig:Fractions}.
These four interactions are also applied in the neutron stars calculations, and all of them satisfy the causality condition.
We list the properties of the neutron stars obtained with these four interactions in Table~\ref{tab:10SetsNSs}, and the corresponding mass-radius relations are shown in Fig.~\ref{fig:MR}.
All of these four interactions meet the mass-radius determinations from astrophysical observations within the 68.3\%CI for both PSR J0030+0451 \cite{Miller:2019cac,Riley:2019yda} and PSR J0740+6620 \cite{Miller:2021qha,Riley:2021pdl}.
Additionally, SP6Lm5 is consistent with the observations within the 68.3\% CI for the CCO in HESS J1731-347 \cite{2022NatAs...6.1444D}.
The squared sound speed $C_{s}^{2}$ as a function of nucleon density for the four interactions are shown in Fig.~\ref{fig:Cs2}, and the interaction SP6Lm5 also predicts a peak in $C_{s}^{2}$.

\section{Interactions with different momentum dependencies}
\label{sec:EffMasses}
\subsection{Interactions with different isoscalar single-nucleon potentials and isoscalar nucleon effective masses}
\begin{figure}[ht]
    \centering
    \includegraphics[width=\linewidth]{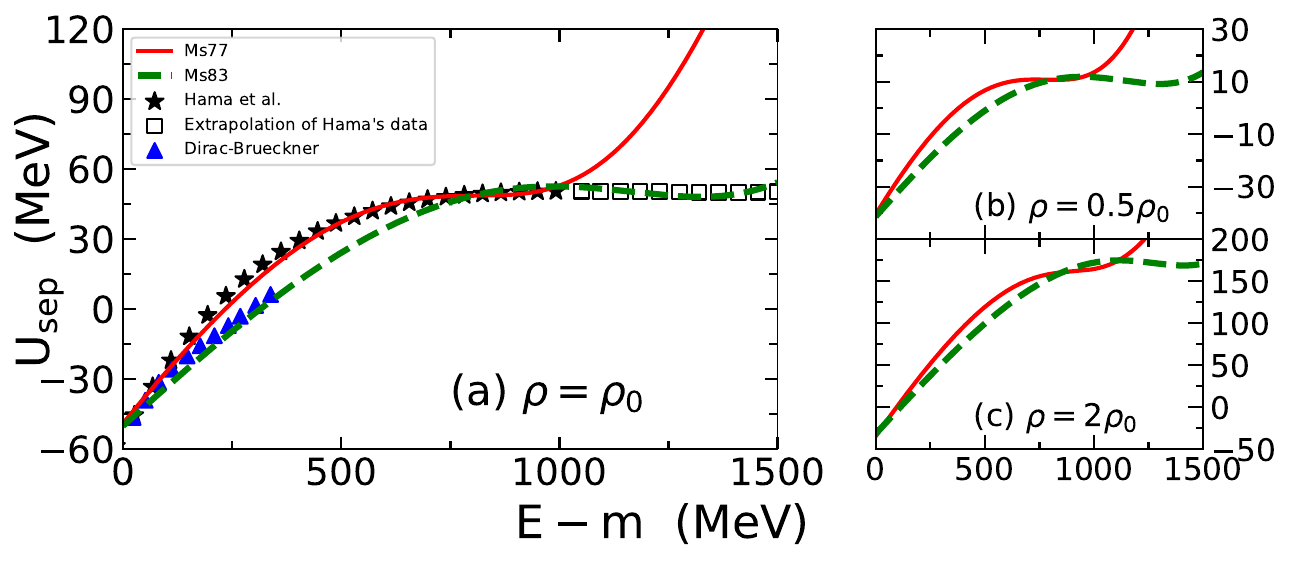}
    \caption{The energy dependence of the single-nucleon potential in cold SNM predicted by Ms77 and Ms83.
    Also shown is the nucleon optical potential (Schr\"{o}dinger equivalent potential) in SNM at $\rho_0$ obtained by Hama \textit{et al.}~\cite{Hama:1990vr,Cooper:1993nx} and the Dirac-Brueckner calculation~\cite{TerHaar:1986xpv}.}
    \label{fig:U0_compare}
\end{figure}

Above we have constructed a total of twelve interaction parameter sets, namely SP6X, with X = Lm5, \dots, L125, which have different density dependencies of the symmetry energy whereas their descriptions of the properties of SNM, including the single-nucleon potential $U_0$ shown in Fig.~\ref{fig:U0_1}, are identical.
The momentum dependence of $U_0$ is determined by fitting the empirical nucleon optical potential data derived from Hama \textit{et al.}'s analysis of proton-nucleus elastic scattering data spanning energies approximately from $0$ to $1\,\mathrm{GeV}$~\cite{Hama:1990vr,Cooper:1993nx}.
In addition, a saturated asymptotic behavior of $U_0$ seems to be anticipated for kinetic energies exceeding $1\,\mathrm{GeV}$, based on an extrapolation from Hama's data.
The single-nucleon potential of the aforementioned twelve interactions, denoted as Ms77 series, corresponding to the isoscalar nucleon effective mass $m^{\ast}_{s,0}=0.773m$, is shown in Fig.~\ref{fig:U0_compare}.
Hama's data \cite{Hama:1990vr,Cooper:1993nx} and the extrapolation are also included in Fig.~\ref{fig:U0_compare} for comparison.
As can be seen in Fig.~\ref{fig:U0_compare}(a), when nucleon kinetic energy exceeds $1\,\mathrm{GeV}$, $U_0$ exhibits a rapid deviation from the saturated behavior.
This behavior can be attributed to the polynomial structure of the $U_0$ in our model, and the apparent deviation could significantly impact the use of the one-body transport model, such as the BUU equation, in studying the HICs with incident energies beyond $1\,\mathrm{GeV}$.
For instance, this impact might be evident in studying the fixed-target Au+Au collision at $E_\mathrm{beam}=1.23\,\mathrm{AGeV}$ \cite{HADES:2020lob} (corresponding to $\sqrt{s_{N N}}=2.4 \,\mathrm{GeV}$) conducted by HADES collaboration.

To maintain the $U_{0}(\rho_0,p)$ saturated over a broader range, we repeat the model parameter optimization procedure in Section~\ref{sec:fitting}.
In this procedure, we keep the values of $\rho_0$, $E_{0}(\rho_0)$, $K_{0}$ and $J_{0}$ unchanged but assign a 20-fold weight to data with kinetic energies between $0.7\,\mathrm{GeV}$ and $1.5\,\mathrm{GeV}$ (approximately corresponding to momenta between $1.3\,\mathrm{GeV}/c $ and $2\,\mathrm{GeV}/c $), and the extrapolation data are also used in the optimization.
Now the coefficients in Eq.~(\ref{eq:U0_a0246}) are determined to be $a_0=-60.27496\,\mathrm{MeV}$, $a_2=4.273110\,\mathrm{MeV}\,\mathrm{fm}^{2}$,$a_4=-0.0523079\,\mathrm{MeV}\,\mathrm{fm}^{4}$ and $a_6=2.0429408 \times 10 ^{-4}\,\mathrm{MeV}\,\mathrm{fm}^{6}$, and $m_{s,0}^{\ast}$ is obtained to be $0.835m$.
We label this new single-nucleon potential as Ms83, whose momentum dependence is shown in Fig.~\ref{fig:U0_compare}.
Additionally, the theoretical calculation results using the Dirac-Brueckner approach \cite{TerHaar:1986xpv} is also shown in Fig.~\ref{fig:U0_compare}(a).
One can see Ms83 displays a rather saturated behavior within the energy range of 1 GeV to 1.5 GeV.
However, within low-energy range, Ms83 predicts a weaker energy dependence than Hama's data and Ms77, while aligning with the Dirac-Brueckner calculations.
Nevertheless, we would like to point out that future comparative study using Ms77 and Ms88 interactions in transport model simulations for heavy-ion collisions will give valuable information on the isoscalar nucleon effective mass and the high energy behavior of the isoscalar nucleon optical potential.

\begin{figure}[ht]
    \centering
    \includegraphics[width=\linewidth]{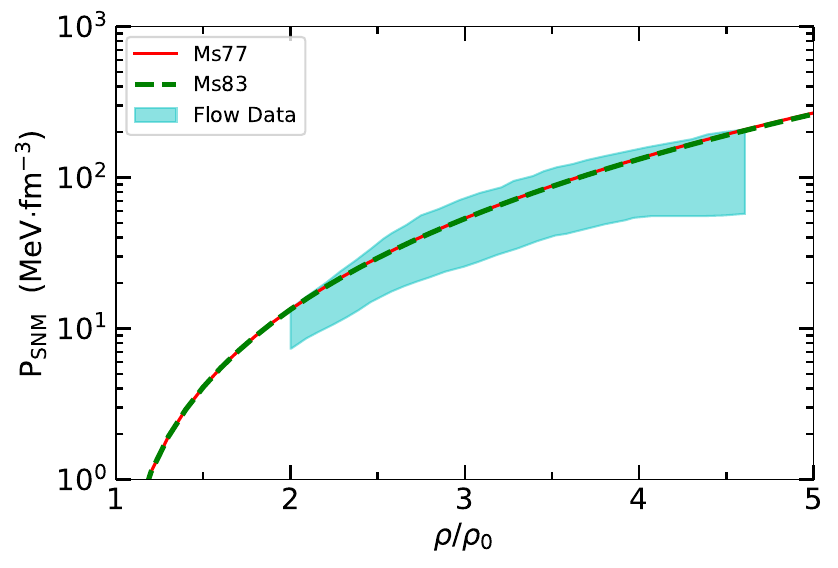}
    \caption{The pressure of SNM ($P_{\mathrm{SNM}}(\rho)$) as a function of nucleon density given by Ms77 and Ms83.
    Also included are the constraints from flow data in HICs~\cite{Danielewicz:2002pu}.}
    \label{fig:flow2}
\end{figure}

The three parameters $a_2$, $a_4$, $a_6$ together with $\rho_0$, $E_{0}(\rho_0)$, $K_{0}$, $J_{0}$ are necessary and sufficient to determine all the Skyrme parameters (and parameter combinations) related to SNM, i.e., $t_0$, $t_{3}^{[1]}$, $t_{3}^{[3]}$, $t_{3}^{[5]}$, $2C^{[2]}+D^{[2]}$, $2C^{[4]}+D^{[4]}$ and $2C^{[6]}+D^{[6]}$.
Table~\ref{tab:Two_SNM} summarizes the parameters (combinations) of SNM with Ms77 and Ms83, as well as the quantities of SNM given by them.
\begin{table*}
\caption{\label{tab:Two_SNM}
The quantities and parameters related to the SNM for Ms77 and Ms83 interaction series.
}
\begin{ruledtabular}
\begin{tabular}{cccccc}
Quantities    & Ms77 & Ms83 & Parameters (combinations)  & Ms77 & Ms83 \\ \hline
$\rho_0\,(\rm{fm}^{-3})$    & 0.160   & 0.160  & $t_0\,(\mathrm{MeV}\, \mathrm{fm}^3)$              & -1840.45 & -1850.36 \\
$E_0(\rho_0)\,(\rm{MeV})$   & -16.0   & -16.0  & $t_{3}^{[1]}\,(\mathrm{MeV}\, \mathrm{fm}^4)$      & 13010.2  & 13407.3  \\
$K_0\,(\rm{MeV})$   & 230.0   & 230.0  & $t_{3}^{[3]}\,(\mathrm{MeV}\, \mathrm{fm}^6)$      & -4036.41 & -3367.90 \\
$J_0\,(\rm{MeV})$   & -383.0  & -383.0 & $t_{3}^{[5]}\,(\mathrm{MeV}\, \mathrm{fm}^8)$      & 2386.36  & 1974.96  \\
$I_0\,(\rm{MeV})$   & 1819  & 1799  &
$2C^{[2]}+D^{[2]}\,(\mathrm{MeV}\, \mathrm{fm}^5)$ & 697.928  & 446.249  \\
$m^\ast_{s,0}/m$            & 0.773   & 0.835 & $2C^{[4]}+D^{[4]}\,(\mathrm{MeV}\, \mathrm{fm}^7)$ & -26.4050  & -10.7665  \\
$a_0\,(\mathrm{MeV})$       & -64.0345  & -60.2750 &
$2C^{[6]}+D^{[6]}\,(\mathrm{MeV}\, \mathrm{fm}^9)$ & 0.0813312  & 0.0204294 \\
$a_2\,(\mathrm{MeV}\,\mathrm{fm}^{2})$ & 6.51778  & 4.27311 \\
$a_4\,(\mathrm{MeV}\,\mathrm{fm}^{4})$ & -0.125955  & -0.0523079 & \\
$a_6\,(\mathrm{MeV}\,\mathrm{fm}^{6})$ & $8.13312\times10^{-4}$  & $2.04294\times10^{-4}$ & \\
\end{tabular}
\end{ruledtabular}
\end{table*}
In Fig.~\ref{fig:flow2}, we present the pressure of SNM, $P_{\mathrm{SNM}}(\rho)$, as a function of nucleon density given by Ms77 and Ms83.
Additionally, we include the constraints on $P_{\mathrm{SNM}}(\rho)$ obtained from collective flow data \cite{Danielewicz:2002pu}.
Both Ms77 and Ms83 are compatible with the constraints given by the flow data.

We also notice that phenomenologically introducing higher-order momentum terms (e.g., $p^8$) into the momentum-dependent terms could enhance the flexibility of the single-nucleon potential \cite{Yang:2023umm}, which can maintain the saturated behavior over a broader energy range.
However, higher-order density gradient terms (e.g., $\nabla^{8}\rho$) will appear in the Skyrme pseudopotential EDF form, which could pose additional challenges in the calculations of finite nuclei.

\subsection{Interactions with different symmetry potentials and isovector nucleon effective masses}
The symmetry potential $U_{\mathrm{sym}}$ as well as the isovector nucleon effective mass $m_{v}^{\ast}$ are closely related to many essential questions in both nuclear physics and astrophysics \cite{Li:2018lpy}, but they still suffer from significant uncertainties.
Thus, we construct two other sets of $b_2$, $b_4$ and $b_6$ in Eq.~(\ref{eq:Usym_b0246}), in addition to the one we have constructed in Section~\ref{sec:fitting}.
Furthermore, we change the sign of the values of $b_2$, $b_4$ and $b_6$, which will reverse the momentum dependence of the symmetry potential and lead to $m_{v}^{\ast}>m_{s}^{\ast}$, as well as negative nucleon effective mass splitting ($\mspl<0$) in neutron-rich matter.
In the present work, we have constructed two different types of momentum dependencies for $U_0$ and six different types of momentum dependencies for $U_{\mathrm{sym}}$, resulting in a total of twelve different momentum dependence types, i.e., twelve different sets of $C^{[n]}$ and $D^{[n]}$ ($n=2,4,6$).
We label these sets with corresponding values of their $m_{s,0}^{\ast}$ and $m_{v,0}^{\ast}$, namely Ms77Mv69, Ms77Mv86, Ms77Mv67, Ms77Mv89, Ms77Mv64, Ms77Mv94, Ms83Mv74, Ms83Mv94, Ms83Mv72, Ms83Mv98, Ms83Mv69 and Ms83Mv104.
We present in Table~\ref{tab:MsMv_quants} the isovector nucleon effective mass at saturation density $m_{v,0}^{\ast}$, the nucleon effective mass splitting at $\rho_0$ with isospin asymmetry being 0.5 ($\mspl(\rho_0,0.5)$) and 1 ($\mspl(\rho_0,1)$), as well as the values of $b_2$, $b_4$ and $b_6$ for these twelve momentum dependencies.
The fourth-order symmetry energy and the linear isospin splitting coefficient are only relevant to $C^{[n]}$ and $D^{[n]}$ ($n=2,4,6$), and the values of $E_{\mathrm{sym},4}(\rho_0)$ and $\Delta m_1^{\ast}(\rho_0)$ are also shown in Table~\ref{tab:MsMv_quants}.
It is seen that the values of $E_{\mathrm{sym},4}(\rho_0)$ predicted by these twelve momentum dependencies are in good agreement with the estimated values obtained from other commonly used non-relativistic models \cite{Pu:2017kjx}.
Additionally, it is observed that all positive values of $\Delta m_1^{\ast}(\rho_0)$ (predicted by the momentum dependencies with $\mspl>0$ in neutron-rich matter) are consistent with the values extracted using different theoretical methods \cite{Li:2014qta,Zhang:2015qdp,Kong:2017nil,Wang:2023owh}.
It is worth noting that despite $\mspl>0$ being valid around the saturation density and the corresponding Fermi momentum, as predicted by analyzing the symmetry potential $U_{\mathrm{sym}}(\rho_0,p)$ using nuclear optical potential data \cite{Li:2008gp,Li:2004zi,Xu:2010fh,Li:2018lpy}, $\mspl<0$ is theoretically possible at high momentum \cite{Chen:2011ag}.
These predictions can be tested in the transport models for HICs by investigating the light particle emission \cite{Zhang:2014sva,Wang:2023buv}.
\begin{table*}
\caption{\label{tab:MsMv_quants}
Isovector nucleon effective mass at saturation density ($m_{v,0}^{\ast}$), nucleon effective mass splitting at saturation density with isospin asymmetry being 0.5 ($\mspl(\rho_0,0.5)$) and 1 ($\mspl(\rho_0,1)$), the values of $b_2$, $b_4$ and $b_6$ as well as values of the fourth-order symmetry energy $E_{\mathrm{sym},4}(\rho_0)$ and the linear isospin splitting coefficient $\Delta m_1^{\ast}(\rho_0)$ at saturation density for the interaction series with twelve momentum dependencies (i.e., twelve different combinations of $m_{s,0}^{\ast}$ and $m_{v,0}^{\ast}$).
}
\begin{ruledtabular}
\begin{tabular}{c|cccccc|cccccc}
&\multicolumn{6}{c|}{Ms77} & \multicolumn{6}{c}{Ms83} \\ \hline
Quantities        & Mv69 & Mv86 & Mv67 & Mv89 & Mv64 & Mv94 & Mv74 & Mv94 & Mv72 & Mv98 & Mv69 & Mv104 \\ \hline
$m_{v,0}^{\ast}/m$  & 0.691 & 0.865 & 0.673 & 0.895 & 0.648 & 0.943 & 0.743 & 0.948 & 0.722 & 0.984 & 0.693 & 1.043  \\
$\mspl(\rho_0,0.5)/m$ &
0.0794 & -0.0794 & 0.101 & -0.101 & 0.142 & -0.142 & 0.0925 & -0.0925 & 0.117 & -0.117 & 0.166 & -0.166
\\
$\mspl(\rho_0,1)/m$ &
0.164  & -0.164  & 0.208 & -0.208 & 0.297 & -0.297 & 0.189  & -0.189  & 0.241 & -0.241 & 0.345 & -0.345
\\
$b_2\,(\mathrm{MeV}\, \mathrm{fm}^{2})$ &
-3.0    & 3.0   & -3.8  & 3.8   & -5.0  & 5.0   & -3.0    & 3.0   & -3.8  & 3.8   & -5.0  & 5.0
\\
$b_4\,(\mathrm{MeV}\, \mathrm{fm}^{4})$ &
0.078   & -0.078    & 0.10  & -0.10  & 0.035    & -0.035    & 0.078   & -0.078    & 0.10  & -0.10  & 0.035    & -0.035
\\
$b_6\,(\mathrm{MeV}\, \mathrm{fm}^{6})$ &
-0.0007 & 0.0007 & -0.001 & 0.001 & $-1\times10^{-7}$ & $1\times10^{-7}$ &
-0.0007 & 0.0007 & -0.001 & 0.001 & $-1\times10^{-7}$ & $1\times10^{-7}$
\\
$E_{\mathrm{sym},4}(\rho_0)\,(\mathrm{MeV})$ &
0.7471  & 0.3183  & 0.8043  & 0.2611  & 0.8699 & 0.1954 &
0.7353  & 0.3066  & 0.7926  & 0.2494  & 0.8582 & 0.1837
\\
$\Delta m_1^{\ast}(\rho_0)$ &
0.1740  & -0.1408 & 0.2158  & -0.1826 & 0.2978 & -0.2646 &
0.1918  & -0.1755 & 0.2406  & -0.2247 & 0.3362 & -0.3199
\end{tabular}
\end{ruledtabular}
\end{table*}
In Table~\ref{tab:12_CDs}, we list the values of $C^{[n]}$, $D^{[n]}$ for these twelve momentum dependencies.
\begin{table*}
\caption{\label{tab:12_CDs}
Values of $C^{[n]}$, $D^{[n]}$ for the interaction series with twelve momentum dependencies (i.e., twelve different combinations of $m_{s,0}^{\ast}$ and $m_{v,0}^{\ast}$).
}
\begin{ruledtabular}
\begin{tabular}{c|cccccc|cccccc}
&\multicolumn{6}{c|}{Ms77} & \multicolumn{6}{c}{Ms83} \\ \hline
Parameters\footnote{The unit of $C^{[n]}$ ($D^{[n]}$) is $\mathrm{MeV} \, \mathrm{fm}^{n+3}$.}    &
Mv69 & Mv86 & Mv67 & Mv89 & Mv64 & Mv94 & Mv74 & Mv94 & Mv72 & Mv98 & Mv69 & Mv104 \\ \hline
$C^{[2]}$&
523.869   & 174.058   & 571.158   & 126.769   & 609.330   & 88.5980   &
398.030   & 48.2192   & 445.319   & 0.930128  & 483.490   & -37.2413
\\
$D^{[2]}$&
-349.811  & 349.811   & -444.389  & 444.389   & -520.732  & 520.732   &
-349.811  & 349.811   & -444.389  & 444.389   & -520.732  & 520.732
\\
$C^{[4]}$&
-21.8732  & -4.53180  & -24.4464  & -1.95864  & -16.7026  & -9.70238  &
-14.0540  & 3.28744   & -16.6271  & 5.86060   & -8.88339  & -1.88314
\\
$D^{[4]}$&
17.3414   &-17.3414   & 22.4877   &-22.4877   & 7.00025   &-7.00025   &
17.3414   &-17.3414   & 22.4877   &-22.4877   & 7.00025   &-7.00025
\\
$C^{[6]}$&
0.07567   & 0.00567   & 0.09067   &-0.00933   & 0.04067   & 0.04066   &
0.04521   &-0.02479   & 0.06021   &-0.03979   & 0.01022   & 0.01021
\\
$D^{[6]}$&
-0.07000  &  0.07000  & -0.10000  & 0.10000   & -0.00001  & 0.00001   &
-0.07000  & 0.07000   & -0.10000  & 0.10000   & -0.00001  & 0.00001
\\
\end{tabular}
\end{ruledtabular}
\end{table*}

\begin{figure}[ht]
    \centering
    \includegraphics[width=\linewidth]{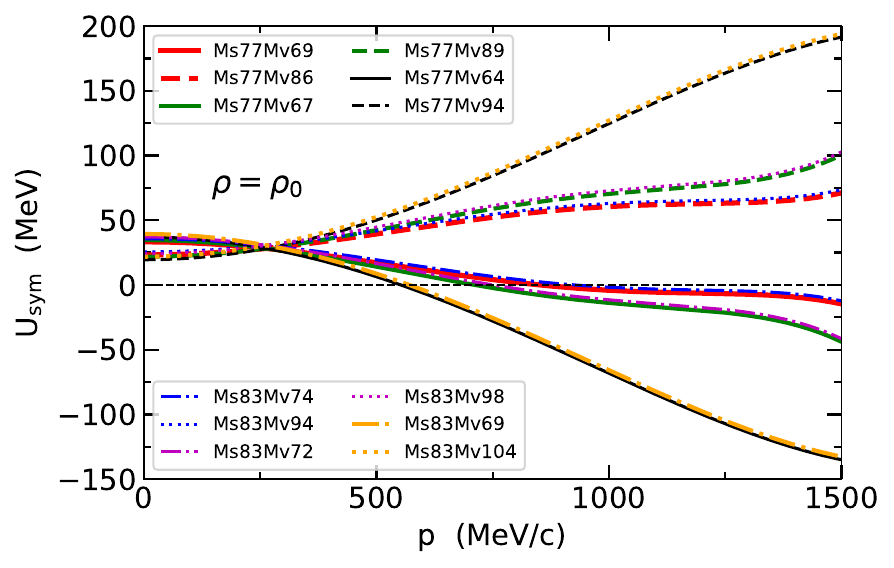}
    \caption{Momentum dependence of the symmetry potential at saturation density with various interactions (see text for the details). Here $E_{\mathrm{sym}}(\rho_0) = 30\,\mathrm{MeV}$ has been assumed.
    }
    \label{fig:Usym_6Types}
\end{figure}

Shown in Fig.~\ref{fig:Usym_6Types} is the different momentum dependencies of the symmetry potential at saturation density.
Note that the momentum dependence of $U_{\mathrm{sym}}$ depends solely on $b_2$, $b_4$ and $b_6$.
As can be seen from Eq.~(\ref{eq:Usym_b0246}) and Eq.~(\ref{eq:HVH_Cor}), fixing $b_2$, $b_4$, $b_6$ and changing $m_{s,0}^{\ast}$ or $E_{\mathrm{sym}}(\rho_0)$ results in a vertical shift of $U_{\mathrm{sym}}$.
In Fig.~\ref{fig:Usym_6Types}, we take $E_{\mathrm{sym}}(\rho_0)=30\,\mathrm{MeV}$ as an example for plotting.
One sees that very different high-energy behaviors can be obtained from the twelve interaction series with different momentum dependencies of the symmetry potentials.
This will be very useful for the determination of the isovector nucleon effective mass and the isospin splitting coefficient of nucleon effective mass in HICs.

\subsection{Interaction parameter family with various symmetry energies and nucleon effective masses}
In the following, we will combine the twelve momentum dependencies, featured by $m_{s,0}^{\ast}$ and $m_{v,0}^{\ast}$, with different density dependencies of the symmetry energy, which are characterized by $L$ (and $E_{\mathrm{sym}}(\rho_0)$, $K_{\mathrm{sym}}$, $J_{\mathrm{sym}}$).
We have obtained twelve different symmetry energy behaviors in previous sections, and the corresponding values of $E_{\mathrm{sym}}(\rho_0)$, $L$, $K_{\mathrm{sym}}$ and $J_{\mathrm{sym}}$ are presented in Table~\ref{tab:10SetsQuants}.
The combination of the twelve different momentum dependencies and twelve different symmetry energies forms a parameter set family consisting of $12 \times 12=144$ parameter sets, and we denote them by the corresponding values of their $L$, $m_{s,0}^{\ast}$ and $m_{v,0}^{\ast}$.
For example, the parameter sets we have obtained in Section~\ref{sec:fitting} and Section~\ref{sec:soft-stiff} are now SP6Lm5Ms77Mv69, SP6L15Ms77Mv69, $\dots$ and SP6L125Ms77Mv69 under the new notation system, respectively.
Our parameter set family provides abundant options for conducting control experiments:
using parameter sets with the same $a_{2}$, $a_{4}$, $a_{6}$ and $b_{2}$, $b_{4}$, $b_{6}$ to study the effects of the symmetry energy;
using parameter sets with the same $a_{2}$, $a_{4}$, $a_{6}$ and $L$ to study the the effects of nucleon effective mass splitting and the symmetry potential (isovector nucleon effective mass);
using parameter sets with the same $b_{2}$, $b_{4}$, $b_{6}$ and $L$ to study the the effects of the single-nucleon potential (isoscalar nucleon effective mass).
For a certain interaction, one can find the values of $t_0$, $t_{3}^{[1]}$, $t_{3}^{[3]}$ and $t_{3}^{[5]}$ in Table~\ref{tab:Two_SNM}, the values of $C^{[n]}$ and $D^{[n]}$ ($n=2,4,6$) in Table~\ref{tab:12_CDs}, and the values of $x_0$, $x_{3}^{[1]}$, $x_{3}^{[3]}$ and $x_{3}^{[5]}$ in Table~\ref{tab:10_x0135}.
It is worth emphasizing again that these 14 parameters are relevant to the properties of uniform nuclear matter.
Only when the values of $E^{[n]}$ and $F^{[n]}$ ($n=2,4,6$) are determined through finite nuclei calculations can all the 20 Skyrme parameters of the new extended Skyrme interaction (Eq.~(\ref{eq:Vsk})) be completely obtained.
\squeezetable
\begin{table*}
\caption{\label{tab:10_x0135}
The values of $x_0$, $x_{3}^{[1]}$, $x_{3}^{[3]}$ and $x_{3}^{[5]}$ corresponding to different symmetry energies in different momentum dependencies.
The values of $t_0$, $t_{3}^{[1]}$, $t_{3}^{[3]}$ and $t_{3}^{[5]}$ corresponding to different $m_{s,0}^{\ast}$ values (different single-nucleon potentials) are shown in Table~\ref{tab:Two_SNM}, and the values of $C^{[n]}$ and $D^{[n]}$ corresponding to different $m_{s,0}^{\ast}$ and $m_{v,0}^{\ast}$ values combinations (different single-nucleon potentials and symmetry potentials) are shown in Table~\ref{tab:12_CDs}.
}
\begin{ruledtabular}
\begin{tabular}{c|cccccc|cccccc}
&\multicolumn{6}{c|}{Ms77} & \multicolumn{6}{c}{Ms83} \\ \hline
&Mv69 & Mv86 & Mv67 & Mv89 & Mv64 & Mv94 & Mv74 & Mv94 & Mv72 & Mv98 & Mv69 & Mv104 \\ \hline
\textbf{Lm5}
&\multicolumn{3}{l}{\bm{$L=-5\mathrm{MeV}$}}&\multicolumn{3}{l}{\bm{$E_{\mathrm{sym}}(\rho_0)=24\mathrm{MeV}$}}&
\multicolumn{3}{l}{\bm{$K_{\mathrm{sym}}=-10\mathrm{MeV}$}}&\multicolumn{3}{l}{\bm{$J_{\mathrm{sym}}=4250\mathrm{MeV}$}} \\ \cline{1-1}
$x_0$         & -0.530282 & -0.491696  & -0.535620 & -0.486349 & -0.536927 & -0.485049 &
                -0.535259 & -0.496875  & -0.540578 & -0.491567 & -0.541875 & -0.490266 \\
$x_{3}^{[1]}$ & -3.00100  & -2.78452   & -3.03074  & -2.75479  & -3.04328  & -2.74225  &
                -2.95415  & -2.74403   & -2.98297  & -2.71519  & -2.99516  & -2.70304  \\
$x_{3}^{[3]}$ & -38.8852  & -39.9549   & -38.7497  & -40.0924  & -38.4149  & -40.4270  &
                -46.4020  & -47.6805   & -46.2346  & -47.8440  & -45.8349  & -48.2480  \\
$x_{3}^{[5]}$ & -84.9707  & -86.1993   & -84.8169  & -86.3592  & -84.8532  & -86.3220  &
                -102.350  & -103.825   & -102.150  & -104.014  & -102.198  & -103.978  \\ \hline
\textbf{L5}
&\multicolumn{3}{l}{\bm{$L=5\mathrm{MeV}$}}&\multicolumn{3}{l}{\bm{$E_{\mathrm{sym}}(\rho_0)=28\mathrm{MeV}$}}&
\multicolumn{3}{l}{\bm{$K_{\mathrm{sym}}=-250\mathrm{MeV}$}}&\multicolumn{3}{l}{\bm{$J_{\mathrm{sym}}=2100\mathrm{MeV}$}} \\ \cline{1-1}
$x_0$         & -0.378151 & -0.339556  & -0.383487 & -0.334220 & -0.384793 & -0.332911 &
                -0.383947 & -0.345554  & -0.389249 & -0.340241 & -0.390545 & -0.338951 \\
$x_{3}^{[1]}$ & -2.13560  & -1.91909   & -2.16532  & -1.88938  & -2.17787  & -1.87683  &
                -2.11437  & -1.90425   & -2.14317  & -1.87540  & -2.15535  & -1.86325  \\
$x_{3}^{[3]}$ & -23.9051  & -24.9739   & -23.7686  & -25.1113  & -23.4340  & -25.4460  &
                -28.4469  & -29.7269   & -28.2813  & -29.8904  & -27.8808  & -30.2926  \\
$x_{3}^{[5]}$ & -43.7167  & -44.9425   & -43.5597  & -45.1016  & -43.5966  & -45.0651  &
                -52.4995  & -53.9783   & -52.3052  & -54.1680  & -52.3507  & -54.1263  \\ \hline
\textbf{L15}
&\multicolumn{3}{l}{\bm{$L=15\mathrm{MeV}$}}&\multicolumn{3}{l}{\bm{$E_{\mathrm{sym}}(\rho_0)=29\mathrm{MeV}$}}&
\multicolumn{3}{l}{\bm{$K_{\mathrm{sym}}=-240\mathrm{MeV}$}}&\multicolumn{3}{l}{\bm{$J_{\mathrm{sym}}=1450\mathrm{MeV}$}} \\ \cline{1-1}
$x_0$         & -0.0684472 & -0.0298492  & -0.0737822 & -0.0245124 & -0.0750818 & -0.0232057 &
                -0.0758945 & -0.0375044  & -0.0812067 & -0.0321990 & -0.0825030 & -0.0308997 \\
$x_{3}^{[1]}$ & -1.08967  & -0.873154  & -1.11938  & -0.843443 & -1.13192  & -0.830897 &
                -1.09940  & -0.889288  & -1.12823  & -0.860458 & -1.14040  & -0.848292 \\
$x_{3}^{[3]}$ & -14.4987  & -15.5673   & -14.3619  & -15.7045  & -14.0271  & -16.0393  &
                -17.1731  & -18.4532   & -17.0080  & -18.6168  & -16.6071  & -19.0192  \\
$x_{3}^{[5]}$ & -26.3216  & -27.5469   & -26.1639  & -27.7056  & -26.2003  & -27.6693  &
                -31.4807  & -32.9596   & -31.2873  & -33.1494  & -31.3320  & -33.1082  \\ \hline
\textbf{L25}
&\multicolumn{3}{l}{\bm{$L=25\mathrm{MeV}$}}&\multicolumn{3}{l}{\bm{$E_{\mathrm{sym}}(\rho_0)=30\mathrm{MeV}$}}&
\multicolumn{3}{l}{\bm{$K_{\mathrm{sym}}=-210\mathrm{MeV}$}}&\multicolumn{3}{l}{\bm{$J_{\mathrm{sym}}=1200\mathrm{MeV}$}} \\ \cline{1-1}
$x_0$         & 0.0420416  & 0.0806320   & 0.0366990  & 0.0859689  & 0.0353932  & 0.0872735  &
                0.0340008  & 0.0723822   & 0.0286826  & 0.0776876  & 0.0273863  & 0.0789890  \\
$x_{3}^{[1]}$ & -0.680842 & -0.464331  & -0.710567 & -0.434621 & -0.723112 & -0.422079 &
                -0.702677 & -0.492581  & -0.731523 & -0.463752 & -0.743691 & -0.451579 \\
$x_{3}^{[3]}$ & -10.5117  & -11.5800   & -10.3747  & -11.7173  & -10.0399  & -12.0521  &
                -12.3944  & -13.6747   & -12.2296  & -13.8384  & -11.8283  & -14.2404  \\
$x_{3}^{[5]}$ & -19.5903  & -20.8148   & -19.4322  & -20.9736  & -19.4684  & -20.9374  &
                -23.3467  & -24.8259   & -23.1538  & -25.0160  & -23.1977  & -24.9739  \\ \hline
\textbf{L35}
&\multicolumn{3}{l}{\bm{$L=35\mathrm{MeV}$}}&\multicolumn{3}{l}{\bm{$E_{\mathrm{sym}}(\rho_0)=30\mathrm{MeV}$}}&
\multicolumn{3}{l}{\bm{$K_{\mathrm{sym}}=-190\mathrm{MeV}$}}&\multicolumn{3}{l}{\bm{$J_{\mathrm{sym}}=670\mathrm{MeV}$}} \\ \cline{1-1}
$x_0$         & 0.150705  & 0.189301   & 0.145368  & 0.194639  & 0.144061  & 0.195939  &
                0.142082  & 0.180467   & 0.136758  & 0.185776  & 0.135485  & 0.187075  \\
$x_{3}^{[1]}$ & -0.123375 & 0.0931500   & -0.153086 & 0.122866  & -0.165636 & 0.135392  &
                -0.161721 & 0.0483831   & -0.190582 & 0.0772171  & -0.202700 & 0.0893879  \\
$x_{3}^{[3]}$ & -4.12449  & -5.19276   & -3.98754  & -5.32981  & -3.65274  & -5.66486  &
                -4.73965  & -6.01970   & -4.57531  & -6.18371  & -4.17370  & -6.58550  \\
$x_{3}^{[5]}$ & -7.26000  & -8.48441   & -7.10190  & -8.64273  & -7.13818  & -8.60696  &
                -8.44867  & -9.92749   & -8.25699  & -10.1182  & -8.30033  & -10.0756  \\ \hline
\textbf{L45}
&\multicolumn{3}{l}{\bm{$L=45\mathrm{MeV}$}}&\multicolumn{3}{l}{\bm{$E_{\mathrm{sym}}(\rho_0)=30\mathrm{MeV}$}}&
\multicolumn{3}{l}{\bm{$K_{\mathrm{sym}}=-110\mathrm{MeV}$}}&\multicolumn{3}{l}{\bm{$J_{\mathrm{sym}}=700\mathrm{MeV}$}} \\ \cline{1-1}
$x_0$         & 0.241259  & 0.279857   & 0.235926  & 0.285195  & 0.234622  & 0.286499  &
                0.232158  & 0.270542   & 0.226842  & 0.275844  & 0.225544  & 0.277147  \\
$x_{3}^{[1]}$ & 0.221723  & 0.438254   & 0.192019  & 0.467970  & 0.179481  & 0.480506  &
                0.173172  & 0.383274   & 0.144323  & 0.412092  & 0.132156  & 0.424271  \\
$x_{3}^{[3]}$ & -1.60841  & -2.67657   & -1.47141  & -2.81363  & -1.13635  & -3.14858  &
                -1.72396  & -3.00408   & -1.55972  & -3.16831  & -1.15824  & -3.56985  \\
$x_{3}^{[5]}$ & -5.59382  & -6.81815   & -5.43577  & -6.97655  & -5.47144  & -6.94052  &
                -6.43530  & -7.91419   & -6.24363  & -8.10525  & -6.28707  & -8.06222  \\ \hline
\textbf{L55}
&\multicolumn{3}{l}{\bm{$L=55\mathrm{MeV}$}}&\multicolumn{3}{l}{\bm{$E_{\mathrm{sym}}(\rho_0)=33\mathrm{MeV}$}}&
\multicolumn{3}{l}{\bm{$K_{\mathrm{sym}}=-100\mathrm{MeV}$}}&\multicolumn{3}{l}{\bm{$J_{\mathrm{sym}}=900\mathrm{MeV}$}} \\ \cline{1-1}
$x_0$         & 0.128971  & 0.171249   & 0.123626  & 0.173254  & 0.122331  & 0.173679  &
                0.120462  & 0.158850   & 0.115152  & 0.164158  & 0.113855  & 0.169259  \\
$x_{3}^{[1]}$ & -0.224252 & 0.0009036   & -0.253982 & 0.0228114  & -0.266501 & 0.0332829  &
                -0.259614 & -0.0495037  & -0.288447 & -0.0206733 & -0.300614 & 0.0004193  \\
$x_{3}^{[3]}$ & -4.93739  & -5.96366   & -4.80058  & -6.13873  & -4.46552  & -6.48348  &
                -5.71383  & -6.99388   & -5.54935  & -7.15814  & -5.14793  & -7.50646  \\
$x_{3}^{[5]}$ & -11.7254  & -12.8957   & -11.5676  & -13.1034  & -11.6035  & -13.0798  &
                -13.8440  & -15.3228   & -13.6519  & -15.5141  & -13.6954  & -15.4045  \\ \hline
\textbf{L65}
&\multicolumn{3}{l}{\bm{$L=65\mathrm{MeV}$}}&\multicolumn{3}{l}{\bm{$E_{\mathrm{sym}}(\rho_0)=34\mathrm{MeV}$}}&
\multicolumn{3}{l}{\bm{$K_{\mathrm{sym}}=-70\mathrm{MeV}$}}&\multicolumn{3}{l}{\bm{$J_{\mathrm{sym}}=650\mathrm{MeV}$}} \\ \cline{1-1}
$x_0$         & 0.239451  & 0.278048   & 0.234112  & 0.283385  & 0.232809  & 0.284689  &
                0.230352  & 0.268739   & 0.225042  & 0.274045  & 0.223744  & 0.275345  \\
$x_{3}^{[1]}$ & 0.184564  & 0.401093   & 0.154847  & 0.430807  & 0.142311  & 0.443344  &
                0.137098  & 0.347205   & 0.108262  & 0.376033  & 0.0960940  & 0.388207  \\
$x_{3}^{[3]}$ & -0.950334 & -2.01846   & -0.813351 & -2.15552  & -0.478294 & -2.49054  &
                -0.935327 & -2.21537   & -0.771047 & -2.37962  & -0.369550 & -2.78117  \\
$x_{3}^{[5]}$ & -4.99398  & -6.21824   & -4.83589  & -6.37665  & -4.87161  & -6.34079  &
                -5.71064  & -7.18927   & -5.51886  & -7.38050  & -5.56230  & -7.33747  \\ \hline
\textbf{L75}
&\multicolumn{3}{l}{\bm{$L=75\mathrm{MeV}$}}&\multicolumn{3}{l}{\bm{$E_{\mathrm{sym}}(\rho_0)=34\mathrm{MeV}$}}&
\multicolumn{3}{l}{\bm{$K_{\mathrm{sym}}=-10\mathrm{MeV}$}}&\multicolumn{3}{l}{\bm{$J_{\mathrm{sym}}=550\mathrm{MeV}$}} \\ \cline{1-1}
$x_0$         & 0.284729  & 0.323327   & 0.279395  & 0.328664  & 0.278087  & 0.329975  &
                0.275389  & 0.313777   & 0.270075  & 0.319078  & 0.268783  & 0.320386  \\
$x_{3}^{[1]}$ & 0.450028  & 0.666561   & 0.420324  & 0.696273  & 0.407777  & 0.708827  &
                0.394703  & 0.604815   & 0.365859  & 0.633629  & 0.353703  & 0.645821  \\
$x_{3}^{[3]}$ & 1.75934   & 0.691306   & 1.89642   & 0.554268  & 2.23145   & 0.219304  &
                2.31229   & 1.03217    & 2.47629   & 0.867835  & 2.87804   & 0.466438  \\
$x_{3}^{[5]}$ & -1.66162  & -2.88572   & -1.50332  & -3.04396  & -1.53912  & -3.00806  &
                -1.68388  & -3.16285   & -1.49256  & -3.35410  & -1.53579  & -3.31093  \\ \hline
\textbf{L85}
&\multicolumn{3}{l}{\bm{$L=85\mathrm{MeV}$}}&\multicolumn{3}{l}{\bm{$E_{\mathrm{sym}}(\rho_0)=36\mathrm{MeV}$}}&
\multicolumn{3}{l}{\bm{$K_{\mathrm{sym}}=10\mathrm{MeV}$}}&\multicolumn{3}{l}{\bm{$J_{\mathrm{sym}}=470\mathrm{MeV}$}} \\ \cline{1-1}
$x_0$         & 0.333630  & 0.372234   & 0.328298  & 0.377563  & 0.326990  & 0.378867  &
                0.324027  & 0.362415   & 0.318716  & 0.367721  & 0.317421  & 0.369023  \\
$x_{3}^{[1]}$ & 0.577452  & 0.793999   & 0.547753  & 0.823693  & 0.535205  & 0.836229  &
                0.518351  & 0.728460   & 0.489512  & 0.757287  & 0.477352  & 0.769467  \\
$x_{3}^{[3]}$ & 3.15289   & 2.08496    & 3.29006   & 1.94784   & 3.62502   & 1.61277   &
                3.98244   & 2.70229    & 4.14634   & 2.53801   & 4.54816   & 2.13660   \\
$x_{3}^{[5]}$ & 0.471173  & -0.752748  & 0.629521  & -0.911146 & 0.593745  & -0.875421 &
                0.893138  & -0.585807  & 1.08426   & -0.777069 & 1.04129   & -0.733884 \\ \hline
\textbf{L105}
&\multicolumn{3}{l}{\bm{$L=105\mathrm{MeV}$}}&\multicolumn{3}{l}{\bm{$E_{\mathrm{sym}}(\rho_0)=37\mathrm{MeV}$}}&
\multicolumn{3}{l}{\bm{$K_{\mathrm{sym}}=150\mathrm{MeV}$}}&\multicolumn{3}{l}{\bm{$J_{\mathrm{sym}}=220\mathrm{MeV}$}} \\ \cline{1-1}
$x_0$         & 0.480330 & 0.518930  & 0.475000 & 0.524273 & 0.473695 & 0.525571 &
                0.469945 & 0.508333  & 0.464634 & 0.513644 & 0.463342 & 0.514938 \\
$x_{3}^{[1]}$ & 1.09245  & 1.30899   & 1.06275  & 1.33871  & 1.05021  & 1.35123  &
                1.01810  & 1.22821   & 0.989261 & 1.25705  & 0.977110 & 1.26921  \\
$x_{3}^{[3]}$ & 6.75277  & 5.68495   & 6.89019  & 5.54809  & 7.22529  & 5.21290  &
                8.29703  & 7.01691   & 8.46072  & 6.85264  & 8.86278  & 6.45120  \\
$x_{3}^{[5]}$ & 1.87073  & 0.646814  & 2.02919  & 0.488642 & 1.99348  & 0.524268 &
                2.58429  & 1.10535   & 2.77530  & 0.914144 & 2.73240  & 0.957265 \\ \hline
\textbf{L125}
&\multicolumn{3}{l}{\bm{$L=125\mathrm{MeV}$}}&\multicolumn{3}{l}{\bm{$E_{\mathrm{sym}}(\rho_0)=39\mathrm{MeV}$}}&
\multicolumn{3}{l}{\bm{$K_{\mathrm{sym}}=220\mathrm{MeV}$}}&\multicolumn{3}{l}{\bm{$J_{\mathrm{sym}}=320\mathrm{MeV}$}} \\ \cline{1-1}
$x_0$         & 0.565458 & 0.604056  & 0.560125 & 0.609397 & 0.558819 & 0.610691 &
                0.554611 & 0.592996  & 0.549301 & 0.598309 & 0.548007 & 0.599607 \\
$x_{3}^{[1]}$ & 1.48534  & 1.70188   & 1.45564  & 1.73160  & 1.44311  & 1.74411  &
                1.39935  & 1.60945   & 1.37051  & 1.63829  & 1.35835  & 1.65047  \\
$x_{3}^{[3]}$ & 11.2433  & 10.1753   & 11.3805  & 10.0386  & 11.7158  & 9.70317  &
                13.6788  & 12.3982   & 13.8423  & 12.2341  & 14.2445  & 11.8330  \\
$x_{3}^{[5]}$ & 8.66906  & 7.44521   & 8.82760  & 7.28724  & 8.79203  & 7.32251  &
                10.7986  & 9.31922   & 10.9890  & 9.12807  & 10.9466  & 9.17170  \\
\end{tabular}
\end{ruledtabular}
\end{table*}

\squeezetable
\begin{table*}
\caption{\label{tab:Isym_MR}
The values of $I_{\mathrm{sym}}$ and properties of neutron stars, including core-crust transition density $\rho_t$ (the zero-pressure density $\rho_z$ for interactions with $L=-5\mathrm{MeV}$), the radius $R_{1.4}$ and dimensionless tidal deformability $\Lambda_{1.4}$ of $1.4M_{\odot}$ neutron star as well as the maximum mass $M_{\mathrm{TOV}}$, predicted by the parameter set family.}
\begin{ruledtabular}
\begin{tabular}{c|cccccc|cccccc}
&\multicolumn{6}{c|}{Ms77} & \multicolumn{6}{c}{Ms83} \\ \hline
&Mv69 & Mv86 & Mv67 & Mv89 & Mv64 & Mv94 & Mv74 & Mv94 & Mv72 & Mv98 & Mv69 & Mv104 \\ \hline
\textbf{Lm5}
&\multicolumn{3}{l}{\bm{$L=-5\mathrm{MeV}$}}&\multicolumn{3}{l}{\bm{$E_{\mathrm{sym}}(\rho_0)=24\mathrm{MeV}$}}&
\multicolumn{3}{l}{\bm{$K_{\mathrm{sym}}=-10\mathrm{MeV}$}}&\multicolumn{3}{l}{\bm{$J_{\mathrm{sym}}=4250\mathrm{MeV}$}} \\ \cline{1-1}
$I_{{\rm{sym}}}\,(\rm{MeV})$
              & -1140 & -1042  & -1154 & -1028 & -1150 & -1032 &
                -1154 & -1056  & -1168 & -1042 & -1164 & -1046 \\
$\rho_z\,(\rm{fm}^{-3})$
              & 0.160  & 0.165   & 0.159  & 0.166  & 0.158  & 0.166  &
                0.160  & 0.165   & 0.159  & 0.166  & 0.159  & 0.167  \\
$R_{1.4}\,(\mathrm{km})$
              & 11.70  & 11.68   & 11.71  & 11.68  & 11.71  & 11.68  &
                11.71  & 11.69   & 11.71  & 11.68  & 11.71  & 11.68  \\
$\Lambda_{1.4}$
              & 714.0  & 714.3  & 714.8  & 713.3  & 715.0  & 712.9  &
                716.4  & 714.9  & 714.5  & 713.8  & 714.7  & 713.3  \\
$M_{\mathrm{TOV}}/M_{\odot}$
              & 2.26  & 2.26   & 2.25  & 2.26  & 2.25  & 2.26  &
                2.25  & 2.25   & 2.25  & 2.25  & 2.25  & 2.25  \\ \hline
\textbf{L5}
&\multicolumn{3}{l}{\bm{$L=5\mathrm{MeV}$}}&\multicolumn{3}{l}{\bm{$E_{\mathrm{sym}}(\rho_0)=28\mathrm{MeV}$}}&
\multicolumn{3}{l}{\bm{$K_{\mathrm{sym}}=-250\mathrm{MeV}$}}&\multicolumn{3}{l}{\bm{$J_{\mathrm{sym}}=2100\mathrm{MeV}$}} \\ \cline{1-1}
$I_{{\rm{sym}}}\,(\rm{MeV})$
              & -293.9 & -196.1  & -307.8 & -182.2 & -304.0 & -185.9 &
                -308.2 & -210.5  & -322.1 & -196.6 & -318.4 & -200.3 \\
$\rho_t\,(\rm{fm}^{-3})$
              & 0.139  & 0.143   & 0.138  & 0.144  & 0.138  & 0.145  &
                0.139  & 0.143   & 0.138  & 0.144  & 0.138  & 0.145  \\
$R_{1.4}\,(\mathrm{km})$
              & 11.67  & 11.57   & 11.68  & 11.55  & 11.70  & 11.53  &
                11.66  & 11.56   & 11.68  & 11.55  & 11.70  & 11.53  \\
$\Lambda_{1.4}$
              & 433.8  & 431.7  & 434.1  & 431.4  & 435.0  & 430.2  &
                433.8  & 431.6  & 434.0  & 431.4  & 435.0  & 430.2  \\
$M_{\mathrm{TOV}}/M_{\odot}$
              & 2.32  & 2.33   & 2.32  & 2.33  & 2.32  & 2.33  &
                2.31  & 2.32   & 2.31  & 2.32  & 2.31  & 2.32  \\ \hline
\textbf{L15}
&\multicolumn{3}{l}{\bm{$L=15\mathrm{MeV}$}}&\multicolumn{3}{l}{\bm{$E_{\mathrm{sym}}(\rho_0)=29\mathrm{MeV}$}}&
\multicolumn{3}{l}{\bm{$K_{\mathrm{sym}}=-240\mathrm{MeV}$}}&\multicolumn{3}{l}{\bm{$J_{\mathrm{sym}}=1450\mathrm{MeV}$}} \\ \cline{1-1}
$I_{{\rm{sym}}}\,(\rm{MeV})$
              & -1220 & -1122  & -1233 & -1108 & -1230 & -1112 &
                -1234 & -1136  & -1248 & -1123 & -1244 & -1126 \\
$\rho_t\,(\rm{fm}^{-3})$
              & 0.115  & 0.117   & 0.115  & 0.117  & 0.114  & 0.117  &
                0.115  & 0.117   & 0.115  & 0.117  & 0.114  & 0.118  \\
$R_{1.4}\,(\mathrm{km})$
              & 11.67  & 11.59   & 11.69  & 11.58  & 11.70  & 11.56  &
                11.67  & 11.59   & 11.68  & 11.57  & 11.70  & 11.56  \\
$\Lambda_{1.4}$
              & 359.3  & 356.3  & 359.6  & 355.9  & 361.0  & 354.3  &
                359.0  & 356.0  & 359.4  & 355.6  & 360.7  & 354.0  \\
$M_{\mathrm{TOV}}/M_{\odot}$
              & 2.31  & 2.32   & 2.31  & 2.32  & 2.31  & 2.32  &
                2.30  & 2.31   & 2.30  & 2.31  & 2.30  & 2.31  \\ \hline
\textbf{L25}
&\multicolumn{3}{l}{\bm{$L=25\mathrm{MeV}$}}&\multicolumn{3}{l}{\bm{$E_{\mathrm{sym}}(\rho_0)=30\mathrm{MeV}$}}&
\multicolumn{3}{l}{\bm{$K_{\mathrm{sym}}=-210\mathrm{MeV}$}}&\multicolumn{3}{l}{\bm{$J_{\mathrm{sym}}=1200\mathrm{MeV}$}} \\ \cline{1-1}
$I_{{\rm{sym}}}\,(\rm{MeV})$
              & -1586 & -1488  & -1600 & -1474 & -1596 & -1478 &
                -1600 & -1503 & -1614 & -1489 & -1610 & -1492 \\
$\rho_t\,(\rm{fm}^{-3})$
              & 0.100  & 0.101   & 0.100  & 0.102  & 0.100  & 0.102  &
                0.101  & 0.102   & 0.100  & 0.102  & 0.100  & 0.102  \\
$R_{1.4}\,(\mathrm{km})$
              & 11.88  & 11.81   & 11.89  & 11.80  & 11.90  & 11.78  &
                11.88  & 11.81   & 11.89  & 11.80  & 11.90  & 11.78  \\
$\Lambda_{1.4}$
              & 355.7  & 352.7  & 356.0  & 352.3  & 357.4  & 350.6  &
                355.3  & 352.3  & 355.6  & 351.9  & 357.0  & 350.3  \\
$M_{\mathrm{TOV}}/M_{\odot}$
              & 2.28  & 2.29   & 2.28  & 2.29  & 2.28  & 2.29  &
                2.28  & 2.28   & 2.27  & 2.28  & 2.27  & 2.28  \\ \hline
\textbf{L35}
&\multicolumn{3}{l}{\bm{$L=35\mathrm{MeV}$}}&\multicolumn{3}{l}{\bm{$E_{\mathrm{sym}}(\rho_0)=30\mathrm{MeV}$}}&
\multicolumn{3}{l}{\bm{$K_{\mathrm{sym}}=-190\mathrm{MeV}$}}&\multicolumn{3}{l}{\bm{$J_{\mathrm{sym}}=670\mathrm{MeV}$}} \\ \cline{1-1}
$I_{{\rm{sym}}}\,(\rm{MeV})$
              & -1896 & -1798 & -1910 & -1784 & -1906 & -1788 &
                -1910 & -1812 & -1924 & -1799 & -1920 & -1802 \\
$\rho_t\,(\rm{fm}^{-3})$
              & 0.0874  & 0.0881   & 0.0873  & 0.0882  & 0.0871  & 0.0884  &
                0.0875  & 0.0883   & 0.0874  & 0.0884  & 0.0873  & 0.0886  \\
$R_{1.4}\,(\mathrm{km})$
              & 11.54  & 11.47   & 11.55  & 11.46  & 11.58  & 11.43  &
                11.54  & 11.46   & 11.55  & 11.45  & 11.57  & 11.43  \\
$\Lambda_{1.4}$
              & 264.5  & 259.8  & 264.9  & 259.2  & 267.6  & 256.1  &
                263.4  & 258.8  & 263.9  & 258.2  & 266.5  & 255.1  \\
$M_{\mathrm{TOV}}/M_{\odot}$
              & 2.24  & 2.25   & 2.23  & 2.25  & 2.23  & 2.25  &
                2.23  & 2.24   & 2.22  & 2.24  & 2.22  & 2.24  \\ \hline
\textbf{L45}
&\multicolumn{3}{l}{\bm{$L=45\mathrm{MeV}$}}&\multicolumn{3}{l}{\bm{$E_{\mathrm{sym}}(\rho_0)=30\mathrm{MeV}$}}&
\multicolumn{3}{l}{\bm{$K_{\mathrm{sym}}=-110\mathrm{MeV}$}}&\multicolumn{3}{l}{\bm{$J_{\mathrm{sym}}=700\mathrm{MeV}$}} \\ \cline{1-1}
$I_{{\rm{sym}}}\,(\rm{MeV})$
              & -2446 & -2348 & -2460 & -2334 & -2456 & -2338 &
                -2460 & -2363 & -2474 & -2349 & -2470 & -2352 \\
$\rho_t\,(\rm{fm}^{-3})$
              & 0.0804  & 0.0810   & 0.0803  & 0.0811  & 0.0802  & 0.0812  &
                0.0806  & 0.0812   & 0.0805  & 0.0813  & 0.0804  & 0.0814  \\
$R_{1.4}\,(\mathrm{km})$
              & 12.23  & 12.18   & 12.23  & 12.17  & 12.24  & 12.16  &
                12.23  & 12.18   & 12.23  & 12.17  & 12.24  & 12.16  \\
$\Lambda_{1.4}$
              & 390.2  & 388.1  & 390.4  & 387.8  & 391.8  & 386.1  &
                389.5  & 387.5  & 389.7  & 387.2  & 391.2  & 385.5  \\
$M_{\mathrm{TOV}}/M_{\odot}$
              & 2.22  & 2.22   & 2.21  & 2.22  & 2.21  & 2.22  &
                2.21  & 2.21   & 2.21  & 2.21  & 2.21  & 2.21  \\ \hline
\textbf{L55}
&\multicolumn{3}{l}{\bm{$L=55\mathrm{MeV}$}}&\multicolumn{3}{l}{\bm{$E_{\mathrm{sym}}(\rho_0)=33\mathrm{MeV}$}}&
\multicolumn{3}{l}{\bm{$K_{\mathrm{sym}}=-100\mathrm{MeV}$}}&\multicolumn{3}{l}{\bm{$J_{\mathrm{sym}}=900\mathrm{MeV}$}} \\ \cline{1-1}
$I_{{\rm{sym}}}\,(\rm{MeV})$
              & -1974 & -1889 & -1988 & -1863 & -1984 & -1864 &
                -1988 & -1891 & -2002 & -1877 & -1998 & -1894 \\
$\rho_t\,(\rm{fm}^{-3})$
              & 0.0797  & 0.0804   & 0.0796  & 0.0804  & 0.0795  & 0.0805  &
                0.0799  & 0.0806   & 0.0798  & 0.0806  & 0.0797  & 0.0808  \\
$R_{1.4}\,(\mathrm{km})$
              & 12.80  & 12.76   & 12.81  & 12.76  & 12.81  & 12.75  &
                12.80  & 12.77   & 12.80  & 12.76  & 12.81  & 12.75  \\
$\Lambda_{1.4}$
              & 468.9  & 468.2  & 468.9  & 468.5  & 469.7  & 467.7  &
                468.5  & 468.2  & 468.5  & 468.1  & 469.3  & 466.9  \\
$M_{\mathrm{TOV}}/M_{\odot}$
              & 2.21  & 2.22   & 2.21  & 2.22  & 2.21  & 2.22  &
                2.20  & 2.21   & 2.20  & 2.21  & 2.20  & 2.21  \\ \hline
\textbf{L65}
&\multicolumn{3}{l}{\bm{$L=65\mathrm{MeV}$}}&\multicolumn{3}{l}{\bm{$E_{\mathrm{sym}}(\rho_0)=34\mathrm{MeV}$}}&
\multicolumn{3}{l}{\bm{$K_{\mathrm{sym}}=-70\mathrm{MeV}$}}&\multicolumn{3}{l}{\bm{$J_{\mathrm{sym}}=650\mathrm{MeV}$}} \\ \cline{1-1}
$I_{{\rm{sym}}}\,(\rm{MeV})$
              & -2340 & -2242 & -2354 & -2228 & -2350 & -2232 &
                -2354 & -2257 & -2368 & -2243 & -2364 & -2246 \\
$\rho_t\,(\rm{fm}^{-3})$
              & 0.0741  & 0.0746   & 0.0740  & 0.0747  & 0.0739  & 0.0748  &
                0.0743  & 0.0749   & 0.0742  & 0.0749  & 0.0741  & 0.0751  \\
$R_{1.4}\,(\mathrm{km})$
              & 12.91  & 12.88   & 12.92  & 12.88  & 12.93  & 12.87  &
                12.91  & 12.88   & 12.91  & 12.87  & 12.92  & 12.87  \\
$\Lambda_{1.4}$
              & 475.9  & 475.8  & 475.9  & 475.8  & 476.7  & 474.9  &
                475.2  & 475.2  & 475.2  & 475.2  & 476.1  & 474.3  \\
$M_{\mathrm{TOV}}/M_{\odot}$
              & 2.19  & 2.19   & 2.19  & 2.19  & 2.19  & 2.19  &
                2.18  & 2.18   & 2.18  & 2.18  & 2.18  & 2.18  \\ \hline
\textbf{L75}
&\multicolumn{3}{l}{\bm{$L=75\mathrm{MeV}$}}&\multicolumn{3}{l}{\bm{$E_{\mathrm{sym}}(\rho_0)=34\mathrm{MeV}$}}&
\multicolumn{3}{l}{\bm{$K_{\mathrm{sym}}=-10\mathrm{MeV}$}}&\multicolumn{3}{l}{\bm{$J_{\mathrm{sym}}=550\mathrm{MeV}$}} \\ \cline{1-1}
$I_{{\rm{sym}}}\,(\rm{MeV})$
              & -2640 & -2542 & -2654 & -2528 & -2650 & -2532 &
                -2654 & -2557 & -2668 & -2543 & -2664 & -2546 \\
$\rho_t\,(\rm{fm}^{-3})$
              & 0.0691  & 0.0696   & 0.0690  & 0.0697  & 0.0689  & 0.0698  &
                0.0693  & 0.0699   & 0.0692  & 0.0699  & 0.0691  & 0.0701  \\
$R_{1.4}\,(\mathrm{km})$
              & 13.15  & 13.12   & 13.15  & 13.12  & 13.16  & 13.11  &
                13.15  & 13.12   & 13.15  & 13.12  & 13.16  & 13.12  \\
$\Lambda_{1.4}$
              & 546.7  & 547.8  & 546.5  & 548.0  & 547.1  & 547.4  &
                546.2  & 547.4  & 546.1  & 547.5  & 546.5  & 574.0  \\
$M_{\mathrm{TOV}}/M_{\odot}$
              & 2.17  & 2.17   & 2.17  & 2.18  & 2.17  & 2.18  &
                2.16  & 2.17   & 2.16  & 2.17  & 2.16  & 2.17  \\ \hline
\textbf{L85}
&\multicolumn{3}{l}{\bm{$L=85\mathrm{MeV}$}}&\multicolumn{3}{l}{\bm{$E_{\mathrm{sym}}(\rho_0)=36\mathrm{MeV}$}}&
\multicolumn{3}{l}{\bm{$K_{\mathrm{sym}}=10\mathrm{MeV}$}}&\multicolumn{3}{l}{\bm{$J_{\mathrm{sym}}=470\mathrm{MeV}$}} \\ \cline{1-1}
$I_{{\rm{sym}}}\,(\rm{MeV})$
              & -2752 & -2654 & -2766 & -2640 & -2762 & -2644 &
                -2766 & -2669 & -2780 & -2655 & -2776 & -2658 \\
$\rho_t\,(\rm{fm}^{-3})$
              & 0.0667  & 0.0673   & 0.0666  & 0.0673  & 0.0665  & 0.0674  &
                0.0669  & 0.0675   & 0.0668  & 0.0675  & 0.0667  & 0.0677  \\
$R_{1.4}\,(\mathrm{km})$
              & 13.36  & 13.33   & 13.35  & 13.33  & 13.36  & 13.33  &
                13.35  & 13.33   & 13.36  & 13.33  & 13.36  & 13.33  \\
$\Lambda_{1.4}$
              & 571.7  & 573.4  & 571.4  & 573.6  & 571.8  & 573.3  &
                571.2  & 572.9  & 570.9  & 573.1  & 571.2  & 572.8  \\
$M_{\mathrm{TOV}}/M_{\odot}$
              & 2.16  & 2.16   & 2.15  & 2.16  & 2.16  & 2.16  &
                2.15  & 2.15   & 2.15  & 2.15  & 2.15  & 2.15  \\ \hline
\textbf{L105}
&\multicolumn{3}{l}{\bm{$L=105\mathrm{MeV}$}}&\multicolumn{3}{l}{\bm{$E_{\mathrm{sym}}(\rho_0)=37\mathrm{MeV}$}}&
\multicolumn{3}{l}{\bm{$K_{\mathrm{sym}}=150\mathrm{MeV}$}}&\multicolumn{3}{l}{\bm{$J_{\mathrm{sym}}=220\mathrm{MeV}$}} \\ \cline{1-1}
$I_{{\rm{sym}}}\,(\rm{MeV})$
              & -3638 & -3540 & -3652 & -3526 & -3648 & -3530 &
                -3652 & -3555 & -3666 & -3541 & -3662 & -3544 \\
$\rho_t\,(\rm{fm}^{-3})$
              & 0.0618  & 0.0622   & 0.0617  & 0.0623  & 0.0619  & 0.0624  &
                0.0619  & 0.0624   & 0.0619  & 0.0625  & 0.0618  & 0.0625  \\
$R_{1.4}\,(\mathrm{km})$
              & 13.80  & 13.79   & 13.80  & 13.79  & 13.80  & 13.79  &
                13.81  & 13.80   & 13.81  & 13.79  & 13.81  & 13.79  \\
$\Lambda_{1.4}$
              & 751.4  & 757.4  & 750.0  & 758.1  & 748.6  & 758.4  &
                752.6  & 757.7  & 752.2  & 758.4  & 751.0  & 758.5  \\
$M_{\mathrm{TOV}}/M_{\odot}$
              & 2.15  & 2.15   & 2.15  & 2.15  & 2.15  & 2.16  &
                2.14  & 2.14   & 2.14  & 2.14  & 2.14  & 2.14  \\ \hline
\textbf{L125}
&\multicolumn{3}{l}{\bm{$L=125\mathrm{MeV}$}}&\multicolumn{3}{l}{\bm{$E_{\mathrm{sym}}(\rho_0)=39\mathrm{MeV}$}}&
\multicolumn{3}{l}{\bm{$K_{\mathrm{sym}}=220\mathrm{MeV}$}}&\multicolumn{3}{l}{\bm{$J_{\mathrm{sym}}=320\mathrm{MeV}$}} \\ \cline{1-1}
$I_{{\rm{sym}}}\,(\rm{MeV})$
              & -3970 & -3872 & -3984 & -3858 & -3980 & -3862 &
                -3984 & -3886 & -3998 & -3873 & -3994 & -3876 \\
$\rho_t\,(\rm{fm}^{-3})$
              & 0.0568  & 0.0572   & 0.0568  & 0.0573  & 0.0567  & 0.0574  &
                0.0570  & 0.0574   & 0.0569  & 0.0574  & 0.0568  & 0.0575  \\
$R_{1.4}\,(\mathrm{km})$
              & 14.06  & 14.06   & 14.07  & 14.05  & 14.07  & 14.05  &
                14.07  & 14.06   & 14.07  & 14.05  & 14.07  & 14.05  \\
$\Lambda_{1.4}$
              & 837.6  & 842.7  & 837.0  & 843.4  & 836.6  & 844.0  &
                837.5  & 842.5  & 836.8  & 843.2  & 836.4  & 843.7  \\
$M_{\mathrm{TOV}}/M_{\odot}$
              & 2.13  & 2.13   & 2.13  & 2.13  & 2.13  & 2.13  &
                2.12  & 2.13   & 2.12  & 2.13  & 2.12  & 2.13  \\
\end{tabular}
\end{ruledtabular}
\end{table*}

\begin{figure*}[ht]
    \centering
    \includegraphics[width=\linewidth]{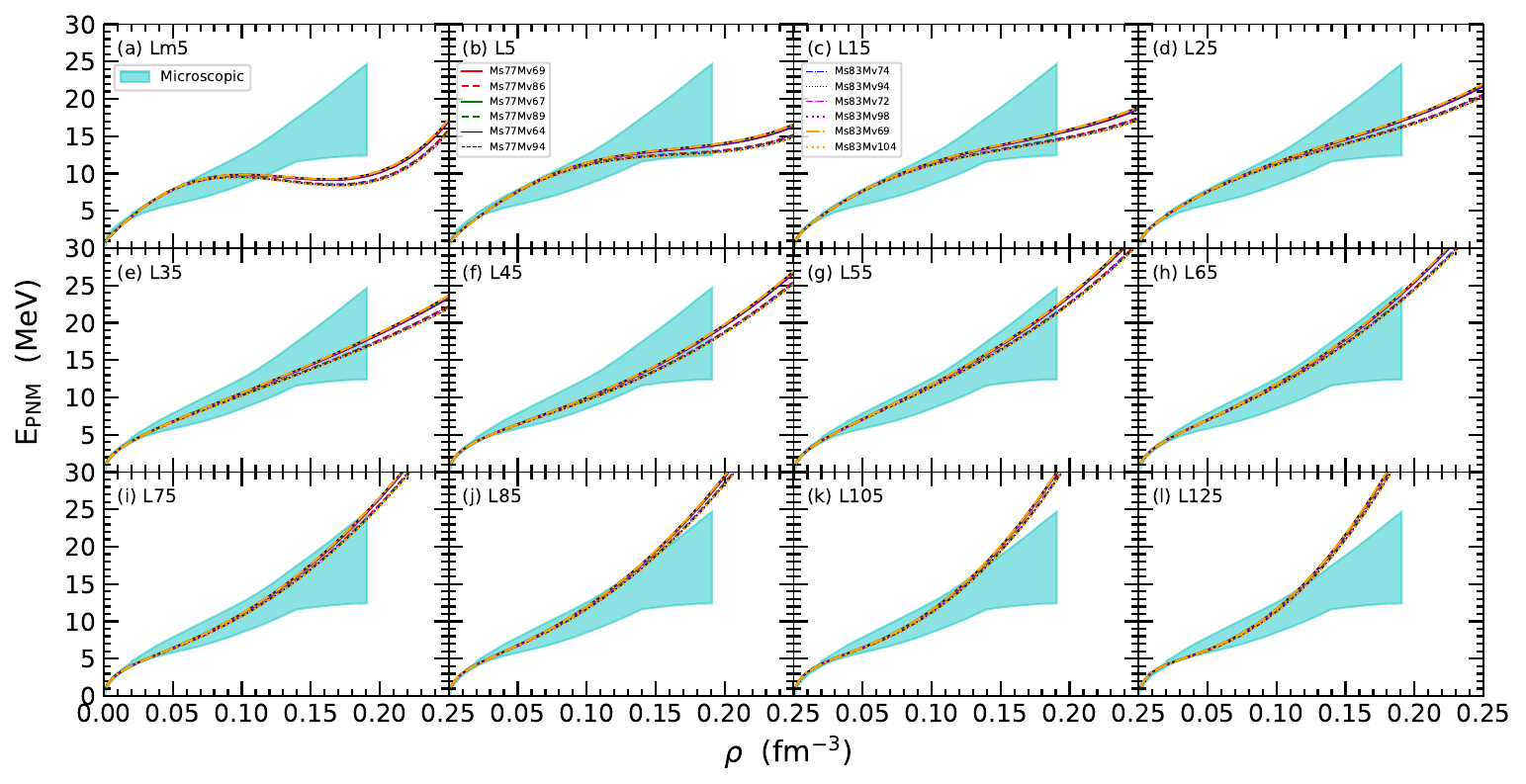}
    \caption{The EOS of PNM ($E_{\mathrm{PNM}}$) predicted by the parameter set family, categorized according to the slope parameter $L$ of the symmetry energy.
    The band represents the microscopic calculation results~\cite{Zhang:2022bni}.}
    \label{fig:Epnm_all}
\end{figure*}

\begin{figure*}[ht]
    \centering
    \includegraphics[width=\linewidth]{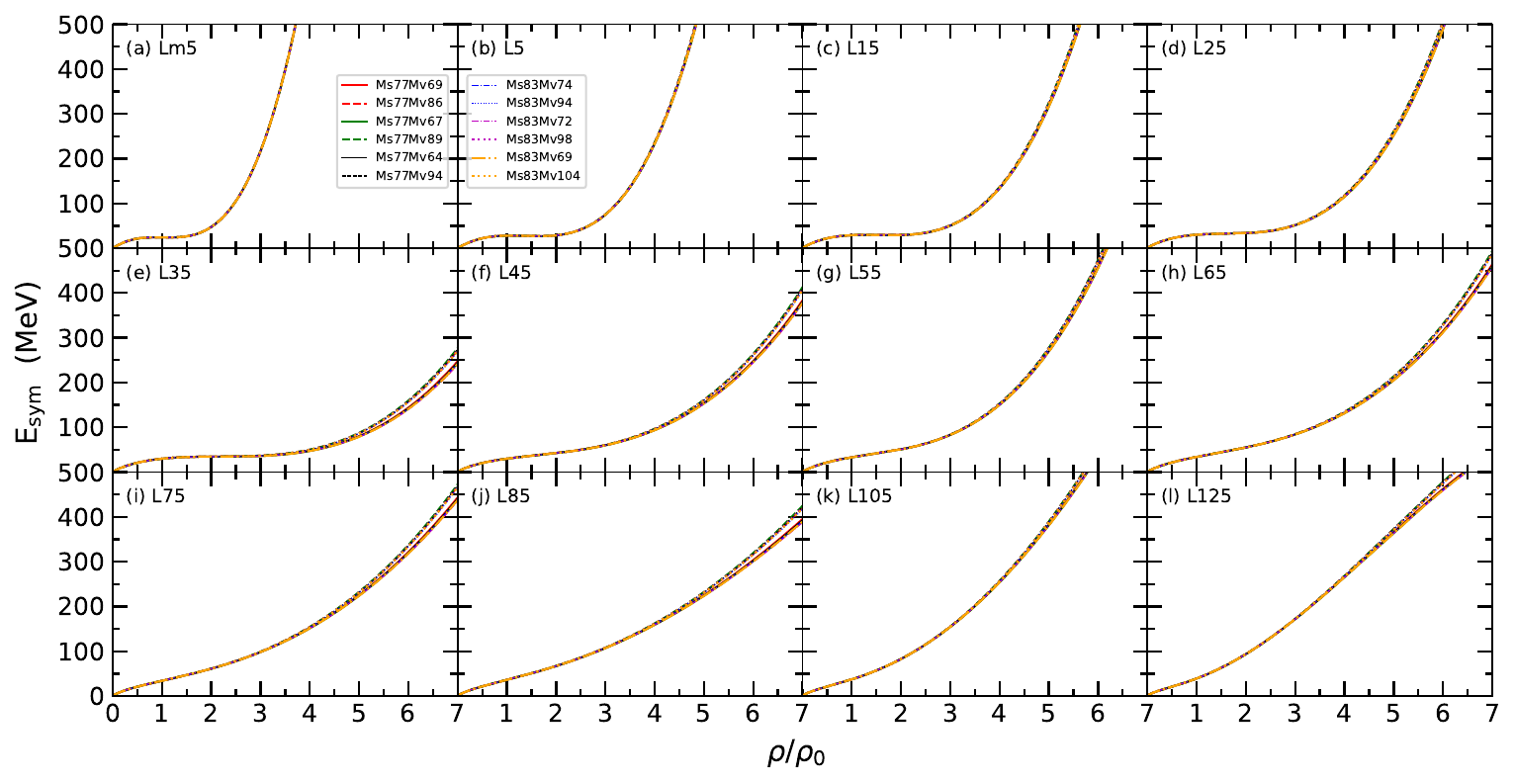}
    \caption{The density dependence of the symmetry energy given by the parameter set family, categorized according to the slope parameter $L$ of the symmetry energy.}
    \label{fig:Esym_all}
\end{figure*}

The EOS of the PNM as a function of nucleon density are exhibited in Fig.~\ref{fig:Epnm_all}, and we categorize the parameter sets according to the slope parameter $L$ of the symmetry energy, for the sake of clarity (same in the following figures).
The results from microscopic calculations \cite{Zhang:2022bni} are also shown in Fig.~\ref{fig:Epnm_all} for comparison.
Except for the parameter sets with $L= -5$, $85$, $105$ and $125\,\mathrm{MeV}$, all other parameter sets in the parameter set family are in line with the microscopic calculations for $E_{\mathrm{PNM}}$.
Shown in Fig.~\ref{fig:Esym_all} is the density dependence of the symmetry energy given by the parameter set family.
Fig.~\ref{fig:Epnm_all} and Fig.~\ref{fig:Esym_all} demonstrate that the momentum dependencies of $U_0$ and $U_{\mathrm{sym}}$, represented by $m^{\ast}_{s,0}$ and $m^{\ast}_{v,0}$, have little impact on $E_{\mathrm{PNM}}(\rho)$ and $E_{\mathrm{sym}}(\rho)$.
However, $E_{\mathrm{PNM}}(\rho)$ and $E_{\mathrm{sym}}(\rho)$ exhibit a bifurcation caused by the isospin splitting of the nucleon effective mass, resulting in two distinct branches in their curves.
Parameter sets with $\mspl<0$ (in neutron-rich matter), i.e., $m^{\ast}_{s,0}<m^{\ast}_{v,0}$, predict smaller $E_{\mathrm{PNM}}(\rho)$ compared to those with $\mspl>0$ around the saturation density as can be seen from Fig.~\ref{fig:Epnm_all}.
Fig.~\ref{fig:Esym_all} illustrates that parameter sets with $\mspl<0$ predict larger $E_{\mathrm{sym}}(\rho)$ at suprasaturation densities (and larger values of the high-order symmetry energy kurtosis parameter, $I_{\mathrm{sym}}$).
The values of $I_{\mathrm{sym}}$ with the parameter set family are listed in Table~\ref{tab:Isym_MR}.

\begin{figure*}[ht]
    \centering
    \includegraphics[width=\linewidth]{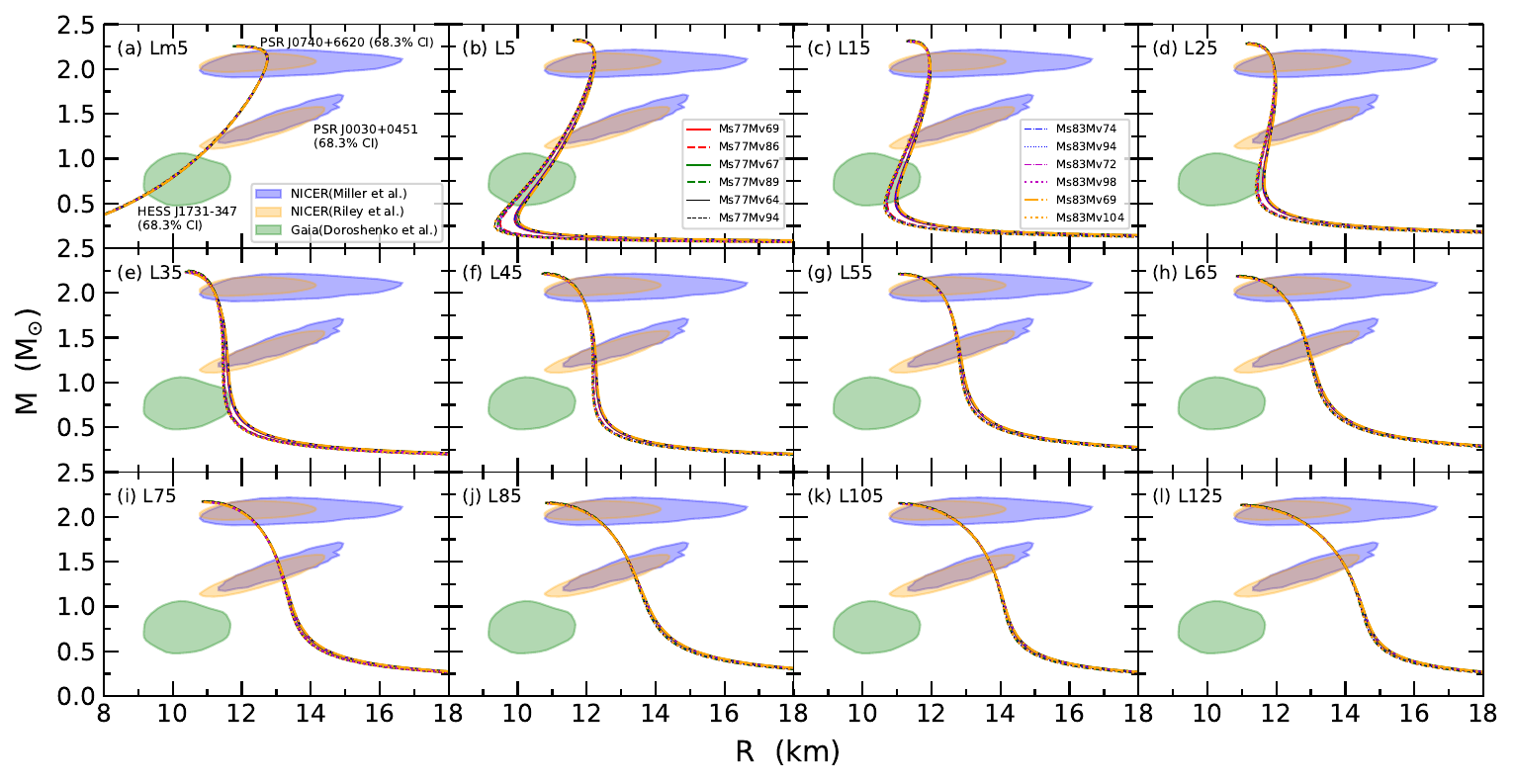}
    \caption{M-R relation for static neutron stars from the parameter set family, categorized according to the slope parameter $L$ of the symmetry energy.
    The NICER (XMM-Newton) constraints for PSR J0030+0451 \cite{Miller:2019cac,Riley:2019yda}, PSR J0740+6620 \cite{Miller:2021qha,Riley:2021pdl} and Gaia constraint for the CCO in HESS J1731-347 \cite{2022NatAs...6.1444D} are also included for comparison.
    All contours are plotted for 68.3\% CI.}
    \label{fig:MR_all}
\end{figure*}

We also calculate the properties of neutron stars using the parameter set family.
Shown in Fig.~\ref{fig:MR_all} is the mass-radius relation of neutron stars predicted by the parameter set family.
The determinations from astrophysical observations of the mass-radius relation of neutron stars \cite{Miller:2019cac,Riley:2019yda,Miller:2021qha,Riley:2021pdl,2022NatAs...6.1444D} within 68.3\% CI are also plotted in Fig.~\ref{fig:MR_all} for comparison.

\begin{figure*}[ht]
    \centering
    \includegraphics[width=\linewidth]{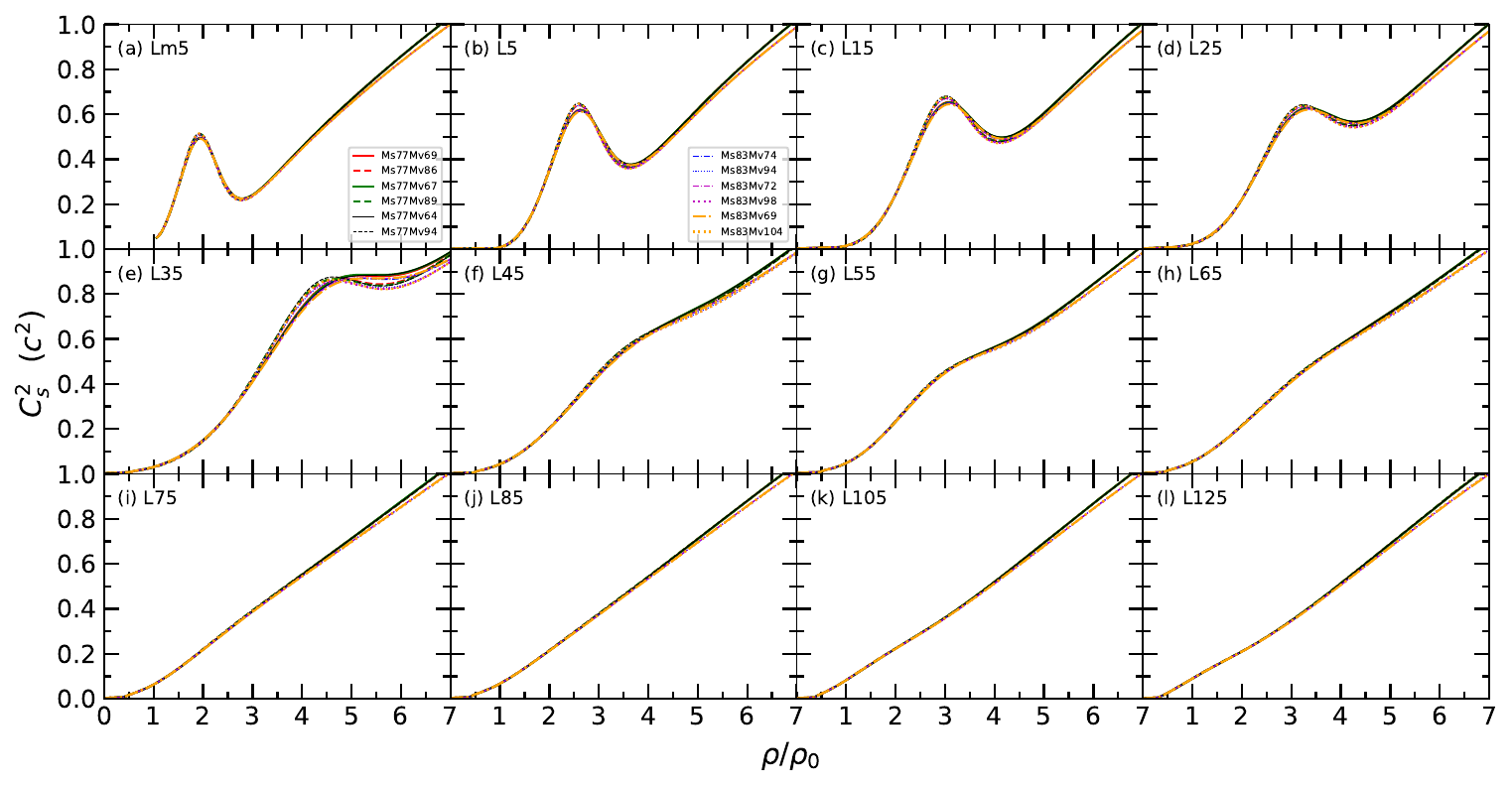}
    \caption{The squared sound speed ($C_{s}^{2}\equiv dP/d\epsilon$) of neutron star matter as a function of nucleon density given by the parameter set family, categorized according to the slope parameter $L$ of the symmetry energy.}
    \label{fig:Cs2_All}
\end{figure*}

One sees that all the interactions predict that the maximum masses of neutron star exceed $2.1\,M_{\odot}$, and the mass-radius relations align with the astrophysical measurements of both PSR J0030+0451 and PSR J0740+6620 within the 68.3\% CI.
In addition, parameter sets with $L= -5$, $5$, $15$, $25$ and $35\,\mathrm{MeV}$ are compatible with the constraint for the CCO in HESS J1731-347 within 68.3\% CI.
Moreover, it is seen from Fig.~\ref{fig:MR_all} that the mass-radius curves split into two branches, especially for the parameter sets with $L= 5$, $15$, $25$, $35$ and $45\,\mathrm{MeV}$, and those with $\mspl<0$ (in neutron-rich matter) predict smaller radii.

In Fig.~\ref{fig:Cs2_All}, we display the squared sound speed $C_{s}^{2}\equiv dP/d\epsilon$ of neutron star matter given by the parameter set family, and the causality condition is satisfied in the calculations.
It is seen that parameter sets with $L= -5$, $5$, $15$, $25\,\mathrm{MeV}$ predict distinct peak structures in $C_{s}^{2}$ between $2\rho_0$ and $3.5\rho_0$, with peak values approximately ranging from $0.5c^2$ to $0.7c^2$.
Additionally, as shown in Fig.~\ref{fig:Cs2_All}, the $C_{s}^{2}$ curves exhibit a distinction into two branches at high densities, where parameter sets with $m^{\ast}_{s,0} = 0.773m$ predict larger values of $C_{s}^{2}$ compared to those with $m^{\ast}_{s,0} = 0.835m$.

Also presented in Table~\ref{tab:Isym_MR} are the properties of neutron stars predicted by the parameter set family, including the core-crust transition density $\rho_t$ (the zero-pressure density $\rho_z$ for interactions with $L=-5\,\mathrm{MeV}$), the radius $R_{1.4}$ and dimensionless tidal deformability $\Lambda_{1.4}$ of $1.4M_{\odot}$ neutron star and the maximum mass $M_{\mathrm{TOV}}$.
The parameter sets with $ 5\,\mathrm{MeV} \leq L \leq 85\,\mathrm{MeV}$ satisfy the constraint of $\Lambda_{1.4} \leqslant 580$.

\section{SUMMARY AND OUTLOOK}
\label{sec:summary}
Based on the N3LO Skyrme pseudopotential, we have obtained new extended Skyrme interactions by modifying the density-dependent term in the spirit of the Fermi momentum expansion.
The Haimiltonian density and single-nucleon potential of the new extended Skyrme effective interactions are derived in the Hartree-Fock approximation.
The momentum dependence of the single-nucleon potential is regulated to fit the empirical nucleon optical potential (and its extrapolation), and so the new Skyrme interactions can be used in transport model simulations for HICs up to $1.5\,\mathrm{GeV}$/nucleon.
At the same time,
the extension of the density-dependent term can provide very flexible freedom to adjust the density dependence of the symmetry energy as well as the high density behavior of symmetric nuclear matter, and thus can essentially satisfy all the constraints on the neutron stars from the multimessenger astrophysical observations.
We have constructed
a series of interactions, denoted as SP6X, with the density slope parameter $L$ of the symmetry energy ranging from $-5\,\mathrm{MeV}$ to $125\,\mathrm{MeV}$, and we find the following results:
(i) To simultaneously satisfy constraints from microscopic calculations on the EOS of pure neutron matter $E_{\mathrm{PNM}}(\rho)$ and astrophysical observations of neutron star mass-radius relations of both PSR J0030+0451 and PSR J0740+6620, $L$ should fall within the range of $5\,\mathrm{MeV} \leq L \leq 75\,\mathrm{MeV}$;
(ii) To further additionally describe the neutron star with small mass and radius in CCO of HESS J1731-347, a range of $-5\,\mathrm{MeV} \leq L \leq 35\,\mathrm{MeV}$ is necessary;
(iii) A distinct peak structure in the squared sound speed for neutron star matter emerges for interactions with a soft symmetry energy around saturation density, especially when $-5\,\mathrm{MeV} \leq L \leq 25\,\mathrm{MeV}$.

In addition, we have constructed a versatile parameter set family consisting of parameter sets with different density dependencies of the symmetry energy as well as various momentum dependencies of the single-nucleon potential and the symmetry potential (i.e., various isoscalar and isovector nucleon effective masses).
This parameter set family will be useful for investigating in a more transparent way the effects of the symmetry energy and nucleon effective masses, and serves as a useful theoretical tool to extract the density dependence of the symmetry energy and the nucleon effective masses by analyzing the data from different nuclear experiments and astrophysical observations.

Besides the neutron stars on which we have focused in this work, the nuclear energy density functional constructed in the present work will next be applied in the large-scale LBUU transport model simulations for HICs as well as nuclear structure calculations.
In addition,
we will also extend our model to finite temperature to explore the thermal effects on the nuclear matter EOS and construct the corresponding EOS tables.
These EOS tables can be used to study the liquid-gas phase transition as well as to simulate core-collapse supernovae and binary neutron star mergers.
Our work will be useful for the determination of the high density behavior of the symmetry energy and in-medium nucleon effective masses by analyzing various data from nuclear experiments and multimessenger observations of neutron stars in a single unified nuclear energy density functional.

\begin{acknowledgments}
This work was supported in part by the National Natural Science Foundation of China under Grant Nos. 12235010 and 11625521, the National SKA Program of China No. 2020SKA0120300, and the Science and Technology Commission of Shanghai Municipality under Grant No. 23JC1402700.
\end{acknowledgments}


\appendix
\begin{widetext}

\section{Expressions of Macroscopic quantities}
\label{sec:Appendix_Exp}
The values of $E_0(\rho)$, $P_{\mathrm{SNM}}(\rho)$, $K_0(\rho)$, $J_0(\rho)$, $a_2(\rho)$, $a_4(\rho)$, $a_6(\rho)$, $b_2(\rho)$,  $b_4(\rho)$, $b_6(\rho)$, $E_{\mathrm{sym}}(\rho)$, $L(\rho)$, $K_{\mathrm{sym}}(\rho)$ and $J_{\mathrm{sym}}(\rho)$ at arbitrary density, can be expressed as a sum of powers of $k_{F}$ ($k_F=a \rho^{1/3}$ with $a=(3\pi^{2}/2)^{1/3}$), where the coefficients can be expressed in terms of linear combinations of 14 parameters: $t_0$, $t_{3}^{[1]}$, $t_{3}^{[3]}$, $t_{3}^{[5]}$, $C^{[2]}$, $D^{[2]}$, $C^{[4]}$, $D^{[4]}$, $C^{[6]}$, $D^{[6]}$, $t_{0}x_{0}$, $t_{3}^{[1]}x_{3}^{[1]}$, $t_{3}^{[3]}x_{3}^{[3]}$ and $t_{3}^{[5]}x_{3}^{[5]}$.
Note that $x_0$ ($x_{3}^{[1]}$, $x_{3}^{[3]}$ and $x_{3}^{[5]}$) are coupled with $t_0$ ($t_{3}^{[1]}$, $t_{3}^{[3]}$ and $t_{3}^{[5]}$).

At arbitrary density $\rho$, the 14 macroscopic quantities can be expressed as linear combinations of the 14 parameters, i.e.,
\begin{equation}
\label{eq:Q=MP+C}
\mathbf{Q}(\rho)=\mathbf{M}(\rho) \cdot \mathbf{P}+\mathbf{C}(\rho),
\end{equation}
with
\begin{equation}
\label{eq:Q_rho}
\mathbf{Q}(\rho)=\left[E_0(\rho),P_{\mathrm{SNM}}(\rho),K_0(\rho),J_0(\rho),a_2(\rho),a_4(\rho),a_6(\rho),b_2(\rho),b_4(\rho),
b_6(\rho),E_{\mathrm{sym}}(\rho),L(\rho),K_{\mathrm{sym}}(\rho),J_{\mathrm{sym}}(\rho) \right]^{\mathrm{T}}
\end{equation}
being the vector of the 14 quantities at $\rho$,
\begin{equation}
\label{eq:P}
\mathbf{P} = \left[t_0,t_{3}^{[1]},t_{3}^{[3]},t_{3}^{[5]},C^{[2]},D^{[2]},C^{[4]},D^{[4]},C^{[6]},D^{[6]},t_{0}x_{0},
t_{3}^{[1]}x_{3}^{[1]},t_{3}^{[3]}x_{3}^{[3]},t_{3}^{[5]}x_{3}^{[5]} \right]^{\mathrm{T}}
\end{equation}
being the vector of the 14 parameters,
\begin{equation}
\label{eq:C_rho}
\mathbf{C}(\rho) = \left[\frac{3 \hbar ^2 k_F^2}{10 m},\frac{2 \hbar ^2 k_F^5}{15 m \pi ^2},-\frac{3 \hbar ^2 k_F^2}{5 m},
\frac{12 \hbar ^2 k_F^2}{5 m}, 0,0,0,0,0,0,\frac{\hbar ^2 k_F^2}{6 m}, \frac{\hbar ^2 k_F^2}{3 m},-\frac{\hbar ^2 k_F^2}{3 m},
\frac{4 \hbar ^2 k_F^2}{3 m} \right]^{\mathrm{T}}
\end{equation}
being the parameter-irrelevant term,
and
\begin{equation}
\label{eq:M_rho}
\mathbf{M}(\rho)=
\renewcommand{\arraystretch}{2}
\left(
\begin{array}{cccccccccccccc}
 \frac{3 \rho }{8} & \frac{\rho ^{4/3}}{16} & \frac{\rho ^2}{16} & \frac{\rho ^{8/3}}{16} & \frac{k_F^5}{20 \pi ^2} & \frac{k_F^5}{40 \pi ^2} & \frac{3 k_F^7}{70 \pi ^2} & \frac{3 k_F^7}{140 \pi ^2} & \frac{8 k_F^9}{45 \pi ^2} & \frac{4 k_F^9}{45 \pi ^2} & 0 & 0 & 0 & 0 \\
 \frac{3 \rho ^2}{8} & \frac{\rho ^{7/3}}{12} & \frac{\rho ^3}{8} & \frac{\rho ^{11/3}}{6} & \frac{k_F^8}{18 \pi ^4} & \frac{k_F^8}{36 \pi ^4} & \frac{k_F^{10}}{15 \pi ^4} & \frac{k_F^{10}}{30 \pi ^4} & \frac{16 k_F^{12}}{45 \pi ^4} & \frac{8 k_F^{12}}{45 \pi ^4} & 0 & 0 & 0 & 0 \\
 0 & \frac{\rho ^{4/3}}{4} & \frac{9 \rho ^2}{8} & \frac{5 \rho ^{8/3}}{2} & \frac{k_F^5}{2 \pi ^2} & \frac{k_F^5}{4 \pi ^2} & \frac{6 k_F^7}{5 \pi ^2} & \frac{3 k_F^7}{5 \pi ^2} & \frac{48 k_F^9}{5 \pi ^2} & \frac{24 k_F^9}{5 \pi ^2} & 0 & 0 & 0 & 0 \\
 0 & -\frac{\rho ^{4/3}}{2} & 0 & 5 \rho ^{8/3} & -\frac{k_F^5}{2 \pi ^2} & -\frac{k_F^5}{4 \pi ^2} & \frac{6 k_F^7}{5 \pi ^2} & \frac{3 k_F^7}{5 \pi ^2} & \frac{144 k_F^9}{5 \pi ^2} & \frac{72 k_F^9}{5 \pi ^2} & 0 & 0 & 0 & 0 \\
 0 & 0 & 0 & 0 & \frac{k_F^3}{12 \pi ^2} & \frac{k_F^3}{24 \pi ^2} & \frac{k_F^5}{12 \pi ^2} & \frac{k_F^5}{24 \pi ^2} & \frac{k_F^7}{4 \pi ^2} & \frac{k_F^7}{8 \pi ^2} & 0 & 0 & 0 & 0 \\
 0 & 0 & 0 & 0 & 0 & 0 & \frac{k_F^3}{24 \pi ^2} & \frac{k_F^3}{48 \pi ^2} & \frac{7 k_F^5}{20 \pi ^2} & \frac{7 k_F^5}{40 \pi ^2} & 0 & 0 & 0 & 0 \\
 0 & 0 & 0 & 0 & 0 & 0 & 0 & 0 & \frac{k_F^3}{12 \pi ^2} & \frac{k_F^3}{24 \pi ^2} & 0 & 0 & 0 & 0 \\
 0 & 0 & 0 & 0 & 0 & \frac{k_F^3}{24 \pi ^2} & 0 & \frac{5 k_F^5}{72 \pi ^2} & 0 & \frac{7 k_F^7}{24 \pi ^2} & 0 & 0 & 0 & 0 \\
 0 & 0 & 0 & 0 & 0 & 0 & 0 & \frac{k_F^3}{48 \pi ^2} & 0 & \frac{7 k_F^5}{24 \pi ^2} & 0 & 0 & 0 & 0 \\
 0 & 0 & 0 & 0 & 0 & 0 & 0 & 0 & 0 & \frac{k_F^3}{24 \pi ^2} & 0 & 0 & 0 & 0 \\
 -\frac{\rho }{8} & -\frac{\rho ^{4/3}}{48} & -\frac{\rho ^2}{48} & -\frac{\rho ^{8/3}}{48} & \frac{k_F^5}{36 \pi ^2} & \frac{k_F^5}{18 \pi ^2} & \frac{k_F^7}{18 \pi ^2} & \frac{k_F^7}{12 \pi ^2} & \frac{2 k_F^9}{5 \pi ^2} & \frac{8 k_F^9}{15 \pi ^2} & -\frac{\rho }{4} & -\frac{\rho ^{4/3}}{24} & -\frac{\rho ^2}{24} & -\frac{\rho ^{8/3}}{24} \\
 -\frac{3 \rho }{8} & -\frac{\rho ^{4/3}}{12} & -\frac{\rho ^2}{8} & -\frac{\rho ^{8/3}}{6} & \frac{5 k_F^5}{36 \pi ^2} & \frac{5 k_F^5}{18 \pi ^2} & \frac{7 k_F^7}{18 \pi ^2} & \frac{7 k_F^7}{12 \pi ^2} & \frac{18 k_F^9}{5 \pi ^2} & \frac{24 k_F^9}{5 \pi ^2} & -\frac{3 \rho }{4} & -\frac{\rho ^{4/3}}{6} & -\frac{\rho ^2}{4} & -\frac{\rho ^{8/3}}{3} \\
 0 & -\frac{\rho ^{4/3}}{12} & -\frac{3 \rho ^2}{8} & -\frac{5 \rho ^{8/3}}{6} & \frac{5 k_F^5}{18 \pi ^2} & \frac{5 k_F^5}{9 \pi ^2} & \frac{14 k_F^7}{9 \pi ^2} & \frac{7 k_F^7}{3 \pi ^2} & \frac{108 k_F^9}{5 \pi ^2} & \frac{144 k_F^9}{5 \pi ^2} & 0 & -\frac{\rho ^{4/3}}{6} & -\frac{3 \rho ^2}{4} & -\frac{5 \rho ^{8/3}}{3} \\
 0 & \frac{\rho ^{4/3}}{6} & 0 & -\frac{5 \rho ^{8/3}}{3} & -\frac{5 k_F^5}{18 \pi ^2} & -\frac{5 k_F^5}{9 \pi ^2} & \frac{14 k_F^7}{9 \pi ^2} & \frac{7 k_F^7}{3 \pi ^2} & \frac{324 k_F^9}{5 \pi ^2} & \frac{432 k_F^9}{5 \pi ^2} & 0 & \frac{\rho ^{4/3}}{3} & 0 & -\frac{10 \rho ^{8/3}}{3} \\
\end{array}
\right)
\end{equation}
being the representation matrix.

In practice, we usually encounter the macroscopic quantities, and subsequently employ them to inversely infer the values of parameters.
From Eq.~(\ref{eq:Q=MP+C}), we obtain
\begin{equation}
\label{eq:P=Mi(Q-C)}
\mathbf{P} = \mathbf{M}^{-1}(\rho) \cdot \mathbf{Q}(\rho) - \mathbf{C}^{\prime}(\rho)
\end{equation}
where
\begin{equation}
\label{eq:C_prime}
\begin{aligned}
\mathbf{C}^{\prime}(\rho) &= \mathbf{M}^{-1}(\rho) \cdot \mathbf{C}(\rho) \\
&= \left[ \frac{64 a^2 \hbar ^2}{25 m \rho ^{1/3}}, -\frac{72 a^2 \hbar ^2}{5 m \rho ^{2/3}},
\frac{24 a^2 \hbar ^2}{5 m \rho ^{4/3}}, -\frac{24 a^2 \hbar ^2}{25 m \rho ^2}, 0,0,0,0,0,0,
-\frac{256 a^2 \hbar ^2}{75 m \rho ^{1/3}},\frac{96 a^2 \hbar ^2}{5 m \rho ^{2/3}},
-\frac{32 a^2 \hbar ^2}{5 m \rho ^{4/3}},\frac{32 a^2 \hbar ^2}{25 m \rho ^2}
\right]^{\mathrm{T}},
\end{aligned}
\end{equation}
and
\begin{gather}
\mathbf{M}^{-1}(\rho)= \notag
\\
\renewcommand{\arraystretch}{2}
\left(
\begin{array}{cccccccccccccc}
 \frac{512}{15 \rho } & -\frac{472}{15 \rho ^2} & \frac{8}{5 \rho } & -\frac{8}{45 \rho } & \frac{8 a^2}{25 \rho ^{1/3}} & -\frac{208 a^3 k_{F} }{175}  & \frac{10936 a^6 \rho }{1125} & 0 & 0 & 0 & 0 & 0 & 0 & 0 \\
 -\frac{288}{\rho ^{4/3}} & \frac{288}{\rho ^{7/3}} & -\frac{16}{\rho ^{4/3}} & \frac{2}{\rho ^{4/3}} & -\frac{36 a^2}{5 \rho ^{2/3}} & \frac{792 a^4}{35} & -\frac{3756 a^4 k_{F}^2}{25}  & 0 & 0 & 0 & 0 & 0 & 0 & 0 \\
 \frac{128}{\rho ^2} & -\frac{128}{\rho ^3} & \frac{8}{\rho ^2} & -\frac{4}{3 \rho ^2} & -\frac{24 a^2}{5 \rho ^{4/3}} & -\frac{48 a^4}{7 \rho ^{2/3}} & \frac{1928 a^6}{15} & 0 & 0 & 0 & 0 & 0 & 0 & 0 \\
 -\frac{144}{5 \rho ^{8/3}} & \frac{144}{5 \rho ^{11/3}} & -\frac{8}{5 \rho ^{8/3}} & \frac{2}{5 \rho ^{8/3}} & \frac{12 a^2}{25 \rho ^2} & -\frac{1032 a^4}{175 \rho ^{4/3}} & -\frac{6684 a^6}{125 \rho ^{2/3}} & 0 & 0 & 0 & 0 & 0 & 0 & 0 \\
 0 & 0 & 0 & 0 & \frac{8}{\rho } & -\frac{16 a^2}{\rho ^{1/3}} & \frac{216 a^3 k_{F}}{5} & -\frac{8}{\rho } & \frac{80 a^2}{3 \rho ^{1/3}} & -\frac{392 a^3 k_{F} }{3}  & 0 & 0 & 0 & 0 \\
 0 & 0 & 0 & 0 & 0 & 0 & 0 & \frac{16}{\rho } & -\frac{160 a^2}{3 \rho ^{1/3}} & \frac{784a^3 k_{F}}{3}  & 0 & 0 & 0 & 0 \\
 0 & 0 & 0 & 0 & 0 & \frac{16}{\rho } & -\frac{336 a^2}{5 \rho ^{1/3}} & 0 & -\frac{16}{\rho } & \frac{112 a^2}{\rho ^{1/3}} & 0 & 0 & 0 & 0 \\
 0 & 0 & 0 & 0 & 0 & 0 & 0 & 0 & \frac{32}{\rho } & -\frac{224 a^2}{\rho ^{1/3}} & 0 & 0 & 0 & 0 \\
 0 & 0 & 0 & 0 & 0 & 0 & \frac{8}{\rho } & 0 & 0 & -\frac{8}{\rho } & 0 & 0 & 0 & 0 \\
 0 & 0 & 0 & 0 & 0 & 0 & 0 & 0 & 0 & \frac{16}{\rho } & 0 & 0 & 0 & 0 \\
 -\frac{256}{15 \rho } & \frac{236}{15 \rho ^2} & -\frac{4}{5 \rho } & \frac{4}{45 \rho } & -\frac{32 a^2}{75 \rho ^{1/3}} & \frac{384 a^3 k_{F} }{175} & -\frac{33728 a^6 \rho }{1125} & -\frac{4 a^2}{5 \rho ^{1/3}} & \frac{24 a^3 k_{F}}{5}  & -60 a^6 \rho  & -\frac{256}{5 \rho } & \frac{236}{15 \rho } & -\frac{12}{5 \rho } & \frac{4}{15 \rho } \\
 \frac{144}{\rho ^{4/3}} & -\frac{144}{\rho ^{7/3}} & \frac{8}{\rho ^{4/3}} & -\frac{1}{\rho ^{4/3}} & \frac{48 a^2}{5 \rho ^{2/3}} & -\frac{1376 a^4}{35} & \frac{10848 a^4 k_{F}^2 }{25}  & \frac{18 a^2}{\rho ^{2/3}} & -92 a^4 & 950 a^6 \rho ^{2/3} & \frac{432}{\rho ^{4/3}} & -\frac{144}{\rho ^{4/3}} & \frac{24}{\rho ^{4/3}} & -\frac{3}{\rho ^{4/3}} \\
 -\frac{64}{\rho ^2} & \frac{64}{\rho ^3} & -\frac{4}{\rho ^2} & \frac{2}{3 \rho ^2} & \frac{32 a^2}{5 \rho ^{4/3}} & \frac{192 a^4}{7 \rho ^{2/3}} & -\frac{6976 a^6}{15} & \frac{12 a^2}{\rho ^{4/3}} & \frac{24 a^4}{\rho ^{2/3}} & -732 a^6 & -\frac{192}{\rho ^2} & \frac{64}{\rho ^2} & -\frac{12}{\rho ^2} & \frac{2}{\rho ^2} \\
 \frac{72}{5 \rho ^{8/3}} & -\frac{72}{5 \rho ^{11/3}} & \frac{4}{5 \rho ^{8/3}} & -\frac{1}{5 \rho ^{8/3}} & -\frac{16 a^2}{25 \rho ^2} & \frac{2336 a^4}{175 \rho ^{4/3}} & \frac{30432 a^6}{125 \rho ^{2/3}} & -\frac{6 a^2}{5 \rho ^2} & \frac{116 a^4}{5 \rho ^{4/3}} & \frac{278 a^6}{\rho ^{2/3}} & \frac{216}{5 \rho ^{8/3}} & -\frac{72}{5 \rho ^{8/3}} & \frac{12}{5 \rho ^{8/3}} & -\frac{3}{5 \rho ^{8/3}} \\
\end{array}
\right). \label{eq:M_inverse}
\end{gather}

\section{Macroscopic quantities at saturation density}
\label{sec:Appendix_Exp_rho0}
The values of the macroscopic quantities at saturation density $\rho_0 = 0.16 \, \mathrm{fm}^{-3}$ are used in the fitting procedure.
Here we provide the values of the elements in $\mathbf{M}(\rho_0)$, $\mathbf{C}(\rho_0)$, $\mathbf{M}^{-1}(\rho_0)$ and $\mathbf{C}^{\prime}(\rho_0)$.

$\mathbf{M}(\rho_0)$ and $\mathbf{C}(\rho_0)$ can be calculated as
\begin{gather}
\mathbf{M}(\rho_0)= \notag
\\
\hspace{-4.0pc}
\renewcommand{\arraystretch}{2}
\setlength{\arraycolsep}{1.5pt}
\left(
\begin{array}{cccccccccccccc}
 6.00\mathrm{E} \text{-} 2 & 5.43\mathrm{E} \text{-} 3 & 1.60\mathrm{E} \text{-} 3 & 4.72\mathrm{E} \text{-} 4 & 2.13\mathrm{E} \text{-} 2 & 1.07\mathrm{E} \text{-} 2 & 3.25\mathrm{E} \text{-} 2 & 1.62\mathrm{E} \text{-} 2 & 2.39\mathrm{E} \text{-} 1 & 1.20\mathrm{E} \text{-} 1 & 0 & 0 & 0 & 0 \\
 9.60\mathrm{E} \text{-} 3 & 1.16\mathrm{E} \text{-} 3 & 5.12\mathrm{E} \text{-} 4 & 2.01\mathrm{E} \text{-} 4 & 5.69\mathrm{E} \text{-} 3 & 2.84\mathrm{E} \text{-} 3 & 1.21\mathrm{E} \text{-} 2 & 6.06\mathrm{E} \text{-} 3 & 1.15\mathrm{E} \text{-} 1 & 5.75\mathrm{E} \text{-} 2 & 0 & 0 & 0 & 0 \\
 0 & 2.17\mathrm{E} \text{-} 2 & 2.88\mathrm{E} \text{-} 2 & 1.89\mathrm{E} \text{-} 2 & 2.13\mathrm{E} \text{-} 1 & 1.07\mathrm{E} \text{-} 1 & 9.09\mathrm{E} \text{-} 1 & 4.55\mathrm{E} \text{-} 1 & 1.29\mathrm{E} 1 & 6.46 & 0 & 0 & 0 & 0 \\
 0 & -4.34\mathrm{E} \text{-} 2 & 0 & 3.77\mathrm{E} \text{-} 2 & -2.13\mathrm{E} \text{-} 1 & -1.07\mathrm{E} \text{-} 1 & 9.09\mathrm{E} \text{-} 1 & 4.55\mathrm{E} \text{-} 1 & 3.88\mathrm{E} 1 & 1.94\mathrm{E} 1 & 0 & 0 & 0 & 0 \\
 0 & 0 & 0 & 0 & 2.00\mathrm{E} \text{-} 2 & 1.00\mathrm{E} \text{-} 2 & 3.55\mathrm{E} \text{-} 2 & 1.78\mathrm{E} \text{-} 2 & 1.89\mathrm{E} \text{-} 1 & 9.47\mathrm{E} \text{-} 2 & 0 & 0 & 0 & 0 \\
 0 & 0 & 0 & 0 & 0 & 0 & 1.00\mathrm{E} \text{-} 2 & 5.00\mathrm{E} \text{-} 3 & 1.49\mathrm{E} \text{-} 1 & 7.46\mathrm{E} \text{-} 2 & 0 & 0 & 0 & 0 \\
 0 & 0 & 0 & 0 & 0 & 0 & 0 & 0 & 2.00\mathrm{E} \text{-} 2 & 1.00\mathrm{E} \text{-} 2 & 0 & 0 & 0 & 0 \\
 0 & 0 & 0 & 0 & 0 & 1.00\mathrm{E} \text{-} 2 & 0 & 2.96\mathrm{E} \text{-} 2 & 0 & 2.21\mathrm{E} \text{-} 1 & 0 & 0 & 0 & 0 \\
 0 & 0 & 0 & 0 & 0 & 0 & 0 & 5.00\mathrm{E} \text{-} 3 & 0 & 1.24\mathrm{E} \text{-} 1 & 0 & 0 & 0 & 0 \\
 0 & 0 & 0 & 0 & 0 & 0 & 0 & 0 & 0 & 1.00\mathrm{E} \text{-} 2 & 0 & 0 & 0 & 0 \\
 -2.00\mathrm{E} \text{-} 2 & -1.81\mathrm{E} \text{-} 3 & -5.33\mathrm{E} \text{-} 4 & -1.57\mathrm{E} \text{-} 4 & 1.18\mathrm{E} \text{-} 2 & 2.37\mathrm{E} \text{-} 2 & 4.21\mathrm{E} \text{-} 2 & 6.32\mathrm{E} \text{-} 2 & 5.39\mathrm{E} \text{-} 1 & 7.18\mathrm{E} \text{-} 1 & -4.00\mathrm{E} \text{-} 2 & -3.62\mathrm{E} \text{-} 3 & -1.07\mathrm{E} \text{-} 3 & -3.14\mathrm{E} \text{-} 4 \\
 -6.00\mathrm{E} \text{-} 2 & -7.24\mathrm{E} \text{-} 3 & -3.20\mathrm{E} \text{-} 3 & -1.26\mathrm{E} \text{-} 3 & 5.92\mathrm{E} \text{-} 2 & 1.18\mathrm{E} \text{-} 1 & 2.95\mathrm{E} \text{-} 1 & 4.42\mathrm{E} \text{-} 1 & 4.85 & 6.46 & -1.20\mathrm{E} \text{-} 1 & -1.45\mathrm{E} \text{-} 2 & -6.40\mathrm{E} \text{-} 3 & -2.51\mathrm{E} \text{-} 3 \\
 0 & -7.24\mathrm{E} \text{-} 3 & -9.60\mathrm{E} \text{-} 3 & -6.29\mathrm{E} \text{-} 3 & 1.18\mathrm{E} \text{-} 1 & 2.37\mathrm{E} \text{-} 1 & 1.18 & 1.77 & 2.91\mathrm{E} 1 & 3.88\mathrm{E} 1 & 0 & -1.45\mathrm{E} \text{-} 2 & -1.92\mathrm{E} \text{-} 2 & -1.26\mathrm{E} \text{-} 2 \\
 0 & 1.45\mathrm{E} \text{-} 2 & 0 & -1.26\mathrm{E} \text{-} 2 & -1.18\mathrm{E} \text{-} 1 & -2.37\mathrm{E} \text{-} 1 & 1.18 & 1.77 & 8.73\mathrm{E} 1 & 1.16\mathrm{E}2 & 0 & 2.90\mathrm{E} \text{-} 2 & 0 & -2.51\mathrm{E} \text{-} 2 \\
\end{array}
\right), \label{eq:M_rho0}
\end{gather}
and
\begin{equation}
\label{eq:C_rho0}
\mathbf{C}(\rho_0) = \left[ 22.1,~2.36,~-44.2,~177,~0,~0,~0,~0,~0,~0,~12.3,~24.6,~-24.6,~98.3 \right]^{\mathrm{T}}.
\end{equation}


$\mathbf{M}^{-1}(\rho_0)$ and $\mathbf{C}^{\prime}(\rho_0)$ can be obtained as
\begin{gather}
\mathbf{M}^{-1}(\rho_0)= \notag
\\
\hspace{-3.3pc}
\renewcommand{\arraystretch}{2}
\setlength{\arraycolsep}{1.5pt}
\left(
\begin{array}{cccccccccccccc}
 2.13\mathrm{E} 2 & -1.23\mathrm{E} 3 & 1.00\mathrm{E} 1 & -1.11 & 3.55 & -2.35\mathrm{E} 1 & 3.41\mathrm{E} 2 & 0 & 0 & 0 & 0 & 0 & 0 & 0 \\
 -3.32\mathrm{E} 3 & 2.07\mathrm{E} 4 & -1.84\mathrm{E} 2 & 2.30\mathrm{E} 1 & -1.47\mathrm{E} 2 & 8.23\mathrm{E} 2 & -9.70\mathrm{E} 3 & 0 & 0 & 0 & 0 & 0 & 0 & 0 \\
 5.00\mathrm{E} 3 & -3.12\mathrm{E} 4 & 3.13\mathrm{E} 2 & -5.21\mathrm{E} 1 & -3.33\mathrm{E} 2 & -8.46\mathrm{E} 2 & 2.82\mathrm{E} 4 & 0 & 0 & 0 & 0 & 0 & 0 & 0 \\
 -3.82\mathrm{E} 3 & 2.39\mathrm{E} 4 & -2.12\mathrm{E} 2 & 5.30\mathrm{E} 1 & 1.13\mathrm{E} 2 & -2.47\mathrm{E} 3 & -3.98\mathrm{E} 4 & 0 & 0 & 0 & 0 & 0 & 0 & 0 \\
 0 & 0 & 0 & 0 & 5.00\mathrm{E} 1 & -1.78\mathrm{E} 2 & 8.53\mathrm{E} 2 & -5.00\mathrm{E} 1 & 2.96\mathrm{E} 2 & -2.58\mathrm{E} 3 & 0 & 0 & 0 & 0 \\
 0 & 0 & 0 & 0 & 0 & 0 & 0 & 1.00\mathrm{E} 2 & -5.92\mathrm{E} 2 & 5.16\mathrm{E} 3 & 0 & 0 & 0 & 0 \\
 0 & 0 & 0 & 0 & 0 & 1.00\mathrm{E} 2 & -7.46\mathrm{E} 2 & 0 & -1.00\mathrm{E} 2 & 1.24\mathrm{E} 3 & 0 & 0 & 0 & 0 \\
 0 & 0 & 0 & 0 & 0 & 0 & 0 & 0 & 2.00\mathrm{E} 2 & -2.49\mathrm{E} 3 & 0 & 0 & 0 & 0 \\
 0 & 0 & 0 & 0 & 0 & 0 & 5.00\mathrm{E} 1 & 0 & 0 & -5.00\mathrm{E} 1 & 0 & 0 & 0 & 0 \\
 0 & 0 & 0 & 0 & 0 & 0 & 0 & 0 & 0 & 1.00\mathrm{E} 2 & 0 & 0 & 0 & 0 \\
 -1.07\mathrm{E} 2 & 6.15\mathrm{E} 2 & -5.00 & 5.56\mathrm{E} \text{-} 1 & -4.74 & 4.33\mathrm{E} 1 & -1.05\mathrm{E} 3 & -8.88 & 9.47\mathrm{E} 1 & -2.10\mathrm{E} 3 & -3.20\mathrm{E} 2 & 9.83\mathrm{E} 1 & -1.50\mathrm{E} 1 & 1.67 \\
 1.66\mathrm{E} 3 & -1.04\mathrm{E} 4 & 9.21\mathrm{E} 1 & -1.15\mathrm{E} 1 & 1.96\mathrm{E} 2 & -1.43\mathrm{E} 3 & 2.80\mathrm{E} 4 & 3.68\mathrm{E} 2 & -3.34\mathrm{E} 3 & 6.14\mathrm{E} 4 & 4.97\mathrm{E} 3 & -1.66\mathrm{E} 3 & 2.76\mathrm{E} 2 & -3.45\mathrm{E} 1 \\
 -2.50\mathrm{E} 3 & 1.56\mathrm{E} 4 & -1.56\mathrm{E} 2 & 2.60\mathrm{E} 1 & 4.44\mathrm{E} 2 & 3.38\mathrm{E} 3 & -1.02\mathrm{E} 5 & 8.33\mathrm{E} 2 & 2.96\mathrm{E} 3 & -1.60\mathrm{E} 5 & -7.50\mathrm{E} 3 & 2.50\mathrm{E} 3 & -4.69\mathrm{E} 2 & 7.81\mathrm{E} 1 \\
 1.91\mathrm{E} 3 & -1.19\mathrm{E} 4 & 1.06\mathrm{E} 2 & -2.65\mathrm{E} 1 & -1.51\mathrm{E} 2 & 5.59\mathrm{E} 3 & 1.81\mathrm{E} 5 & -2.83\mathrm{E} 2 & 9.71\mathrm{E} 3 & 2.07\mathrm{E} 5 & 5.73\mathrm{E} 3 & -1.91\mathrm{E} 3 & 3.18\mathrm{E} 2 & -7.95\mathrm{E} 1 \\
\end{array}
\right), \label{eq:M_inv_rho0}
\end{gather}
and
\begin{equation}
\label{eq:C_prime_rho0}
\mathbf{C}^{\prime}(\rho_0) = \left[
1180,~12200,~-13800,~9380,0~,0~,0~,0~,0~,0~,1570~,16300~,18400~,-12500
\right]^{\mathrm{T}}.
\end{equation}
\\[1pc]

Readers can solve for the corresponding set of parameter values based on a given set of macroscopic quantities through Eq.~(\ref{eq:P=Mi(Q-C)}), and vice versa through Eq.~(\ref{eq:Q=MP+C}).
The units of the parameters and macroscopic quantities are presented in Table~\ref{tab:10SetsParas} and Table~\ref{tab:10SetsQuants}.

\section{The fourth-order symmetry energy, kurtosis coefficients, linear isospin splitting coefficient and the nucleon effective masses}
\label{sec:Appendix_Other}
In addition to the quantities used in the fitting procedure, for completeness, we hereby present the expressions of $E_{\mathrm{sym},4}(\rho)$, $I_{0}$, $I_{\mathrm{sym}}$, $\Delta m_1^{\ast}(\rho)$, $m_{s}$, and $m_{v}$. Furthermore, we discuss the relationship between $E_{\mathrm{sym},4}(\rho)$, $\Delta m_1^{\ast}(\rho)$ and $m_{s}$, $m_{v}$.

The fourth-order symmetry energy defined in Eq.~(\ref{eq:Esym4}) can be expressed as
\begin{equation}
\label{eq:Esym4_exp}
\begin{aligned}
E_{\mathrm{sym},4}(\rho) & \equiv \left.\frac{1}{4 !} \frac{\partial^4 E(\rho, \delta)}{\partial \delta^4}\right|_{\delta=0} \\
&= \frac{\hbar^2}{162m} a^{2} \rho^{2/3}
+ \frac{1}{648} a^{2} \rho^{5/3} \left( C^{[2]} - D^{[2]} \right)
+ \frac{1}{648} a^{4} \rho^{7/3} \left(8 C^{[4]} + 3 D^{[4]} \right)
+ \frac{2}{135} a^{6} \rho^{3} \left(13 C^{[6]} + 9 D^{[6]} \right) .
\end{aligned}
\end{equation}
The kurtosis coefficients of $E_{0}(\rho)$ and $E_{\mathrm{sym}}(\rho)$ are expressed as
\begin{equation}
\label{eq:I0_exp}
\begin{aligned}
I_{0}(\rho)  \equiv \, & 81 \rho^4 \frac{d^4 E_0(\rho)}{d \rho^4} \\
=& -\frac{84 \hbar^2}{5m} a^2 \rho^{2/3}
+ \sum_{n=1,3,5} \left[ \frac{t_3^{[n]}}{16}(n-6)(n-3)n(n+3) \rho^{\frac{n}{3} +1} \right] \\
&+ \frac{3}{2} a^2 \rho^{5/3} \left( 2 C^{[2]} + D^{[2]} \right)
- \frac{9}{5} a^4 \rho^{7/3} \left( 2 C^{[4]} + D^{[4]} \right) ,
\end{aligned}
\end{equation}
and
\begin{equation}
\label{eq:Isym_exp}
\begin{aligned}
I_{\mathrm{sym}}(\rho) \equiv \, & 81 \rho^4 \frac{d^4 E_{\mathrm{sym}}(\rho)}{d \rho^4} \\
=& -\frac{28 \hbar^2}{3m} a^2 \rho^{2/3}
-  \sum_{n=1,3,5} \left[ \frac{t_3^{[n]}}{48} \left( 2 x_3^{[n]} + 1 \right) (n-6)(n-3)n(n+3) \rho^{\frac{n}{3} +1} \right] \\
& + \frac{5}{3} a^2 \rho^{5/3} \left( C^{[2]} + 2 D^{[2]} \right)
- \frac{7}{3} a^4 \rho^{7/3} \left( 2 C^{[4]} + 3 D^{[4]} \right) ,
\end{aligned}
\end{equation}
respectively.
The isospin splitting coefficients $\Delta m_{2 n-1}^{\ast}(\rho)$ are defined in Eq.~(\ref{eq:spl_coes}), while the linear coefficient can be expressed as
\begin{equation}
\label{eq:spl_linear}
\begin{aligned}
\Delta m_1^{\ast}(\rho) & \equiv \left. \frac{\partial \mspl(\rho,\delta)}{\partial \delta} \right|_{\delta=0}\\
&= -\frac{80 m \hbar^2 \left[ 15 \rho D^{[2]} + 10 a^2 \rho^{5/3} \left( 2 C^{[4]} + 5 D^{[4]}  \right)
+ 72 a^4 \rho^{7/3} \left( 4 C^{[6]} + 7 D^{[6]}  \right)
\right]}
{3 \left[ 40\hbar^2 + 5m \rho \left(2 C^{[2]} + D^{[2]} \right) + 10 m a^2 \rho^{5/3} \left( 2 C^{[4]} + D^{[4]}  \right) + 72 m a^4 \rho^{7/3} \left( 2 C^{[6]} + D^{[6]}  \right) \right]^2 } .
\end{aligned}
\end{equation}

The isoscalar nucleon effective mass $m_{s}^{\ast}$ and isovector nucleon effective mass $m_{v}^{\ast}$ are momentum dependent in the N3LO Skyrme pseudopotential interactions.
We define $\Tilde{M}_{s} \equiv m/m_{s}$ and $\Tilde{M}_{v} \equiv m/m_{v}$.
Thus we have
\begin{equation}
\begin{aligned}
\label{eq:Ms_exp}
\Tilde{M}_{s}(\rho,p) =  1+ \frac{m}{p} \frac{d U_0(\rho,p)}{d p}
 = & 1 + \frac{m}{8 \hbar^2} \rho \left( 2 C^{[2]} + D^{[2]}  \right)
+ \frac{m}{8 \hbar^2} a^2 \rho^{5/3} \left( 2 C^{[4]} + D^{[4]}  \right)
+ \frac{3 m}{8 \hbar^2} a^4 \rho^{7/3} \left( 2 C^{[6]} + D^{[6]}  \right) \\
& + \frac{p^2}{\hbar^2}
\left[
\frac{m}{8 \hbar^2} \rho \left( 2 C^{[4]} + D^{[4]} \right)
+  \frac{21 m}{20 \hbar^2} a^2 \rho^{5/3} \left( 2 C^{[6]} + D^{[6]}  \right)
\right] \\
& + \frac{p^4}{\hbar^4} \left[ \frac{3 m}{8 \hbar^2} \rho \left( 2 C^{[6]} + D^{[6]}  \right) \right] ,
\end{aligned}
\end{equation}
and
\begin{equation}
\begin{aligned}
\label{eq:Mv_exp}
\Tilde{M}_{v}(\rho,p) = &   1+ \frac{m}{p} \frac{d U_{\tau}(\rho,-\tau,p)}{d p} =1 + \frac{m}{4 \hbar^2} \rho C^{[2]} + \frac{m}{4 \hbar^2} 2^{2/3} a^2 \rho^{5/3} C^{[4]}
+ \frac{m}{4 \hbar^2} 2^{4/3} a^4 \rho^{7/3} C^{[6]} \\
& + \frac{p^2}{\hbar^2}
\left[
\frac{m}{4 \hbar^2} \rho C^{[4]} + \frac{21 m}{20 \hbar^2} 2^{2/3} a^2 \rho^{5/3} C^{[6]}
\right]
+ \frac{p^4}{\hbar^4} \left[ \frac{3 m}{4 \hbar^2} \rho C^{[6]}  \right],
\end{aligned}
\end{equation}
with $\tau = 1\, [-1]$ for neutron [proton].
The derivatives of $\Tilde{M}_{s}$ and $\Tilde{M}_{v}$ with respect to momentum are expressed as
\begin{equation}
\begin{aligned}
\label{eq:d2Ms_exp}
\frac{d^2 \Tilde{M}_{s}(\rho,p)}{d p^2} =
\frac{2!}{\hbar^2}
\left[
\frac{m}{8 \hbar^2} \rho \left( 2 C^{[4]} + D^{[4]} \right)
+  \frac{21 m}{20 \hbar^2} a^2 \rho^{5/3} \left( 2 C^{[6]} + D^{[6]}  \right)
\right]
+ \frac{12 p^2}{\hbar^4} \left[ \frac{3 m}{8 \hbar^2} \rho \left( 2 C^{[6]} + D^{[6]}  \right) \right],
\end{aligned}
\end{equation}
\begin{equation}
\begin{aligned}
\label{eq:d4Ms_exp}
\frac{d^4 \Tilde{M}_{s}(\rho,p)}{d p^4} =
\frac{4!}{\hbar^4} \left[ \frac{3 m}{8 \hbar^2} \rho \left( 2 C^{[6]} + D^{[6]}  \right) \right],
\end{aligned}
\end{equation}
\begin{equation}
\begin{aligned}
\label{eq:d2Mv_exp}
\frac{d^2 \Tilde{M}_{v}(\rho,p)}{d p^2} =
\frac{2!}{\hbar^2}
\left[
\frac{m}{4 \hbar^2} \rho C^{[4]} + \frac{21 m}{20 \hbar^2} 2^{2/3} a^2 \rho^{5/3} C^{[6]}
\right]
+ \frac{12 p^2}{\hbar^4} \left[ \frac{3 m}{4 \hbar^2} \rho C^{[6]}  \right],
\end{aligned}
\end{equation}
and
\begin{equation}
\begin{aligned}
\label{eq:d4Mv_exp}
\frac{d^4 \Tilde{M}_{v}(\rho,p)}{d p^4} =
\frac{4!}{\hbar^4} \left[ \frac{3 m}{4 \hbar^2} \rho C^{[6]}  \right] .
\end{aligned}
\end{equation}

Compare Eqs.~(\ref{eq:Ms_exp})-(\ref{eq:d4Mv_exp}) with Eq.~(\ref{eq:Esym4_exp}) and Eq.~(\ref{eq:spl_linear}), we can obtain
\begin{equation}
\label{eq:Esym4_msmv}
\begin{aligned}
E_{\mathrm{sym},4}(\rho) = &
\frac{\hbar^2}{162 m} a^2 \rho^{2/3} \left[ 3\Tilde{M}_{v}(\rho,p=0) - 2 \Tilde{M}_{s}(\rho,p=p_{F}) \right] \\
& + \frac{\hbar^4}{324 m} a^4 \rho^{4/3}
\left[
(2-3 \cdot 2^{2/3}) \left. \frac{d^2 \Tilde{M}_{v}(\rho,p)}{d p^2} \right|_{p=0}
+ 10 \left. \frac{d^2 \Tilde{M}_{s}(\rho,p)}{d p^2} \right|_{p=p_{F}}
\right] \\
& + \frac{\hbar^6}{2430 m} a^6 \rho^2
\left[
\frac{27 \cdot 2^{1/3} - 14 \cdot 2^{2/3} -40}{4} \frac{d^4 \Tilde{M}_{v}(\rho,p)}{d p^4}
- 13 \frac{d^4 \Tilde{M}_{s}(\rho,p)}{d p^4}
\right] ,
\end{aligned}
\end{equation}
and
\begin{equation}
\label{eq:Dm1_msmv}
\begin{aligned}
\Delta m_1^{\ast}(\rho) =& \frac{1}{3  \left[ \Tilde{M}_{s}(\rho,p=p_{F}) \right]^2 }
\Bigg\{
6\left[ \Tilde{M}_{v}(\rho,p=0) -  \Tilde{M}_{s}(\rho,p=p_{F})  \right]  \\
&+ \hbar^2 a^2 \rho^{2/3} \left[
(8-3 \cdot 2^{2/3}) \left. \frac{d^2 \Tilde{M}_{v}(\rho,p)}{d p^2} \right|_{p=0}
-4 \left. \frac{d^2 \Tilde{M}_{s}(\rho,p)}{d p^2} \right|_{p=p_{F}}
\right] \\
&+ \left. \hbar^4 a^4 \rho^{4/3} \left[
\frac{27 \cdot 2^{1/3} - 56 \cdot 2^{2/3} +60}{30} \frac{d^4 \Tilde{M}_{v}(\rho,p)}{d p^4}
+ \frac{4}{3} \frac{d^4 \Tilde{M}_{s}(\rho,p)}{d p^4}
\right]
  \right\},
\end{aligned}
\end{equation}
where $p_F= \hbar (3 \pi^2 \rho/2 )^{1/3}$.

In standard SHF and eSHF models, where $C^{[4]}$, $C^{[6]}$, $D^{[4]}$ and $D^{[6]}$ all equal zero, $m_{s}^{\ast}$ and $m_{v}^{\ast}$ are independent of momentum.
Consequently, Eq.~(\ref{eq:Esym4_msmv}) and Eq.~(\ref{eq:Dm1_msmv}) reduce to the very straightforward forms (i.e., Eq.~(33) in Ref.~\cite{Pu:2017kjx} and Eq.~(8) in Ref.~\cite{Zhang:2015qdp}, respectively):
\begin{equation}
\label{eq:Esym4_msmv_reduce}
E_{\mathrm{sym},4}(\rho) = \frac{\hbar^2}{162 m} a^2 \rho^{2/3} \left[\frac{3 m}{m_v(\rho)} - \frac{2 m}{m_s(\rho)}\right] ,
\end{equation}
and
\begin{equation}
\label{eq:Dm1_msmv_reduce}
\begin{aligned}
\Delta m_1^{\ast}(\rho) = 2 \frac{m_s(\rho)}{m}
\left[ \frac{m_s(\rho)}{m_v(\rho)} - 1 \right] .
\end{aligned}
\end{equation}

\end{widetext}


\bibliography{ImSkyEP}

\end{document}